\documentclass[11pt,a4paper]{article}

\usepackage{graphicx}
\usepackage{afterpage}
\usepackage{epsfig,cite}
\usepackage{amssymb}
\usepackage{amsmath}
\usepackage{dsfont}
\usepackage{multirow}
\usepackage{url,hyperref}

\usepackage{url}
\usepackage{hyperref}

\def\smallfrac#1#2{\hbox{${{#1}\over {#2}}$}}
\def\half{\hbox{${1\over 2}$}}
\newcommand{\be}{\begin{equation}}
\newcommand{\ee}{\end{equation}}
\newcommand{\bea}{\begin{eqnarray}}
\newcommand{\eea}{\end{eqnarray}}
\newcommand{\bi}{\begin{itemize}}
\newcommand{\ei}{\end{itemize}}
\newcommand{\ben}{\begin{enumerate}}
\newcommand{\een}{\end{enumerate}}

\def\frac#1#2{{{#1}\over {#2}}}
\def\gsim{\mathrel{\rlap{\lower4pt\hbox{\hskip1pt$\sim$}}
    \raise1pt\hbox{$>$}}}         
\def\lsim{\mathrel{\rlap{\lower4pt\hbox{\hskip1pt$\sim$}}
    \raise1pt\hbox{$<$}}}         

\newcommand{\rep}{\mathrm{rep}}

\newcommand{\draft}[1]{}

\def\beq{\begin{equation}}  
\def\eeq{\end{equation}}  

\def \n0{N_j^{(0)}}

\def\lapprox{\lower .7ex\hbox{$\;\stackrel{\textstyle <}{\sim}\;$}}
\def\gapprox{\lower .7ex\hbox{$\;\stackrel{\textstyle >}{\sim}\;$}}
\def\half{\smallfrac{1}{2}}

\begin{document}
\begin{flushright}
Edinburgh 2011/15\\
IFUM-981-FT\\
FR-PHENO-2011-013\\
RWTH TTK-11-32\\
\end{flushright}
\begin{center}
{\Large \bf Reweighting and Unweighting of Parton Distributions \\ 
and the LHC W lepton asymmetry data}
\vspace{0.4cm}
\end{center}

\begin{center}
{\bf The NNPDF Collaboration:}\\
Richard~D.~Ball$^{1}$, Valerio~Bertone$^2$, Francesco~Cerutti$^3$,
Luigi~Del~Debbio$^1$,\\ Stefano~Forte$^4$, Alberto~Guffanti$^{2,5}$, 
Nathan~P.~Hartland$^1$, Jos\'e~I.~Latorre$^3$, Juan~Rojo$^4$ 
and Maria~Ubiali$^6$.

\vspace{.3cm}
{\it ~$^1$ Tait Institute, University of Edinburgh,\\
JCMB, KB, Mayfield Rd, Edinburgh EH9 3JZ, Scotland\\
~$^2$  Physikalisches Institut, Albert-Ludwigs-Universit\"at Freiburg,\\ 
Hermann-Herder-Stra\ss e 3, D-79104 Freiburg i. B., Germany  \\
~$^3$ Departament d'Estructura i Constituents de la Mat\`eria, 
Universitat de Barcelona,\\ Diagonal 647, E-08028 Barcelona, Spain\\
~$^4$ Dipartimento di Fisica, Universit\`a di Milano and
INFN, Sezione di Milano,\\ Via Celoria 16, I-20133 Milano, Italy\\
~$^5$ The Niels Bohr International Academy and Discovery Center, \\
The Niels Bohr Institute, Blegdamsvej 17, DK-2100 Copenhagen, Denmark\\
~$^6$ Institut f\"ur Theoretische Teilchenphysik und Kosmologie, RWTH 
Aachen University,\\ D-52056 Aachen, Germany\\}
\end{center}

\vspace{0.6cm}

\begin{center}
{\bf \large Abstract:}
\end{center}

We develop in more detail our reweighting method for incorporating new
datasets in parton fits based on a Monte Carlo representation of
PDFs.  After revisiting the derivation of the reweighting formula,
we show how to construct an unweighted PDF replica set which is
statistically equivalent to a given reweighted set. We then use  
reweighting followed by unweighting to test the consistency 
of the method, specifically by verifying that results  do not depend on
the order in which new data are included in the fit via reweighting.
We apply the reweighting method to study the impact of LHC W lepton
asymmetry data on the NNPDF2.1 set. We show how these data reduce the
PDF uncertainties of light quarks in the medium and small $x$ region,
providing the first solid constraints on PDFs from LHC data.

\clearpage

\tableofcontents

\clearpage

\section{Introduction}

\label{sec-intro}

In a series of previous papers~\cite{Forte:2002fg,DelDebbio:2004qj,DelDebbio:2007ee,Ball:2008by,Rojo:2008ke,Ball:2009mk,Ball:2010de,Ball:2011mu,Ball:2011uy}, 
we constructed increasingly accurate 
sets of parton distributions (PDFs), using a Monte Carlo
approach coupled to the use of neural networks as underlying
interpolating functions. By definition, a PDF set provides a
representation of a probability density in the space of parton
distributions, i.e. a probability density in a space of 
functions~\cite{Giele:2001mr,phystat,Forte:2010dt}. We have performed
various tests that confirm that NNPDF parton sets do indeed behave in
a way which is consistent with the desired statistical properties of
functional probability densities.

An advantage of providing a Monte Carlo representation of this PDF
probability density is that
new information (such as might be provided by new experimental data) can be
included, using Bayes' theorem, by reweighting an existing PDF set,
without having to perform a new PDF
fit~\cite{Giele:1998gw,Ball:2010gb}: it is possible to determine
a reweighting factor for each Monte Carlo replica in such a way that
the information contained in the new data is included by simply
computing weighted averages.
This approach was first 
successfully developed and implemented in Ref.~\cite{Ball:2010gb},
where it was explicitly shown, in studies involving CDF and D0 inclusive 
jet data, that results obtained by reweighting are equivalent to
those found by including the new data in the fit. 

Reweighting takes a set of equally likely 
PDF replicas generated by importance sampling, and 
assigns to them weights reflecting their relative probabilities in the light of 
new data not included in the original fit. 
In this paper we develop a second technique, 
which we call `unweighting', which takes the reweighted set
and replaces it with a new set of 
replicas which are again all equally probable. This new 
set of replicas can then be used in precisely the same way as a 
fitted set. 
Even though no new information is gained by unweighting,
presenting reweighted PDFs in the same form as a corresponding 
refitted set has various obvious practical advantages. 

Furthermore, unweighting
allows us to perform a highly nontrivial test of the reweighting procedure:
namely, we take {\em two} 
new independent datasets, and use them to
sequentially improve an existing set of replicas. This may then be done
in either order, or  
indeed by treating them as one (combined) dataset.
All three methods
should yield equivalent 
results. Checking that this is the case provides a strong test of the
method. However this can only be done if after each reweighting we
unweight, because our simple closed-form
expression for the weights can only be used for the reweighting of an
equally probable (i.e. unweighted) set of PDFs. 

 We perform this check by first taking the NNPDF2.0
NLO DIS+DY fit~\cite{Ball:2010de}, based on deep-inelastic and
Drell-Yan data only, 
and taking as  
new datasets the CDF~\cite{Abulencia:2007ez} and D0~\cite{D0:2008hua} 
Run II inclusive jet data. This completes
and refines the studies of Ref.~\cite{Ball:2010gb}, where it was
verified that the inclusion of the combined CDF+D0 jet 
data by reweighting or refitting
gives equivalent results.
 We then perform a second check using 
as the prior the NNPDF2.1 DIS fit~\cite{Ball:2011mu}, based on
deep-inelastic data only,  
and taking as new datasets the E605~\cite{Moreno:1990sf}
Drell-Yan and Tevatron inclusive jet data. This provides a somewhat 
different
test, because while the D0 and CDF data used in the previous test 
measure the same observable in  the same kinematic region, the
Drell-Yan and jet data affect different PDFs in different kinematic 
regions.

Besides its practical usefulness, the combined reweighting plus unweighting procedure 
is important because it allows one, at least in principle, to perform a global PDF fit 
by sequentially including
new data by reweighting a generic prior distribution of PDFs~\cite{Giele:1998gw}. If the
information contained in the new data is sufficiently precise, and the prior distribution 
sufficently broad,
the results will the be largely independent of the
prior one starts from: this would then give completely unbiased
PDFs. In practice, this procedure is unlikely to be viable because, in order
to get accurate results, the prior set of PDF replicas would have to be
huge. However, the equivalence of  PDFs obtained from reweighting with
those determined using a fitting procedure (such as the NNPDF sets)
confirms that the latter are also unbiased.

Following the success of these consistency tests, we use reweighting 
to evaluate the impact on 
the NNPDF2.1 NLO fit Ref.\cite{Ball:2011mu} of recent LHC data on the
$W$-lepton asymmetry  
from the ATLAS and CMS collaborations. Using unweighting, we are able
to produce a new PDF set, NNPDF2.2, which 
incorporates the effect of these data and the older $W$-lepton asymmetry from D0.

The outline of this paper is as follows. In Sec.~\ref{sec-rw} we
revisit the derivation of the reweighting method,
in particular the determination 
of the weights in terms of the $\chi^2$ of the fit of the new data
to each replica, and we discuss some subtle issues that were not
tackled in Ref.~\cite{Ball:2010gb}, related to the definition of the
measure in data space and to the inclusion by reweighting of multiple
data sets. Then, 
in Sec.~\ref{sec-unw} we present our method of unweighting reweighted
PDF sets, to give a set of replicas which are all equally probable, 
and show that indeed the unweighted set is
equivalent to the original reweighted set. 
We follow this in Sec.~\ref{sec-con} with a study
of the consistency of the combined reweighting and unweighting
procedure, when applied to more  
than one dataset in turn. After this theoretical study, we turn
to phenomenology by using the method to investigate the impact
of LHC measurement of the $W$ lepton asymmetry on PDFs. 
First, we show in Sec.~\ref{sec-lhc} how these data reduce the
PDF uncertainties of light quarks in the medium and small--$x$ region,
providing the first solid constraints on PDFs from LHC data, and then
in Sect.~\ref{sec-tev+lhc}
we construct a new set of NLO PDFs, NNPDF2.2, which includes, on top
of all the data used to determine NNPDF2.1 PDFs, also the D0 $W$ 
asymmetry data already discussed in Ref.~\cite{Ball:2010gb} and the
LHC data discussed in Sect.~\ref{sec-lhc}.

\section{Reweighting}

\label{sec-rw}

In this section we revisit  the derivation 
of the weight formula for reweighting 
ensembles of PDFs. In particular we discuss some of the more subtle issues in 
the formal proof presented in Ref.~\cite{Ball:2010gb}. 
The derivation of the formula for the computation of the weights is nontrivial because we are
dealing with probability densities in multidimensional spaces. In
particular we need to avoid the ambiguities that can appear when dealing
with conditional probabilities with respect to an event of probability
zero, the so-called Borel-Kolmogorov paradox~\cite{Jaynes}. The conditional
probabilities need to be defined carefully as integrations of conditional 
probability densities over finite volumes, in the limit when these volumes are taken to zero.

\subsection{Integration over the data space}
Bayes' theorem can be stated in terms of probability densities:
\begin{equation}
  \label{eq:1}
  {\mathcal P}(f|y) \mathcal Df \, {\mathcal P}(y) d^ny = {\mathcal P}(y|f) d^ny\, {\mathcal P}(f) \mathcal Df\, ,
\end{equation}
where $\mathcal Df$ is the integration measure in the space of PDFs, and
$d^ny$ is the integration measure in the space of data. The latter is
an $n$-dimensional real space, where $n$ is the number of data points
used for reweighting. ${\mathcal P}(f)$ is the prior density in the space of
the PDFs: it is represented by the set $\{f_k\}$ of PDF
replicas. These are all equally probable, e.g., the expected PDF is
simply determined as the average over the set $\{f_k\}$, and are
determined by importance sampling by starting from experimental
data~\cite{phystat}. ${\mathcal P}(f|y)$ is instead the new probability density, given 
the $n$ data points $y$. Note that here, unlike in
Ref.~\cite{Ball:2010gb}, we do not make explicit the dependence  
of conditional probabilities on generic prior information $K$ 
(which includes the data used to 
determine the prior PDF, external parameters such as $\alpha_s$, and 
theoretical assumptions such as the use of perturbative QCD at a given
order).
${\mathcal P}(y)$ is the prior density in the space of data, and we do not need
to specify its explicit form, since it can be fixed by requiring
${\mathcal P}(f|y)$ to be correctly normalised. The only 
relevant property of ${\mathcal P}(y)$ is that it does not depend on the PDFs $f$. 

In order to {\em define} the probability density ${\mathcal P}(f|y)$ at a given point $y$,
we can integrate Eq.~(\ref{eq:1}) in a small sphere $S_\epsilon$ of radius $\epsilon$ centered
at $y$. Integrating the left-hand side  
of Eq.~(\ref{eq:1}) over $S_\epsilon$ we obtain
\begin{equation}
  \label{eq:2L}
  \int_{S_\epsilon} {\mathcal P}(f|y^\prime) \mathcal Df \, {\mathcal P}(y^\prime) d^ny^\prime = 
  \left[n^{-1}\epsilon^n \Omega_n {\mathcal P}(y)\right]\, {\mathcal P}(f|y) \mathcal Df  \, ,
\end{equation}
where $\Omega_n$ is the solid angle in $n$ dimensions. Integrating the
right-hand side similarly, we can cancel 
the volume factors on each side and thus take the limit $\epsilon\to 0$, to give
\begin{equation}
  \label{eq:3}
  {\mathcal P}(f|y) \mathcal Df =  \frac{{\mathcal P}(y|f)}{{\mathcal P}(y)} \, {\mathcal P}(f) \mathcal Df\, .
\end{equation}

Now ${\mathcal P}(y|f)$ is the likelihood density for the data $y$: assuming these data to be normally distributed 
about central values $y[f]$ (which of course depend on the PDF $f$),
\begin{equation}
  \label{eq:4}
  {\mathcal P}(y|f) = (2\pi)^{-n/2}({\rm det}\sigma)^{-1} \exp\big(-\half(y-y[f])\sigma^{-1}(y-y[f])\big)\, ,
\end{equation}
where $\sigma$ is the experimental covariance matrix. The only dependence on $f$ is through the value of 
\begin{equation}
  \label{eq:5}
  \chi^2(y,f) \equiv (y-y[f])\sigma^{-1}(y-y[f])\, . 
\end{equation}
It now follows from Eqs.~(\ref{eq:3}-\ref{eq:5}) that
\begin{equation}
  \label{eq:12}
  {\mathcal P}(f|y)\mathcal Df \propto \exp(-\half\chi(y,f)^2)\, {\mathcal P}(f)\mathcal Df,
\end{equation}
with a constant of proportionality that depends on $y$, but not on
$f$, and can thus be fixed 
if necessary through the normalization condition $\int {\mathcal P}(f|y)\mathcal Df =1$.

\subsection{Weights for a given $\chi$}

This is all fine so far as it goes, but is not sufficient to give us 
a reweighting of our ensemble of PDFs equivalent to a refitting.
The reason for this is that when we fit PDFs, we do not demand that the 
predictions $y[f]$ coincide with the data points $y$, but rather that 
the figure of merit $\chi^2(y,f)$ is optimized. Thus rather than integrating 
both sides of Eq.~(\ref{eq:1}) over the small spheres $S_\epsilon$, we 
should integrate over all $y$ subject only to the single constraint that $\chi^2(y,f)=\chi^2$, 
for some fixed value $\chi$. It is convenient to choose  as a
parameter $\chi$, rather
than $\chi^2$, because we can interpret $\chi$ as the radial
co-ordinate in a system of spherical polar co-ordinates in function
space, centered at
$y'=y[f]$.

The left-hand side of Eq.~(\ref{eq:1}) thus becomes 
\begin{equation}
  \label{eq:2Lchi}
  \int \delta(\chi-\chi(y',f)) {\mathcal P}(f|y^\prime) \mathcal Df \, {\mathcal P}(y^\prime) d^ny^\prime \propto 
  {\mathcal P}(f|\chi) \mathcal Df  \, ,
\end{equation}
thus defining ${\mathcal P}(f|\chi)$ up to an overall constant (independent of $f$). 
We can evaluate it by performing the same 
integration over the right-hand side of Eq.~(\ref{eq:1}), since the dependence on ${\mathcal P}(f)$ factorises:
\begin{equation}
  \label{eq:2Rchi}
  \int \delta(\chi-\chi(y',f)) {\mathcal P}(y^\prime|f) d^ny^\prime {\mathcal P}(f)\, \mathcal Df = 
  2^{1-n/2}(\Gamma(n/2))^{-1} \Omega_n\chi^{n-1}e^{-\half\chi^2}\, {\mathcal P}(f) \mathcal Df\, ,
\end{equation}
where we have used Eq.~(\ref{eq:4}) for the likelihood, and performed
the integral over $y'$ in spherical co-ordinates. 
 Comparing Eq.~(\ref{eq:2Lchi}) and 
Eq.~(\ref{eq:2Rchi}) we thus find 
\begin{equation}
  \label{eq:6x}
  {\mathcal P}(f|\chi) \mathcal Df \propto \chi^{(n-1)} e^{-\half\chi^2} {\mathcal P}(f)\mathcal Df \, .
\end{equation}

In order to define the weight to be associated to each replica, we need to define the probability 
for each replica by integrating the probability density
over a finite volume, and then send that volume to zero. For a given replica $f_k$ we thus
integrate $\chi'$ over the region $\chi_k<\chi^\prime<\chi_k+\epsilon$, where $\chi_k=\chi(y,f_k)$: 
\begin{equation}
  \label{eq:shelldef}
\int_{\chi_k}^{\chi_k+\epsilon} d\chi^\prime {\mathcal P}(f_k|\chi^\prime)=\epsilon {\mathcal P}(f_k|\chi_k)\, .
\end{equation}
Note that this corresponds to integrating Eq.~(\ref{eq:2Lchi}) over a
spherical shell, centered on $y[f_k]$, of radius $\chi_k$ and thickness
$\epsilon$. The thickness of the shell is independent of the choice of
replica: if it were not, we would bias the result.  

It is easy to see using Eq.~(\ref{eq:6x}) that Eq.~(\ref{eq:shelldef})
gives the formula derived in Ref.\cite{Ball:2010gb} for the weights:
since the replicas in the prior distribution all have equal
probability, ${\mathcal P}(f_k)$ is independent of the choice of
replica $f_k$, and the weights are
\begin{equation}
  \label{eq:13}
  w_k \propto {\mathcal P}(f_k|\chi_k) \propto \chi_k^{n-1} e^{-\half\chi_k^2}\, .
\end{equation}
The constant of proportionality may be fixed 
by normalizing the sum of the weights to the number of replicas. 

The factor of $\chi_k^{n-1}$
takes account of the fact that when there are many data points,
larger values of $\chi_k$ have a larger phase space available to them, while very small 
values are phase space suppressed: however good the model it is always very unlikely 
that the theoretical prediction will give exactly the right result for a large 
number of measurements.
This is not a trivial result: it depends critically on choosing the correct volume upon
which to integrate in the space of the new data $y$. Starting from the
same probability density, but using a different integration
volume would produce a different result. Hence we need to
justify our particular choice of volume.  

In this
respect, we note that our choice includes all points in the space of
$y$ with a particular $\chi^2$, and that the thickness of the shell is
independent of its radius $\chi(y,f)$ or centre $y[f]$, in the same way that in
Eq.~(\ref{eq:2L}) the radius of the little sphere was also independent
of $y[f]$. The ultimate justification in both cases is that the probability measure $d^ny$ on the space $y$ is
uniform, i.e. that equal volumes have equal probability: this assumption is of course 
implicit from the start, since without it the likelihood Eq.~(\ref{eq:4}) would not be Gaussian. 

Note that although the above argument is most naturally expressed
using $\chi$ as a co-ordinate in function space
 we would
get the same weights $w_k$ if we were to instead use $\chi^2$, or 
indeed a conditional dependence on any other monotonic function of
$\chi$, so long
as we use the same {\em volume} in the space of data to define the weights.
To see this, note that for example
\begin{equation}
  \label{eq:2Lchi2}
{\mathcal P}(f|\chi^2) \mathcal Df \propto \int \delta(\chi^2-\chi^2(y',f)) {\mathcal P}(f|y^\prime) \mathcal Df \, {\mathcal P}(y^\prime) d^ny^\prime \, ,
\end{equation}
so that, comparing with Eq.~(\ref{eq:2Lchi}),
\begin{equation}
  \label{eq:8}
  {\mathcal P}(f|\chi^2) = {\mathcal P}(f|\chi)/(2\chi)\, .
\end{equation}
As expected, we thus have ${\mathcal P}(f|\chi) d\chi = {\mathcal P}(f|\chi^2) d\chi^2$.
If we work with ${\mathcal P}(f|\chi^2)$, in order to be sure to use the same volume in the space of data 
(i.e. a spherical shell of thickness $\epsilon$) we must now integrate over the interval
$\chi_k^2<(\chi^\prime)^2 < \chi_k^2 + 2 \chi_k \epsilon$:
\begin{equation}
  \label{eq:11}
  w_k \propto
\int_{\chi_k^2}^{\chi_k^2+2\chi_k\epsilon} d\chi^{\prime 2}
  {\mathcal P}(f_k|\chi^{\prime 2})\, , 
\end{equation}
which then yields exactly the same weight Eq.~(\ref{eq:13}) as 
obtained using Eq.~(\ref{eq:shelldef}). 

\subsection{Multiple experiments}
\label{sec:mult-exp}

Let us now discuss the implications of the above prescription for
reweighting with more than one set of data. Suppose we are given 
a set of new data
$\{y\}$, which is made of two {\em independent} subsets $\{y_1\}$
and $\{y_2\}$, containing respectively  $n_1$ and $n_2$ data
points, such as for  example a dataset which includes results from two
independent experimental measurements  (of the same, or of
different observables).

When the two sets of data are used for reweighting simultaneously, the
only quantity that matters is the total $\chi^2$ of the two
experiments. Since we assumed the experiments to be independent, 
$\chi^2 = \chi_1^2 + \chi_2^2$, where $\chi_i\equiv\chi(y_i,f)$,
and the probability density is therefore given 
by Eq.~(\ref{eq:6x}) above:
\begin{equation}
  \label{eq:15}
  {\mathcal P}(f|\chi) \propto (\chi_1^2 + \chi_2^2)^{\half(n_1+n_2-1)} e^{-\half(\chi_1^2 + \chi_2^2)}\, .
\end{equation}

Clearly the individual values of $\chi^2$ of the two sets need not each be fixed
to $\chi_1^2$ and $\chi_2^2$. Hence even though the likelihood factorizes,
\begin{equation}
  \label{eq:likefact}
  {\mathcal P}(y_1 y_2|f) = {\mathcal P}(y_2|f) {\mathcal P}(y_1|f)\, ,
\end{equation}
the weights do not:
\begin{equation}
  \label{eq:16}
  {\mathcal P}(f|\chi) \neq {\mathcal P}(f|\chi_2) {\mathcal P}(f|\chi_1)\, .
\end{equation}
Instead they are determined through the more complicated relation (see Eqs.~(\ref{eq:2Lchi}) and~(\ref{eq:2Rchi})) 
\begin{equation}
  \label{eq:2Lchi12}
 {\mathcal P}(f|\chi)\propto \int \delta(\chi-(\chi^2_1+\chi^2_2)^{1/2})  
\, {\mathcal P}(y_2^\prime|f) d^{n_2}y_2^\prime  
\, {\mathcal P}(y_1^\prime|f) d^{n_1}y_1^\prime \, .
\end{equation}
With Gaussian likelihoods Eq.~(\ref{eq:4}), the integrals can be evaluated to give Eq.~(\ref{eq:15}). 

This means that if we wish to proceed sequentially, then after weighting with 
the first data set, with the usual weights $\chi_1^{n_1-1}\exp(-\half\chi^2_1)$, the weights 
for the second data set are not given by 
\begin{equation}
w_{2\,k}\propto\chi_{2\, k}^{n_2-1} \exp(-\half\chi^2_{2\,k}), 
\label{eq:2weight}
\end{equation}
but rather by 
\begin{equation}
w_{2|1\,k}\propto(\chi_{1\,k}^2+\chi_{2\,k}^2)^{(n_1+n_2-1)/2} \chi_{1\,k}^{-n_1+1}
  \exp(-\half\chi^2_{2\,k}). 
\label{eq:21weight}
\end{equation}
This perhaps appears odd at first sight, but is as it should be: the first dataset
has altered the probability distribution of the PDFs, and thus the
probabilities of the replicas before the second dataset can be
considered must necessarily change. This is taken into account of by
the dividing out the phase space factor of the first dataset, and
multiplying by that of the combined dataset.

Nevertheless, it is possible to factorize the reweightings due to more than one dataset, if rather 
than attempting successive reweightings of the same set of replicas, one first turns the original 
weighted set into an unweighted set, and then computes the second set of weights using this set. 
This procedure will be discussed in detail in  Sec.~\ref{sec-con}: however before 
we can do this we must first develop a procedure for unweighting.

\section{Unweighting}

\label{sec-unw}

In this section we present a method to unweight reweighted PDF sets so
that they can be used without the need for including weights for
individual replicas.  The starting point is a set of $N_{\rm rep}$
reweighted replicas. Each replica, identified by the index
$k=1,\ldots,N_\mathrm{rep}$, carries a weight $w_k$ defined
in Eq.~(\ref{eq:13}), determined by comparing each of the replicas of the 
original unweighted distribution  to the new experimental
information.  Our goal is to unweight this PDF set in order to
obtain a new set of $N'_{\rm rep}$ replicas with all weights equal to
unity, but with the same probability distribution of the original
weighted set, i.e. such that any moment of the probability
distribution computed from the weighted and unweighted set would be
the same in the limit in which $N'_{\rm rep}\to\infty$.

\begin{figure}[th]
  \centering
  \epsfig{width=0.5\textwidth,figure=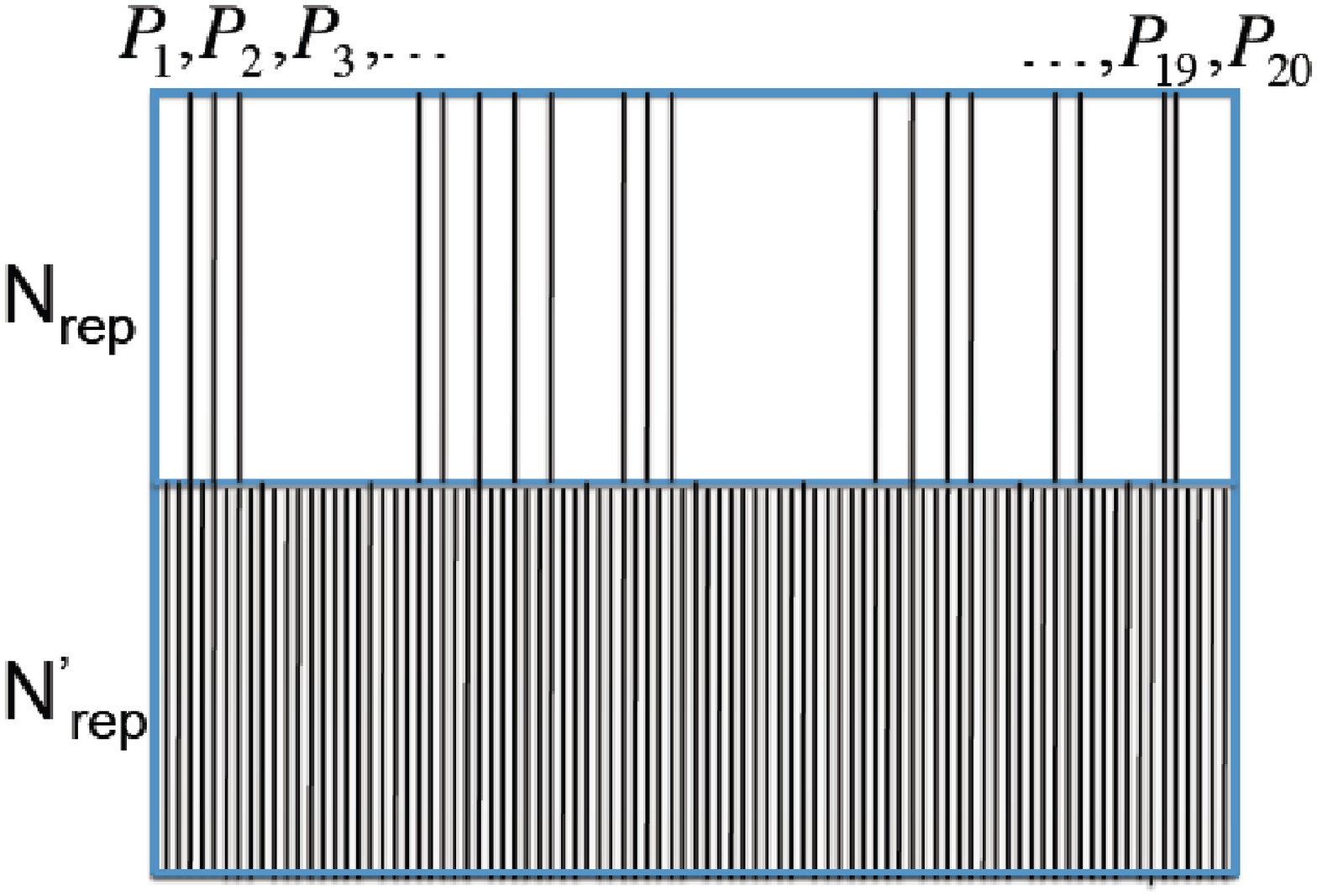,angle=-90}\\
  \vspace{-.5truein}
  \epsfig{width=0.5\textwidth,figure=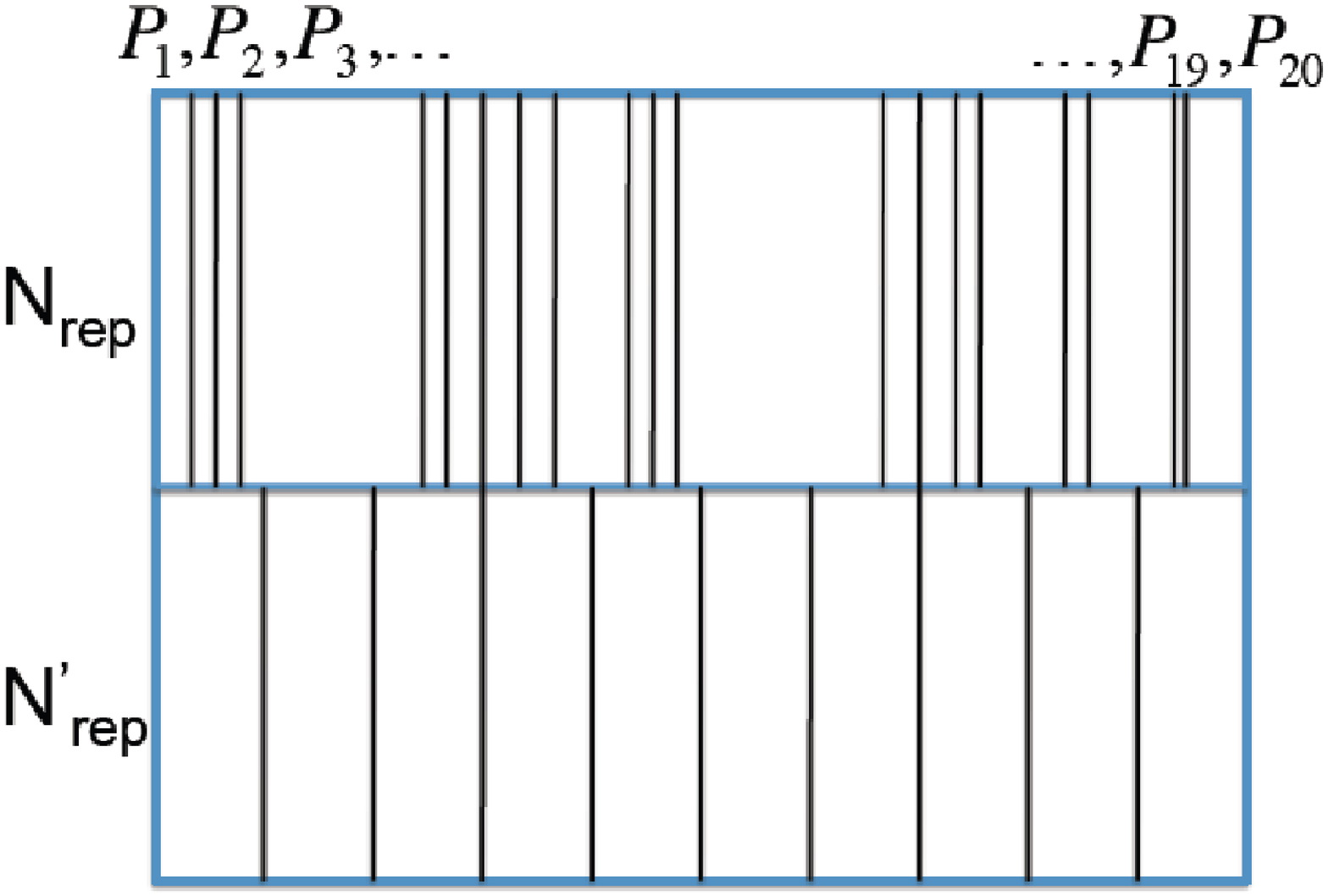,angle=-90}
  \caption{\small Graphical representation of the construction of a set
    of $N'_{\rep}$ unweighted replicas from a set of $N_{\rep}=20$
    weighted ones. Each segment is in one-to-one correspondence to a
    replica, and its length is proportional to the weight of the
    replica. The cases of  $N'_{\rep}\gg N_{\rep}$ (top) and
    $N'_{\rep}=10$ (bottom) are shown.}
  \label{fig:unwplot}
\end{figure}

\subsection{The unweighting method}
\label{sec-unwmet}

The basic idea for constructing the unweighted set consists of
selecting replicas from the weighted set of $N_\mathrm{rep}$ replicas in 
such a way that replicas
carrying a relatively high weight are chosen repeatedly, while those with
vanishingly small weight disappear from the final unweighted set.
The method is depicted graphically in Fig.~\ref{fig:unwplot}.  We start
by subdividing a line of unit length into $N_\mathrm{rep}$ segments, in 
such a way that for each replica the length of the corresponding
segment is proportional to the weight of the replica, and thus to its probability. 
The ordering of the segments is random.
In order to extract a set of 
$N'_\mathrm{rep}$ replicas that faithfully represents this distribution, we
draw another unit interval directly below the first, and subdivide it into
$N'_\mathrm{rep}$ 
segments all of equal length 
$1/N'_\mathrm{rep}$. We then select replicas from the original weighted set
by taking a number of copies of each replica equal to
the number of lower segments whose right edge is contained in the
upper segment corresponding to that specific replica. A little thought shows that
the (all equally probable) $N'_\mathrm{rep}$ replicas in the lower set are then chosen 
according to the probabilities of the $N_\mathrm{rep}$ replicas in the upper set.

To see this, note that, 
if the number of $N'_\mathrm{rep}$ replicas is large enough, (top plot in
Fig.~\ref{fig:unwplot}) then at least one lower segment (width
$1/N'_\mathrm{rep}$) will be contained in each upper segment, and the
original probability distribution is reproduced. This case is
however unrealistic, as it would require $N'_\mathrm{rep}$ to be as large as
the ratio between the highest and lowest weight, which can be very
large indeed. It is also unnecessary, because the amount of information
carried by the weighted set is measured by its Shannon entropy, which
can be used to determine the effective number of unweighted replicas
$N_{\rm eff}$ which carry the same
information~\cite{Ball:2010gb}. Hence, it is pointless to include a
number of replicas $N'_\mathrm{rep}$ significantly larger than $N_{\rm eff}$, as no
information is then gained. Because by construction $N_{\rm eff}\leq
N_\mathrm{rep}$ the more realistic situation is depicted in the bottom plot
of Fig.~\ref{fig:unwplot}: for the larger weights several unweighted
segments are contained in a weighted one, but for the smaller
weights there are often none at all, since we only select a replica
if the edge of a lower segment is contained in the upper
segment corresponding to that replica. Which replica is chosen among 
many all with equally small weight is of course entirely random, since the ordering of the 
replicas is random.

We can now formulate the unweighting algorithm quantitatively.
We start with a set of $N_{\rm rep}$ replicas, each carrying a weight
$w_k$ Eq.~\ref{eq:13};
as in Ref.~\cite{Ball:2010gb}, we normalize the weights according to
\begin{equation}
\label{eq:wnorm}
\sum_{k=1}^{N_{\rm rep}} w_k =N_{\rm rep}.
\end{equation}
The  probability of each replica is determined given its weight as
\begin{equation}
\label{eq:wprob}
p_k= \frac{w_k}{N_{\rm rep}}.
\end{equation}
We then  define probability cumulants 
\be
P_k\equiv P_{k-1} + p_k = \sum_{j=0}^kp_j 
 \ ,
\label{eq:cumulants}
\ee
where in the last step we take $P_0=0$. 
By construction, $0\le P_k\le 1$ and  $P_{k-1} \le
P_{k}$. Indeed, 
 the cumulants provide the co-ordinate
of the edge of the $k$-th upper segment in the plot of Fig.~\ref{fig:unwplot}, 
with origin at the left edge of the unit interval.

The unweighted set is then constructed as follows. We start with
$N_{\rm rep}$ weights $w_k$, and we determine  $N_{\rm rep}$ new weights
\bea 
\label{eq:unweights}
 w'_k=\sum_{j=1}^{N'_{\rm rep}}\theta\Big( \frac{j}{N'_{\rm rep}}-P_{k-1}\Big)
\theta\Big(P_k-\frac{j}{N'_{\rm rep}}\Big) .
\eea
The weights   $w'_k$ are either zero or positive integers,
and they satisfy the normalization condition
\be
N'_{\rm rep}\equiv \sum_{k=1}^{N_{\rm rep}} w'_k:
\label{eq:nprimedef}
\ee 
in fact, they correspond to the graphical counting procedure described
previously.  The unweighted set is then simply constructed by taking
$w^\prime_k$ copies of the $k$-th replica, for all $k=1,\ldots,N_{\rm
  rep}$.  The probability of replica $k$ in the new unweighted set is
then given by
\begin{equation}
\label{eq:wprobprime}
p'_k= \frac{w'_k}{N'_{\rm rep}}.
\end{equation}
As a consequence we have
\be
\label{eq:limit}
\lim_{N'_\mathrm{rep}\to \infty} p'_k=p_k,
\ee
i.e. the unweighted set reproduces the probabilities of the weighted
set in the limit of large sample size, as it ought to.

As already mentioned, even though exact identity of the reweighted and
unweighted probability distribution holds in the limit
Eq.~(\ref{eq:limit}), the amount of information contained in the
weighted set corresponds to $N_{\rm eff} \leq N_{\rm rep}$ unweighted
replicas, with $N_{\rm eff}$ determined  as in Eq.~(10) of 
Ref.~\cite{Ball:2010gb} from the Shannon entropy.  Therefore
for practical applications it is advisable to take $N'_{\rm rep} <
N_{\rm eff}$ --- though there is nothing in principle wrong with
taking $N'_{\rm rep} > N_{\rm eff}$, this would just lead to a highly
redundant replica set. We will study the dependence of unweighted
results on $N'_{\rm rep}$ in an explicit example below.  \clearpage

\begin{figure}[t]
  \centering
  \epsfig{width=0.99\textwidth,figure=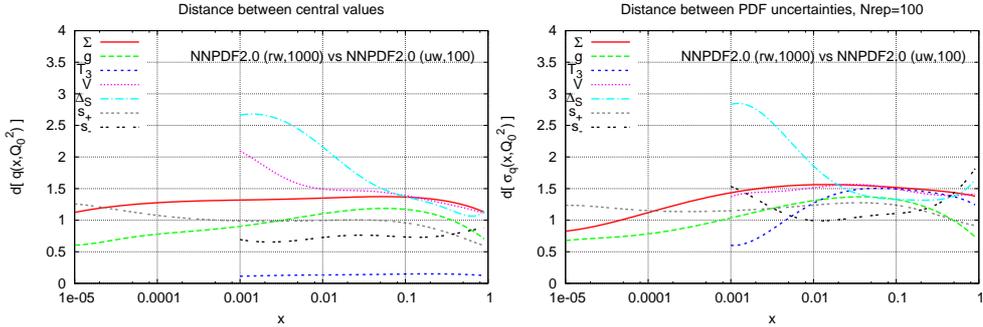}
  \caption{\small Distance between central values (left) and
    uncertainties (right) of the reweighted and unweighted PDFs
    determined from $N_{\rm rep}=1000$ replicas of NNPDF2.0 DIS+DY 
    reweighted with Tevatron jet data, as described in the text. 
     The corresponding distances between
    refitted and reweighted PDFs were shown in Fig.~2 of Ref.~\cite{Ball:2010gb}.}
  \label{fig:distances-uw-jet}
\end{figure}
\begin{figure}[htb]
  \centering
  \epsfig{width=0.49\textwidth,figure=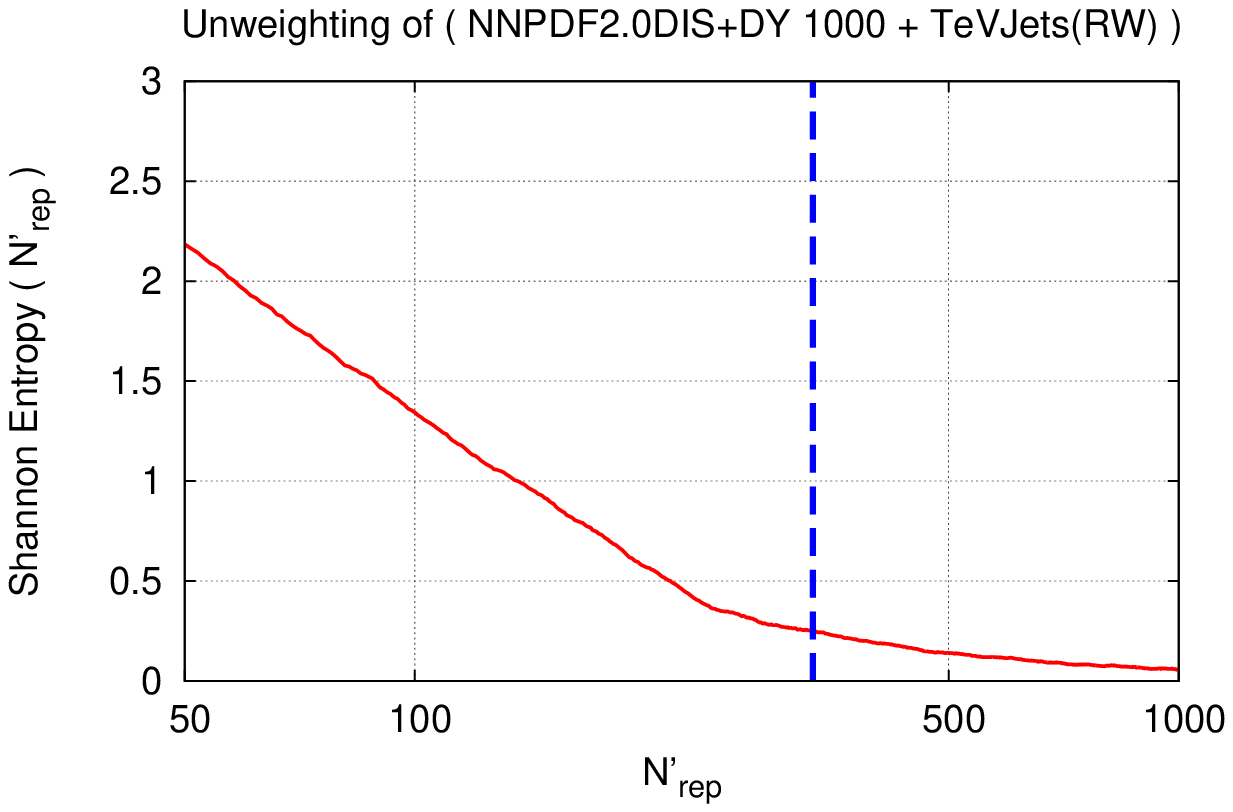}
  \epsfig{width=0.49\textwidth,figure=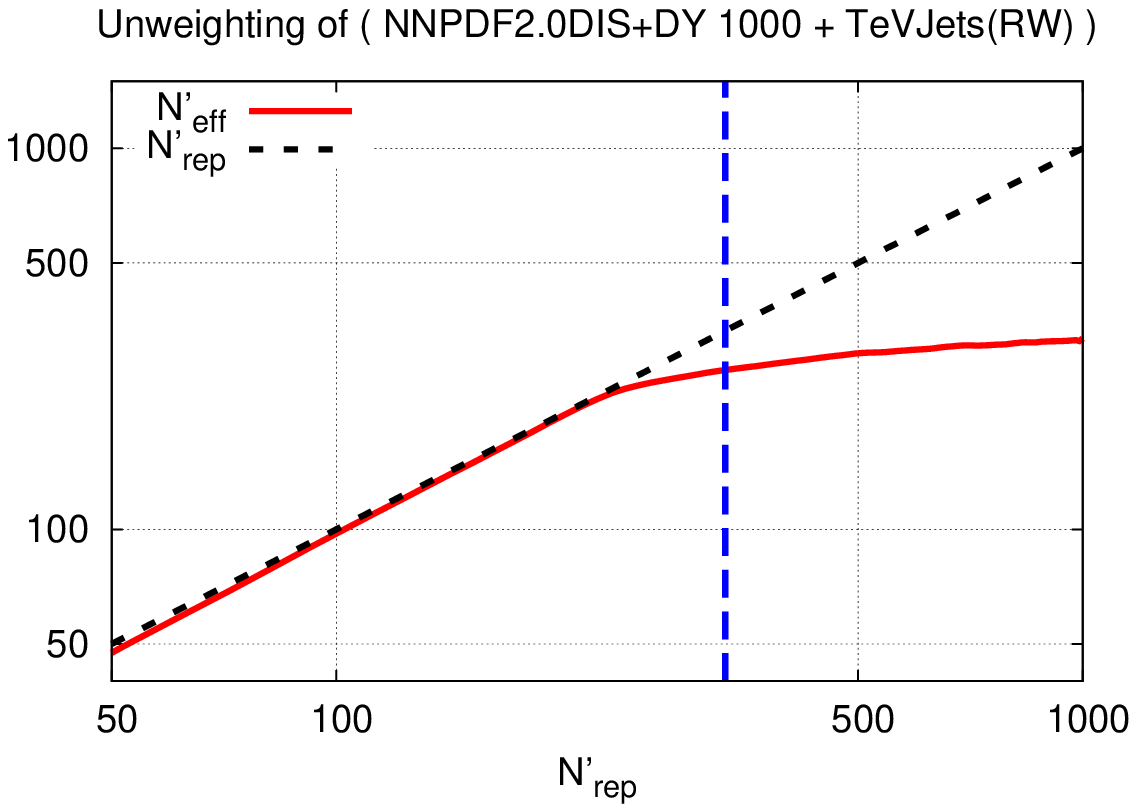}\\

  \caption{\small Left: the relative Shannon entropy $H_R(N'_{\rm rep})$
    Eq.~(\ref{eq:relshannon}) 
    as function of $N'_{\rm rep}$ for the reweighted and
    unweighted PDFs described in the caption of Fig.~\ref{fig:distances-uw-jet}. 
Right: the effective number of replicas of the unweighted set $N'_{\rm
  eff}$ as function of $N'_{\rm rep}$.
The dashed vertical line denotes the
    value $N'_{\rm rep}=N_{\rm eff}$.
    In all plots a moving average of $25$
    replicas has been performed to smooth out random fluctuations. }
  \label{fig:entropy-jets}
\end{figure}

\subsection{Testing unweighting}
\label{sec-unwtest}

As a proof of concept of the unweighting technique, we will apply it
to the two cases 
discussed in Ref.~\cite{Ball:2010gb}: the reweighting of
NNPDF2.0 DIS+DY with Tevatron inclusive jet 
data and the reweighting of NNPDF2.0 with the D0 muon and inclusive
electron $W$ lepton 
asymmetry data. 

First, we consider the reweighting of
NNPDF2.0 DIS+DY~\cite{Ball:2010de} with the Tevatron inclusive jet 
data~\cite{Aaltonen:2008eq,Abulencia:2007ez}. As discussed in
Ref.~\cite{Ball:2010gb}, starting with $N_{\rm rep}=1000$
NNPDF2.0 DIS+DY replicas, after reweighting with jet data
the effective number of replicas is
 $N_{\rm eff}=334$. A reasonable choice for the size of the unweighted set 
would be any number less than 
this: here we chose $N'_{\rm rep}=100$. 
We perform the unweighting following the procedure discussed
above. 
The comparison between the reweighted PDFs and the unweighted set can be made 
quantitative by determining the distances between PDFs and uncertainties. Distances
were defined
in Appendix~A of Ref.~\cite{Ball:2010de}, and in
Ref.~\cite{Ball:2010gb} in the weighted case; recall that distances
$d\sim1$ correspond to statistically identical distributions, while
(with $N_{\rm rep}=100$ replicas)
$d\sim 7$ corresponds to distributions which are statistically
inequivalent, but agree to one sigma.
The distances between 
the reweighted PDF set and the same PDF set after unweighting are
shown in Fig.~\ref{fig:distances-uw-jet}. The corresponding distances
between reweighted and refitted PDFs were shown in Fig.~2 of
Ref.~\cite{Ball:2010gb}. It is clear that the distances between
reweighted and unweighted sets are generally smaller than those between the
reweighted and the refitted sets, and they all fluctuate about $d\sim
1$, showing statistical equivalence (with the possible exception of the
light sea asymmetry at small $x$, which is subject to very large
uncertainties). We conclude that there is no significant loss of 
accuracy in the reweighting due to the unweighting.

We can now study the information contained in the unweighted set as
the number of unweighted replicas $N'_{\rm rep}$ is varied. To this
purpose, we compute the relative Shannon entropy between the
unweighted set and the original weighted set, defined as
\begin{equation}
\label{eq:relshannon}
H_R(N'_{\rm rep})=\sum_{k=1}^{N_{\rm rep}} p_k'\ln \frac{p_k'} {p_k},
\end{equation}
where $p_k'$ are the probabilities Eq.~(\ref{eq:wprobprime}), defined for
each value of $N'_{\rm rep}$. If the starting number of replicas
$N_{\rm rep}$ is large enough
that $N'_{\rm rep}\sim N_{\rm rep}$ is already in the asymptotic
region where
Eq.~(\ref{eq:unweights}) holds, then clearly for large  $N'_{\rm
  rep}\sim N_{\rm rep}$ the relative entropy $H_R(N'_{\rm rep})$ should
fall to zero. For lower values of $N'_{\rm rep}$ $H_R(N'_{\rm rep})$ measures the
information loss between the original weighted set and the unweighted
one. 

\begin{figure}[t]
  \centering
  \epsfig{width=0.99\textwidth,figure=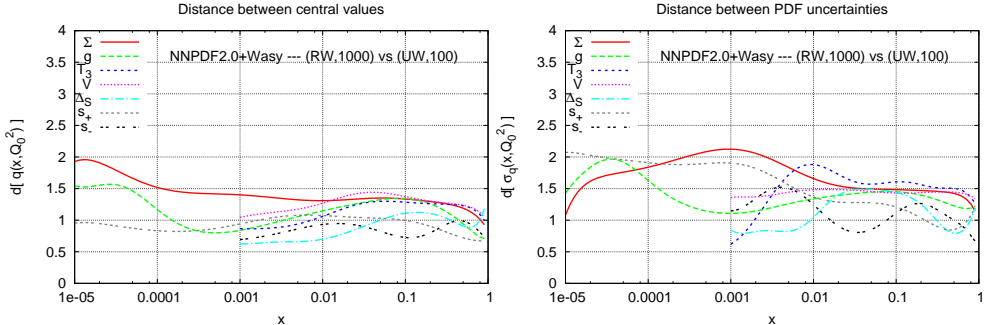}
  \caption{\small  Distance between central values (left) and
    uncertainties (right) of the reweighted and unweighted PDFs
    determined from $N_{\rm rep}=1000$ replicas of NNPDF2.0 DIS+DY 
    reweighted with D0 $W$-lepton asymmetry data, as described in the text. 
   }
  \label{fig:distances-uw-wlasy}
\end{figure}

\begin{figure}[htb]
  \centering
  \epsfig{width=0.49\textwidth,figure=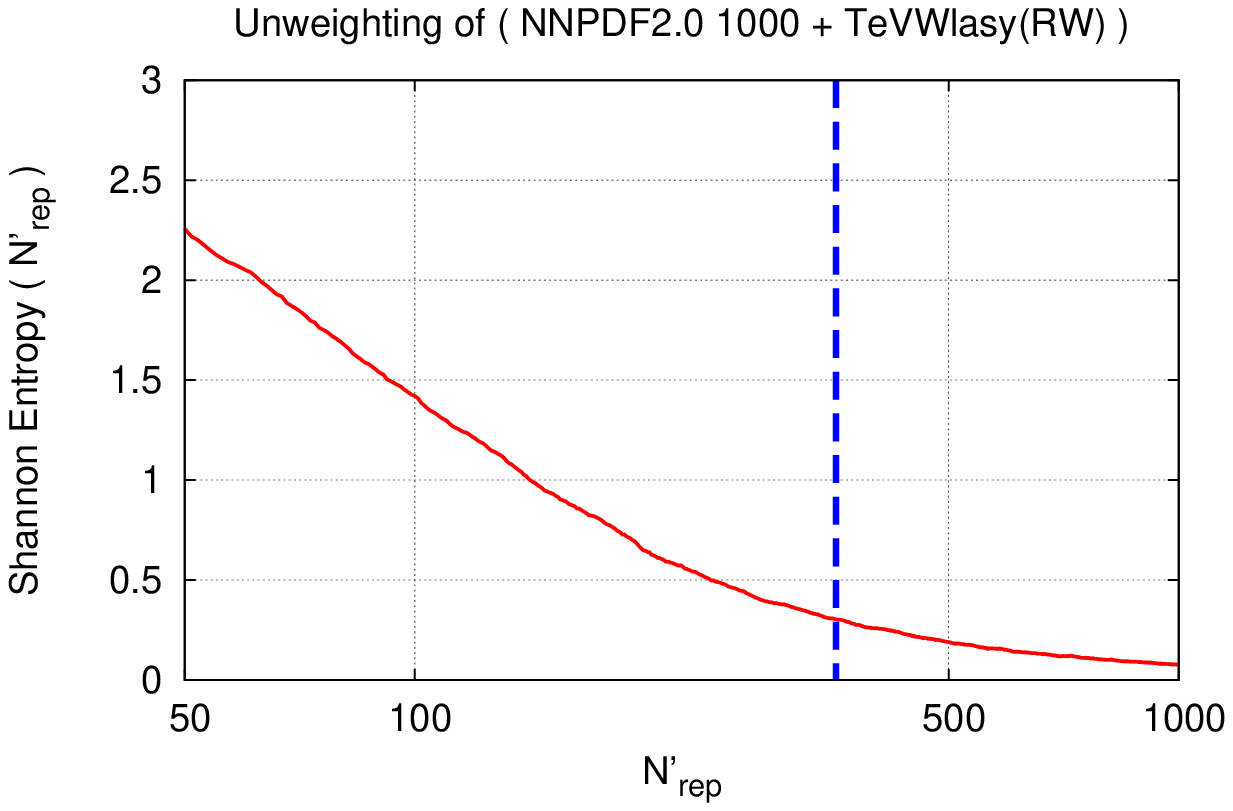}
  \epsfig{width=0.49\textwidth,figure=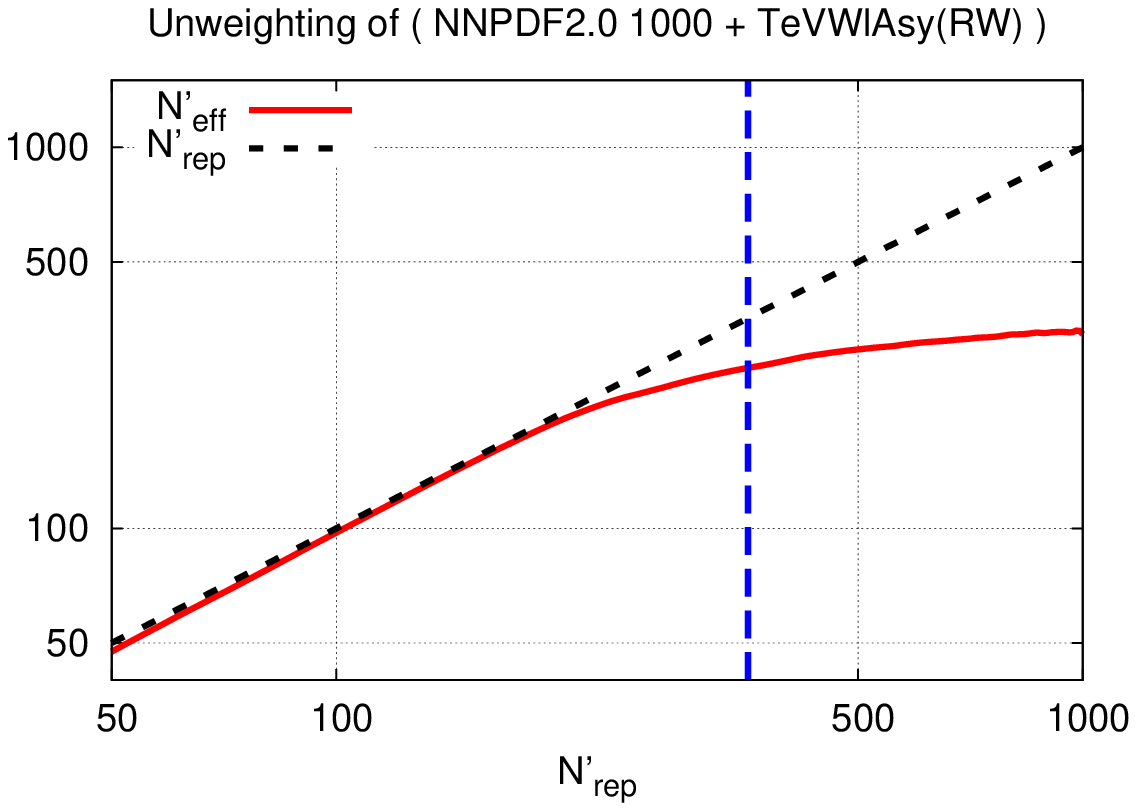}
  \caption{\small Same as Fig.~\ref{fig:entropy-jets}, but for  the pair
    of reweighted and
    unweighted PDFs described in the caption of Fig.~\ref{fig:distances-uw-wlasy}.}
  \label{fig:entropy-wlasy}
\end{figure}

In Fig.~\ref{fig:entropy-jets}  we display $H_R(N'_{\rm rep})$. It is clear 
that $H_R(N'_{\rm rep})$ falls linearly as a
function of $N'_{\rm rep}$ up to $N_{\rm eff }$, as more and more of 
the information in the weighted set is included. Around $N'_{\rm
  rep}\sim N_{\rm eff }$ the slope of the fall changes abruptly,  
and $H_R(N'_{\rm rep})$ then falls slowly to zero as $N'_{\rm
  rep}$ increases, being already close to zero when 
$N'_{\rm rep}\sim N_{\rm eff}$. 
This can also be seen by computing directly the effective number of
replicas $N'_{\rm eff}$ of the unweighted set as a function of
$N'_{\rm rep}$,
which can be determined using Eq.~(10)
of Ref.~\cite{Ball:2010gb}, with the weights $w'_k$
Eq.~(\ref{eq:unweights}) and $N=N_{\rm rep}$. Note that the result
is nontrivial because some of the $w'_k$ are zero, others are
integers larger than one, and the dependence on $N'_{\rm rep}$ comes about only 
through the definition of the weights Eq.~(\ref{eq:unweights}). The result is also shown in
Fig.~\ref{fig:entropy-jets}:  at first $N'_{\rm eff}$  grows linearly as a
function of $N'_{\rm rep}$, and is in fact very nearly equal to it. 
However when it reaches $N'_{\rm rep}\approx
N_{\rm eff}$, the linear growth breaks off abruptly, and saturates at the 
value  $N'_{\rm eff}= N_{\rm eff}$, which is reached asymptotically.
Hence our expectation is borne out by
these plots: the amount
of information in the unweighted set increases with the number of
unweighted replicas $N'_{\rm rep}$, but only up to the point $N'_{\rm
  rep}\approx N_{\rm eff} $, after which nothing is gained by
further increasing $N'_{\rm rep}$.

We now repeat the same analysis for the unweighting of the NNPDF2.0
set, reweighted with the inclusive electron and muon D0 Run--II $W$
lepton asymmetry data~\cite{d0electron,d0muon}.  The reweighting
procedure for these data was presented in detail in
Ref.~\cite{Ball:2010gb}.  The effective number of replicas, after
reweighting a starting set of $N_{\rm eff}=1000$ replicas, is in this case
$N_{\rm eff}=356$.  Again, we can choose the size of the unweighted set
to be $N'_{\rm rep}=100$, as in the case above, and we perform the
unweighting following the same procedure as before.

In Fig.~\ref{fig:distances-uw-wlasy} we show
the distance between the reweighted and unweighted sets, and in
Fig.~\ref{fig:entropy-wlasy} we plot the relative entropy between
these two sets and the effective number of replicas in the unweighted
set
as a function of the number of unweighted replicas. The
conclusions are the same as before: the unweighted set is
indistinguishable from the reweighted one, provided that the number of
unweighted replicas $N'_{\rm rep}$ is of the same order as the
effective number of reweighted replicas $N_{\rm eff}$. 
In the sequel
we will thus feel free to use unweighted replica sets instead of their weighted
counterparts, to which they are essentially equivalent.

\section{Consistency}

\label{sec-con}

\subsection{Multiple Reweighting}

As we discussed in Sec.~\ref{sec:mult-exp}, when adding two new
datasets to a set of prior PDFs, one way to proceed is to treat them
as a single combined dataset, as in Eq.~(\ref{eq:15}), i.e., with weights
$\chi^{(n-1)} \exp(-\chi^2/2)$ with $\chi^2 = \chi^2_1+\chi^2_2$ and
$n=n_1+n_2$. However, it should also be possible to treat them separately,
weighting with first one dataset, then the other. If we do this using
Eq.~(\ref{eq:21weight}) then by construction we get the same answer
that we would get by including the two sets at once, but this is
trivial, because in the weights Eq.~(\ref{eq:21weight}) the effect of
the first weighting is divided out.

However, we can test non-trivially that two subsequent weightings by two 
independent datasets commute by incorporating the unweighting procedure. Formally 
we define the operation $\hat{R}$ as reweighting with the weights given by
Eq.~(\ref{eq:13}), and an unweighting operation $\hat{U}$, as described in
Sec.~\ref{sec-unwmet}. 
Note that because the unweighting operator is a projection operator, it has no inverse. 
Weighting an existing PDF set by incorporating information from a new dataset 
then consists of the combined `weighting' operation $\hat{W} = \hat{U}\hat{R}$.
The weighting operation takes a set of replicas $\{f_k\}$, all equally probable,
and replaces it with a subset which are again all equally probable,
but the selection of which reflects information contained in the new
dataset that was used in the reweighting $\hat R$. Clearly $\hat{W}$ has
no inverse, since it projects onto a lower dimensional space.

Now consider two datasets: the set of replicas produced by the action
of weighting with the first dataset,  
$\hat{W_1}$,  can be subject to a further weighting with the second
dataset $\hat{W_2}$. Now of course the formula used to evaluate the weights used 
for the second reweighting must again be given by Eq.~(\ref{eq:13}): the subset 
of replicas produced by $\hat{W_1}$ are again all equally probable, so the 
second reweighting must work in precisely the same way as the first. The only difference 
is that $\hat{W_2}$ acts only on those replicas 
produced by the action of $\hat{W_1}$. 

Now for consistency it cannot matter in what order we perform these  
two weightings, and indeed their combined effect must be the same as
for a single weighting $\hat{W}_{12}$ , which 
treats the two datasets as a single dataset: 
$\hat{W}_{12} = \hat{W}_2 \hat{W}_1 = \hat{W}_1 \hat{W}_2$, or more explicitly
\begin{equation}
\label{eq:con}
    \hat{U}\hat{R}_{12} = \hat{U}\hat{R_2}\hat{U} \hat{R}_1 = \hat{U}\hat{R}_1 \hat{U}\hat{R}_2.
\end{equation}  
So, for weighting to be consistent it must satisfy two nontrivial conditions: 
the combination property, and the commutation property. Clearly the
first always implies the second  
(if $\hat{W}_1 \hat{W}_2=\hat{W}_{12}$, clearly $\hat{W}_2 \hat{W}_1 =
\hat{W}_1 \hat{W}_2$, because  
$\hat{R}_{12}$ is performed using weights determined through the total
$\chi^2=\chi_1^2+\chi_2^2$),  
but not the reverse (we might have $\hat{W}_2 \hat{W}_1 = \hat{W}_1
\hat{W}_2\neq \hat{W}_{12}$  
if the formula Eq.~(\ref{eq:13}) was incorrect).

In the remaining part of this Section we present two tests of 
the combination and commutation properties when two datasets are
included. First, we consider sets of data for the same observable (the one-jet
inclusive cross-section) in the same kinematic region by two different
experiments. Then, we consider data for two different observables (a
jet cross-section and a Drell-Yan cross section) which affect different PDFs in different
kinematic regions.

\subsection{Tevatron Inclusive Jets}

The first exercise we present is an extension of the reweighting
proof-of-concept in Section~4 of~\cite{Ball:2010gb}. There, Run II
Tevatron inclusive jet  
data production were included by reweighting a PDF set extracted from a
NLO fit to DIS and  
Drell-Yan data (NNPDF2.0 DIS+DY) and the results compared to those
obtained from a fit  
which included the same DIS, Drell-Yan and inclusive jet datasets 
all treated in the same way (NNPDF2.0).

\begin{table}[t]
  \centering
  \begin{tabular}[c]{|c|c|c|c|}
    \hline
    & CDF & D0 & CDF+D0 \\
    \hline
    Data points & 76 & 110 & 186 \\
    \hline
    $N_{\rm eff}$ & 290.8 & 565.8 & 334.5\\
    \hline
  \end{tabular}
  \caption{Datasets used in the Tevatron Run II inclusive jet
    reweighting exercise. For each set the number of data points and the effective number of
    replicas of the reweighted set of $N_{\rm rep}=1000$ replicas are given.}
  \label{tab:check-jets}
\end{table}

\begin{figure}[t]
  \centering
  \includegraphics[width=0.45\textwidth]{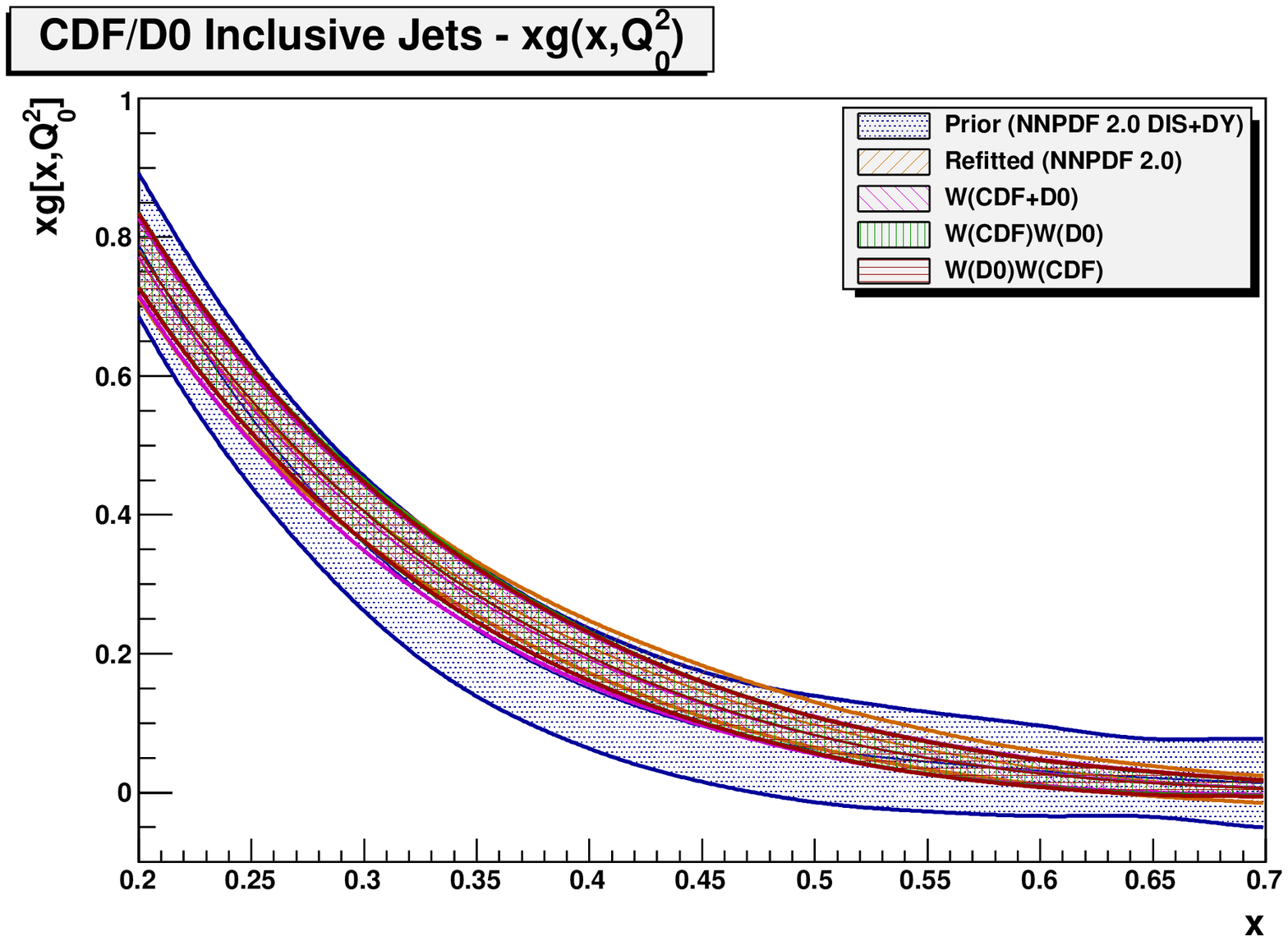}
  \includegraphics[width=0.45\textwidth]{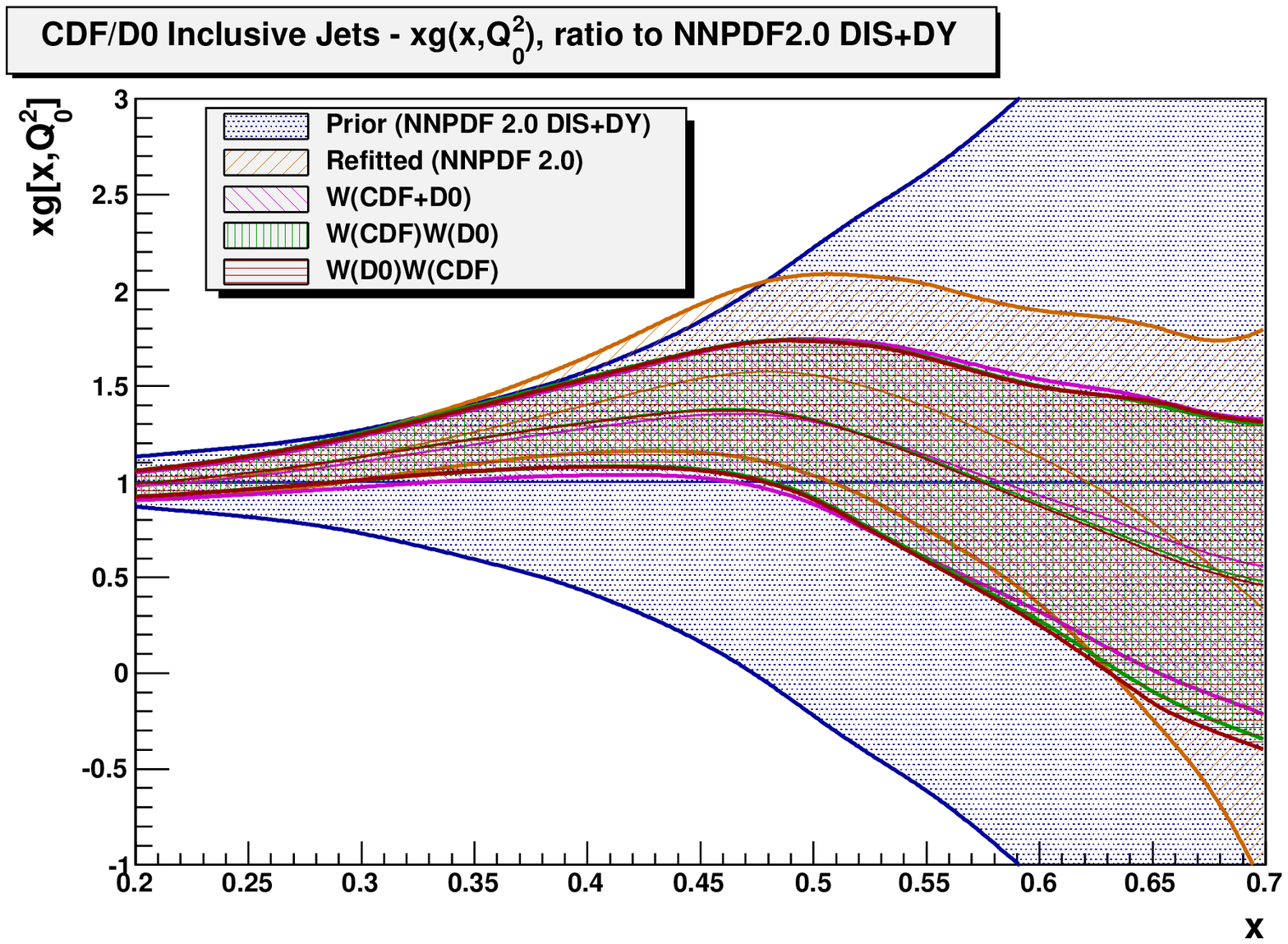}
  \caption{Comparison of the large-$x$ gluon PDF for prior set, 
reweighted sets with
    different successive reweighting orders and refitted set, when the 
    jet data of Table~\ref{tab:check-jets} are included in the
    NNPDF2.0 NLO DIS+DY fit. Results are shown at $Q^2=2$~GeV$^2$, both
    in absolute scale (left) and as a ratio to the prior (right).}
  \label{fig:jets-gluon_lin}
\end{figure}

\begin{figure}[t]
  \centering
  \includegraphics[width=0.8\textwidth]{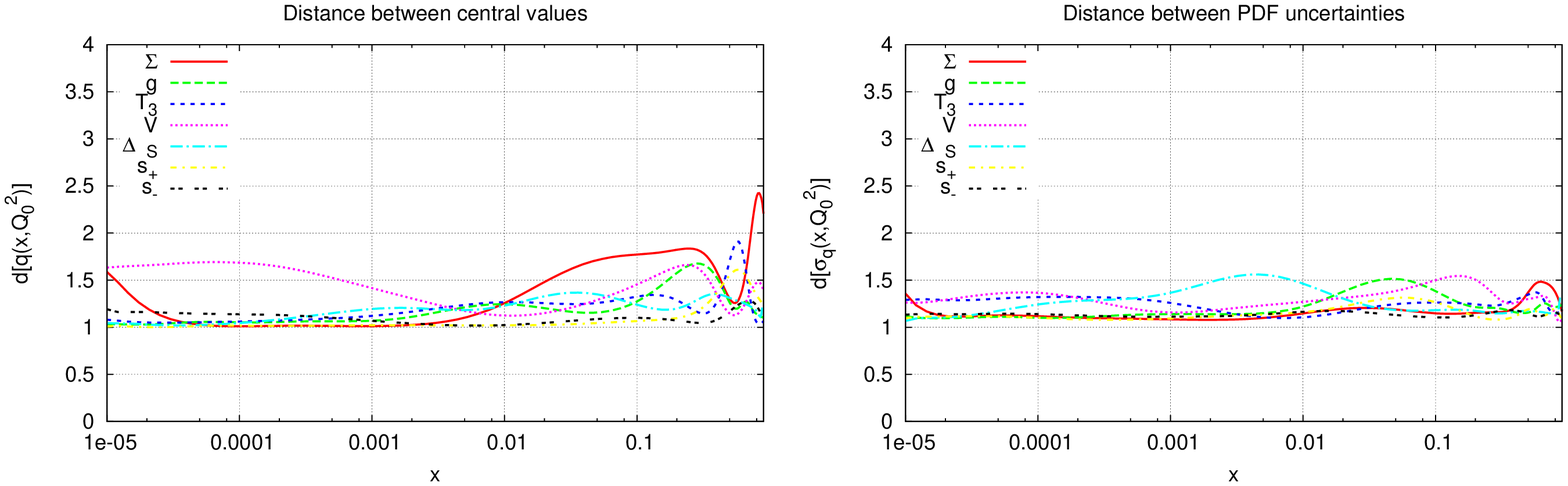}
  \caption{Distances between central values (left) and uncertainties
    (right) of PDFs from
reweighting with the combined CDF+D0 dataset  and PDFs from
reweighting first with CDF data and then with D0 data. }
  \label{fig:jets-d_rw_srw}
  \centering
  \includegraphics[width=0.8\textwidth]{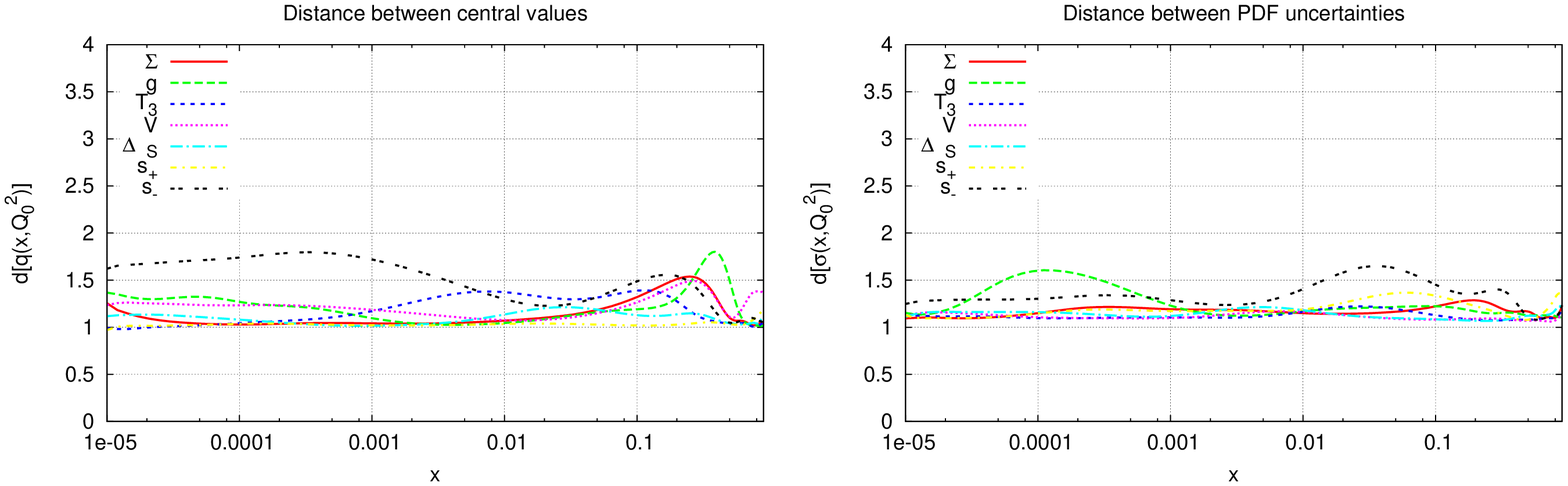}
  \caption{Distances between central values (left) and uncertainties
    (right) of PDFs obtained by
reweighting with CDF and D0 jet data included in either order. }
  \label{fig:jets-d_srw_srw}
\end{figure}

In this Section we look again at the inclusion via reweighting of
the same datasets, namely the CDF Run II-$k_t$ and D0 Run II-cone
inclusive jet data in the NNPDF2.0 DIS+DY fit, but we now focus on
comparing the results obtained in the following two cases:
\begin{enumerate}
\item[(a)] the two new datasets are
  included by reweighting the prior fit in a single step with both
  datasets;
\item[(b)] one of the datasets is included by reweighting, 
  an unweighted set of PDFs is constructed using the procedure detailed in
  Section~\ref{sec-unw}, and finally the latter set is reweighted again with the second
  dataset.  
\end{enumerate}
We will carry out the successive reweighting procedure (b) twice,
exchanging the order in which the CDF and D0 datasets are included, in
order to test the commutativity of the procedure.  A final
unweighting is performed for all the reweighted sets and the PDF
comparisons and computations of distances are performed using these
unweighted sets.

The number of data points and the effective number of replicas $N_{\rm
  eff}$  after reweighting with these data 
of a set of $N_{\rm rep}=1000$ replicas are summarized in  
Table~\ref{tab:check-jets}. In each case, we construct a final set of
$N'_{\rm rep}=100$ unweighted replicas.  When the reweighting is
performed in two steps, we first construct a (redundant) set of $1000$
unweighted replicas, which is then reweighted and unweighted again to
obtain the final set of 100 unweighted replicas.

As discussed in Refs.~\cite{Ball:2010de,Ball:2010gb},  Tevatron jet data 
mostly affect the  gluon at large $x$, leaving all other PDFs
essentially unchanged.
The impact of the inclusion of these data in the fit is shown in 
Fig.~\ref{fig:jets-gluon_lin} 
where we compare the gluon for the prior set, the refitted one, and
sets obtained reweighting  
the prior in the three different ways described above. As in the
previous Section, a more quantitative
assessment can be made by computing distances between various pairs of
PDF sets. In Fig.~\ref{fig:jets-d_rw_srw} we show the distance
between PDFs obtained by reweighting with the two sets at once and
those found including CDF data first and D0 data next, while in 
Fig.~\ref{fig:jets-d_srw_srw} we show distances between sets obtained
by including the CDF and D0 data in either order. It is clear that the
three reweighting procedures lead to completely equivalent results.

\begin{table}[t]
  \centering
  \begin{tabular}[c]{|c|c|c|c|}
    \hline
    & (CDF+D0) & E605 & (CDF+D0)+E605 \\
    \hline
    Data points & 186 & 119 & 305 \\
    \hline
    $N_{\rm eff}$ & 627.1 & 59.5 & 63.7\\
    \hline
  \end{tabular}
  \caption{As Tab.~\ref{tab:check-jets}, but now for the E605 and inclusive jet reweighting exercise.}
  \label{tab:check-e605-jets}
\end{table}

\begin{figure}[h!]
  \centering
  \includegraphics[width=0.45\textwidth]{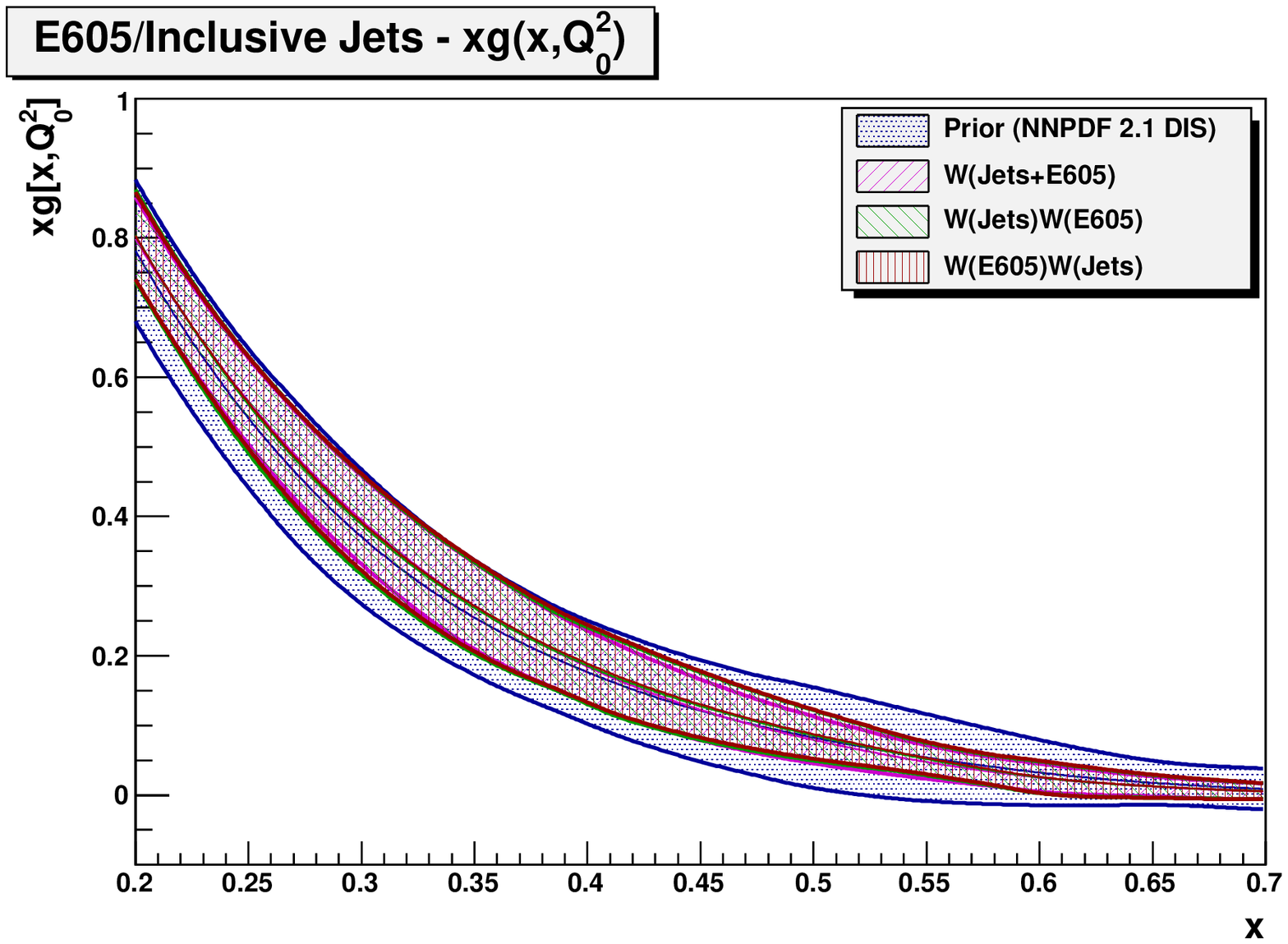}
  \includegraphics[width=0.45\textwidth]{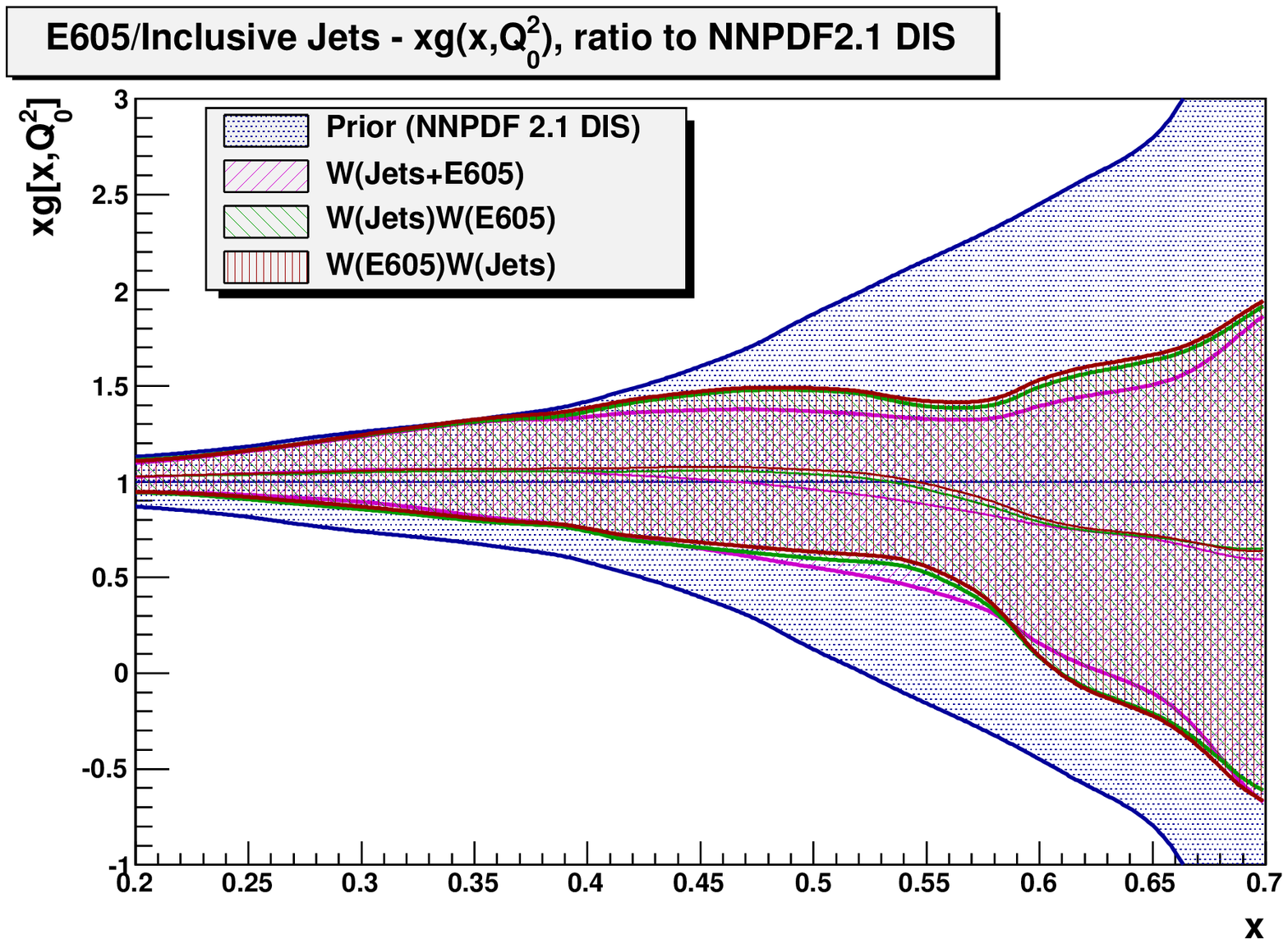}\\
  \includegraphics[width=0.45\textwidth]{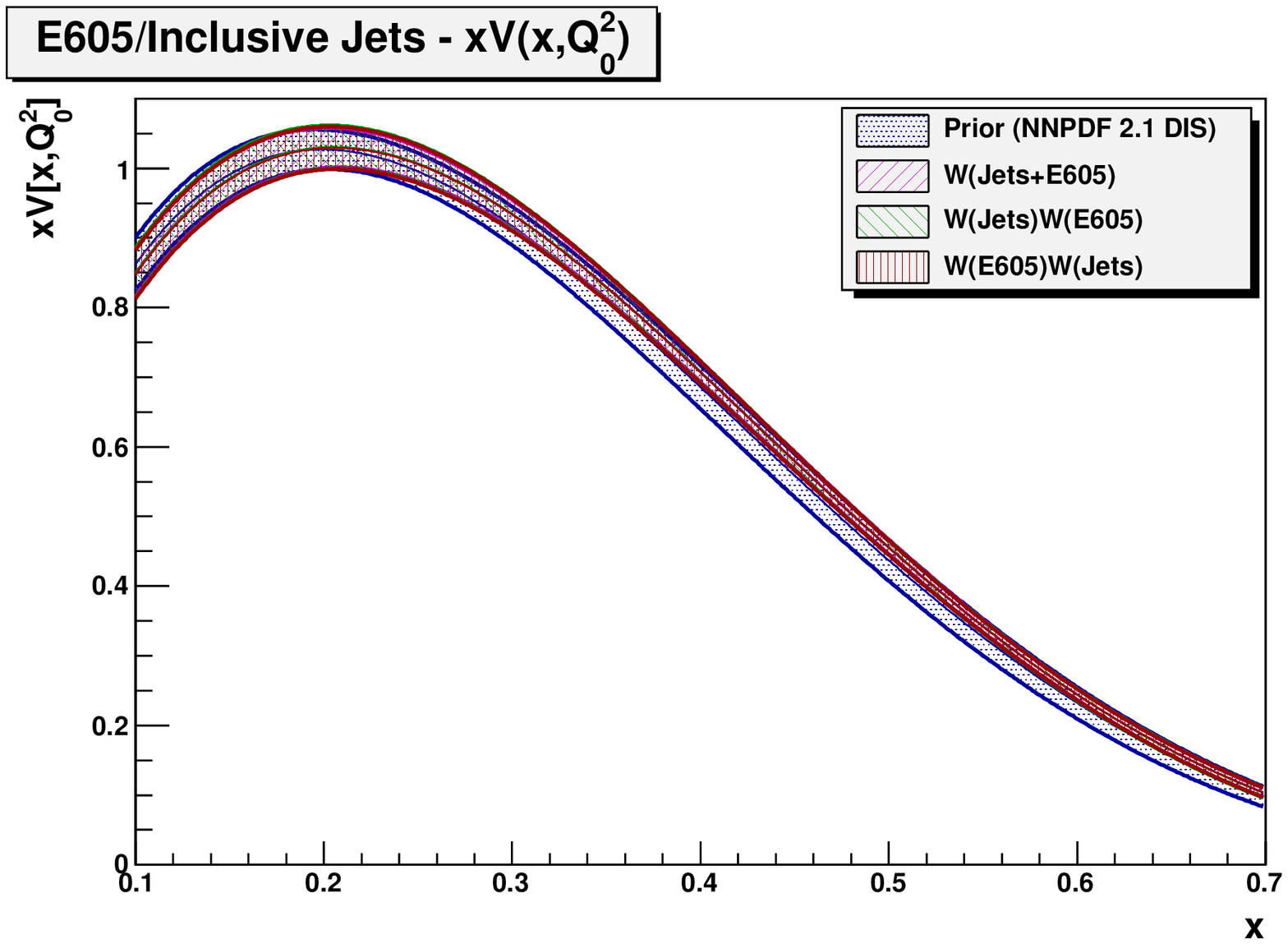}
  \includegraphics[width=0.45\textwidth]{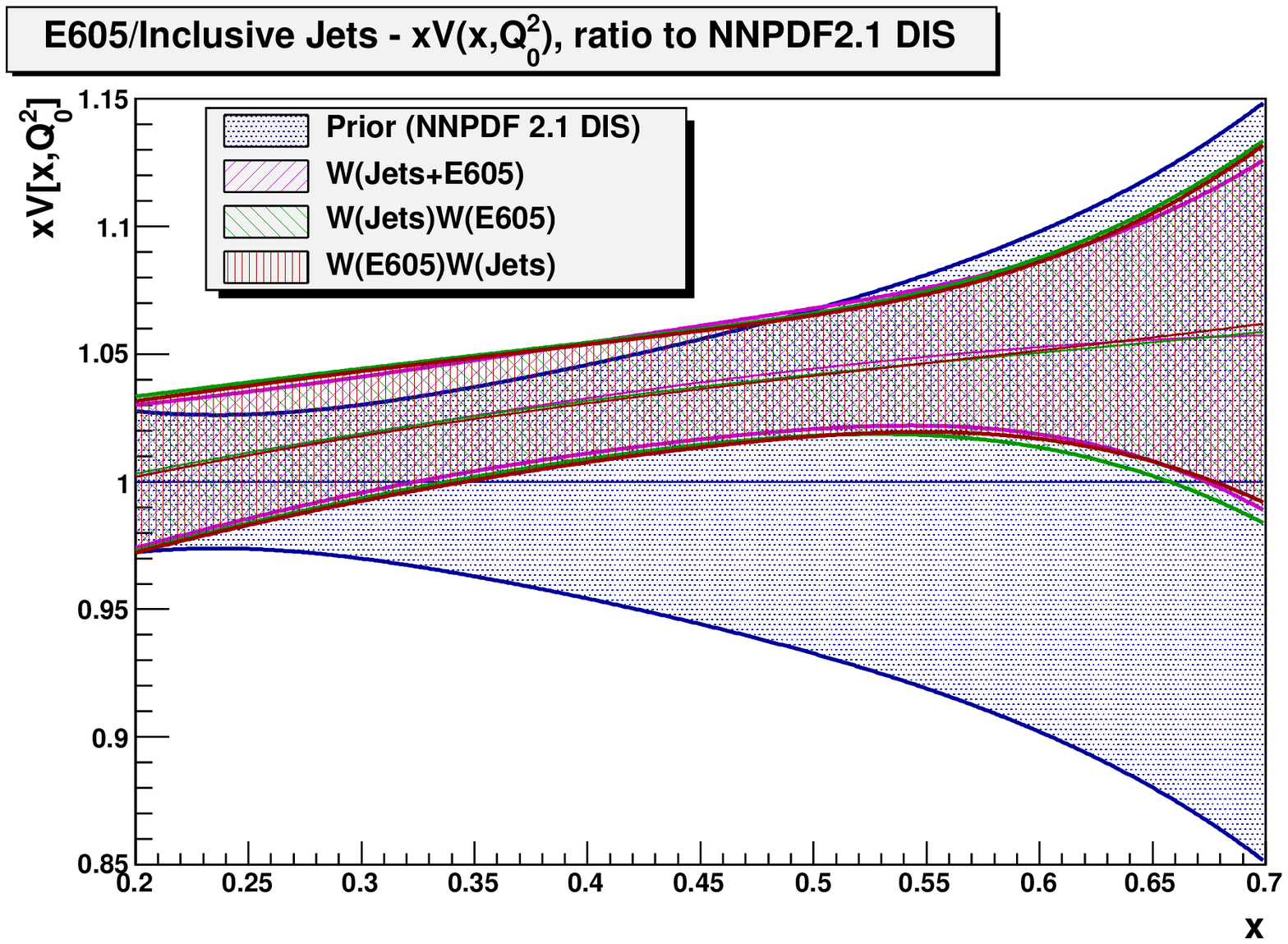}
  \caption{Comparison of the large-$x$ gluon and quark valence PDFs 
 for prior set and reweighted sets with
    different successive reweighting orders, when the 
    jet and Drell-Yan data of Table~\ref{tab:check-e605-jets} are included in the
    NNPDF2.1 NLO DIS fit. Results are shown at $Q^2=2$~GeV$^2$, both
    in absolute scale (left) and as a ratio to the prior (right).}
  \label{fig:e605-jets_pdfs_zoom}
\end{figure}

\begin{figure}[t]
  \centering
  \includegraphics[width=0.8\textwidth]{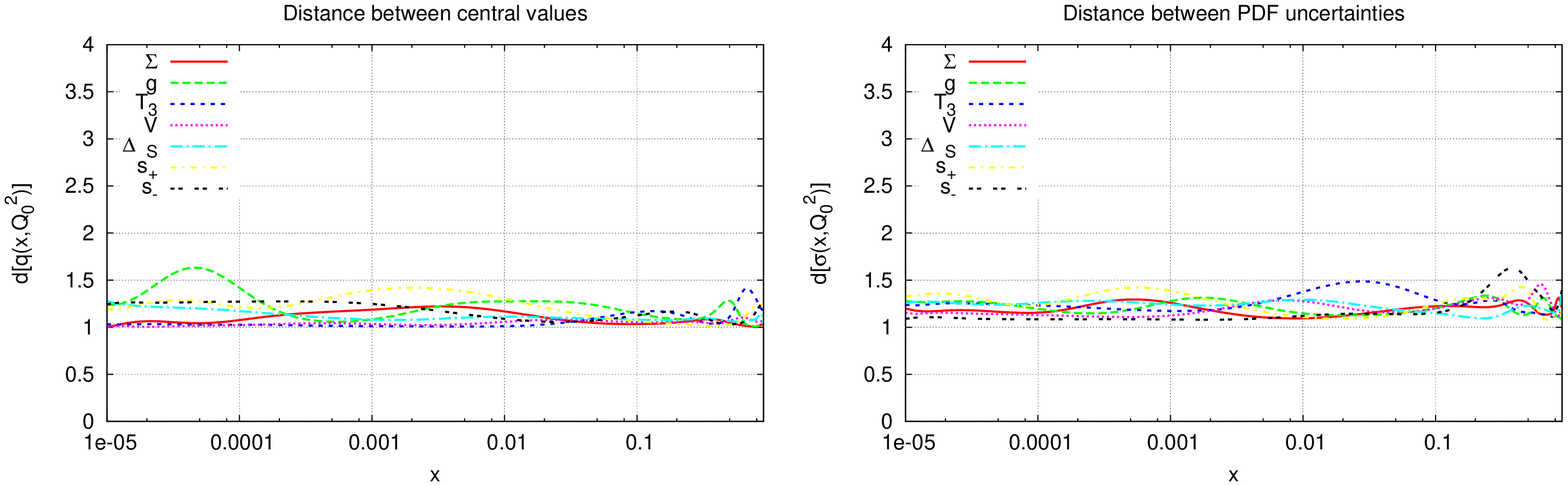}
  \caption{Distances between central values (left) and uncertainties
    (right) of PDFs from
reweighting with the combined jet+Drell-Yan dataset  and PDFs from
reweighting first with jet data and then with Drell-Yan data. }
  \label{fig:e605-jets-d_rw_srw}
  \centering
  \includegraphics[width=0.8\textwidth]{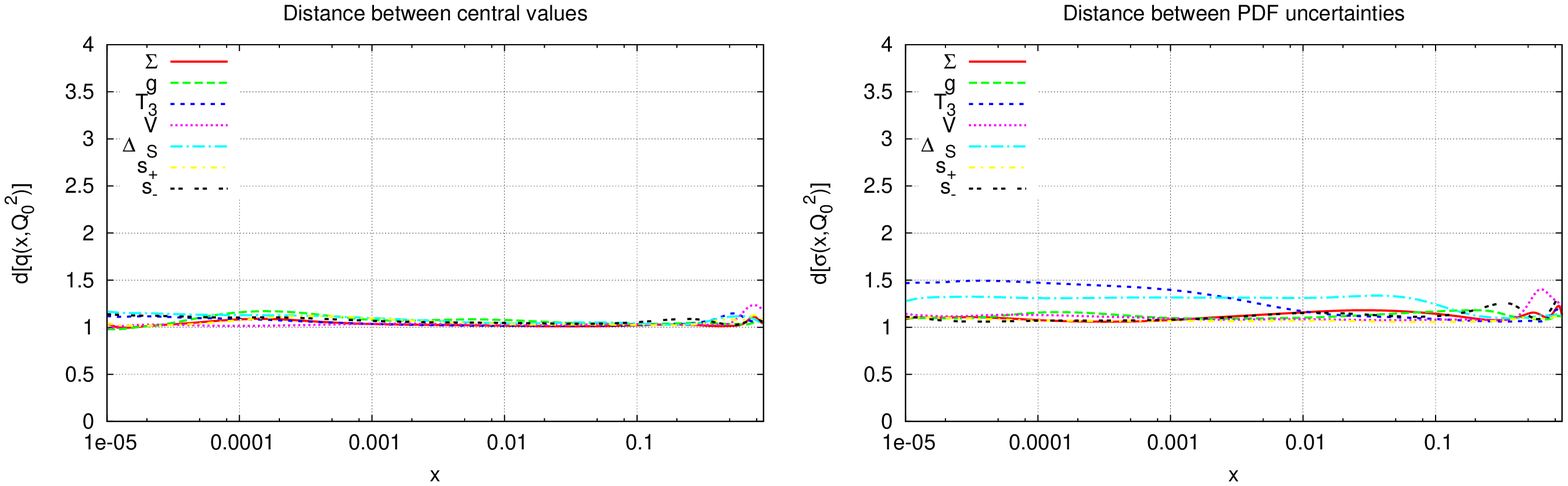}
  \caption{Distances between central values (left) and uncertainties
    (right) of PDFs obtained 
reweighting jet data and Drell-Yan data included in either order. }
  \label{fig:e605-jets-d_srw_srw}
\end{figure}

\subsection{Jet and Drell-Yan data}

In this second exercise we start from a NLO fit to DIS data,
NNPDF2.1 NLO DIS~\cite{Ball:2011mu}, and include the Tevatron 
inclusive jet data discussed in the previous section (D0 and CDF as a single
dataset) and data from one of the Drell-Yan experiments which are included in the 
NNPDF2.1 global analysis (the E605 fixed target  
experiment~\cite{Moreno:1990sf}).

The number of data points and the effective number of replicas $N_{\rm
  eff}$  in this case are summarized in  
Table~\ref{tab:check-e605-jets}. Also in this case, we construct a  set of
$N'_{\rm rep}=100$ unweighted replicas, with 
 $N'_{\rm rep}=1000$ unweighted replicas in the intermediate step if any. 
Note that this is a much less symmetric example than the previous one: the 
Drell-Yan data have a much greater impact than the jet data (in fact for the 
Drell-Yan data $N'_{\rm rep} > N_{\rm eff}$).
 
 As already mentioned, the jet data affect mostly the large $x$ gluon,
 while the Drell-Yan data have mostly an impact on the quark flavour
 and antiflavour separation. The impact of these data on the gluon and
 the total quark valence distribution are shown in
 Fig.~\ref{fig:e605-jets_pdfs_zoom}, where we show the results
 obtained by reweighting with the two sets included together, or one
 after another in either order. Note that in this case we do not
 have a refitted set. Distances between PDFs obtained by
 reweighting in the combined set, or first with jets then with Drell-Yan are
 shown in Fig.~\ref{fig:e605-jets-d_rw_srw}. Distances between PDFs
 obtained reweighting in either order are shown in
 Fig.~\ref{fig:e605-jets-d_srw_srw}. The test is clearly as successful
 here as it was in the previous case, despite being perhaps more challenging.

\section{The W asymmetry at the LHC}
\label{sec-lhc}

In this section we will use the reweighting technique presented 
here and in Ref.~\cite{Ball:2010gb} to study the effect of
including in the NNPDF2.1 NLO global fit the $W$ lepton asymmetry
measurements produced by the experimental collaborations at the LHC, 
and based on data collected in the 2010 run.

The $W$ leptonic charge asymmetry is defined in terms of the $W^{\pm}\to l^\pm\nu_l$ 
differential cross-sections $d\sigma_{l^\pm}/d\eta_l$, with $\eta_l$ being the pseudorapidity 
of the lepton coming from the decay of the $W$ boson, as

\begin{equation}
  \label{eq:wasy}
  A^l_W=\frac{d\sigma_{l^{+}}/d\eta_{l}-d\sigma_{l^{-}}/d\eta_{l}}
  {d\sigma_{l^{+}}/d\eta_{l}+d\sigma_{l^{-}}/d\eta_{l}}
\end{equation}
where the cross-sections are computed inside the acceptance cuts used to select 
the $W\to l\nu_l$ events. 

The ATLAS Collaboration published a first measurement of the muon
charge asymmetry from $W$ boson production in the pseudorapidity range
$|\eta|<2.4$, based on 31pb$^{-1}$ of accumulated
luminosity~\cite{atlas}, while CMS published a measurement of the muon
and the electron charge asymmetries in the pseudorapidity range
$|\eta|<2.2$, based on 36pb$^{-1}$ of data~\cite{cms}.  The data
provide a constraint for the above combination of PDFs in the region
$10^{-3}\lsim x \lsim 10^{-1}$, where they are only partially constrained by the data already
included in the NNPDF global analysis.  In particular, while $u$ is
very well determined by fixed target DIS data, $d$ and the light sea
$(\bar{d}-\bar{u})$ are currently much less constrained.

The LHCb collaboration presented preliminary results for a measurement
of the muon charge asymmetry in the pseudorapidity range
$2<|\eta|<4.5$, covered by the LHCb detector.  This measurement probes
PDFs in the small and large $x$ regions, where data included so far in
the global analyses provide much looser constraints. For this reason
they might eventually have a substantially larger impact on global fits than the
ATLAS or CMS data. However, at the time of writing these experimental
results have only been presented in preliminary form~\cite{lhcb}, and
are therefore not included in this study.

\subsection{Inclusion of individual experiments}

We begin by checking the compatibility of the individual ATLAS and CMS datasets
for the charge lepton asymmetry with the data included in the NNPDF2.1 global fit,
and by studying their impact when they are included separately in the fit using the 
reweighting technique presented in this paper. 

\begin{figure}[htb]
  \begin{center}
    \epsfig{width=0.32\textwidth,figure=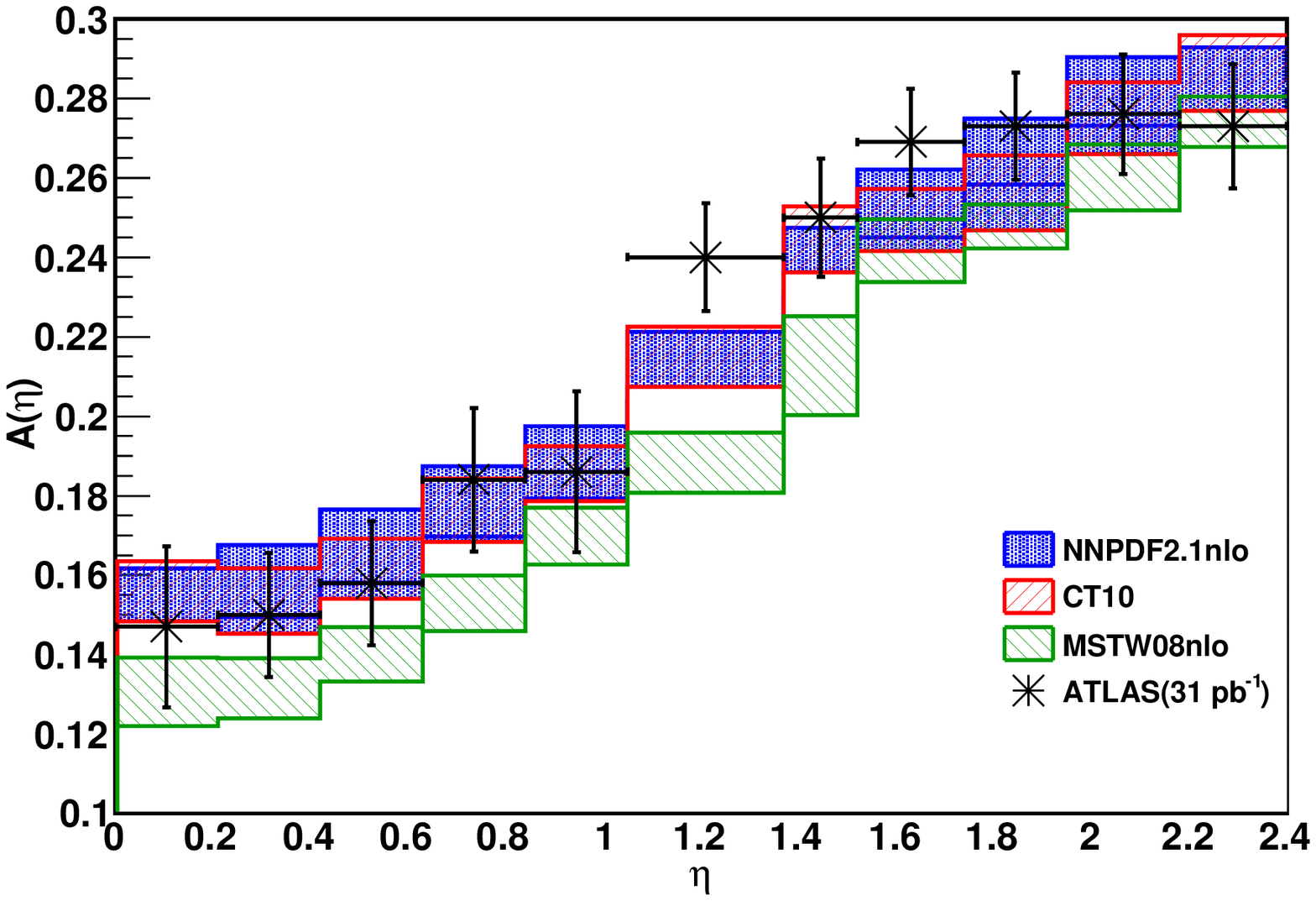}
    \epsfig{width=0.32\textwidth,figure=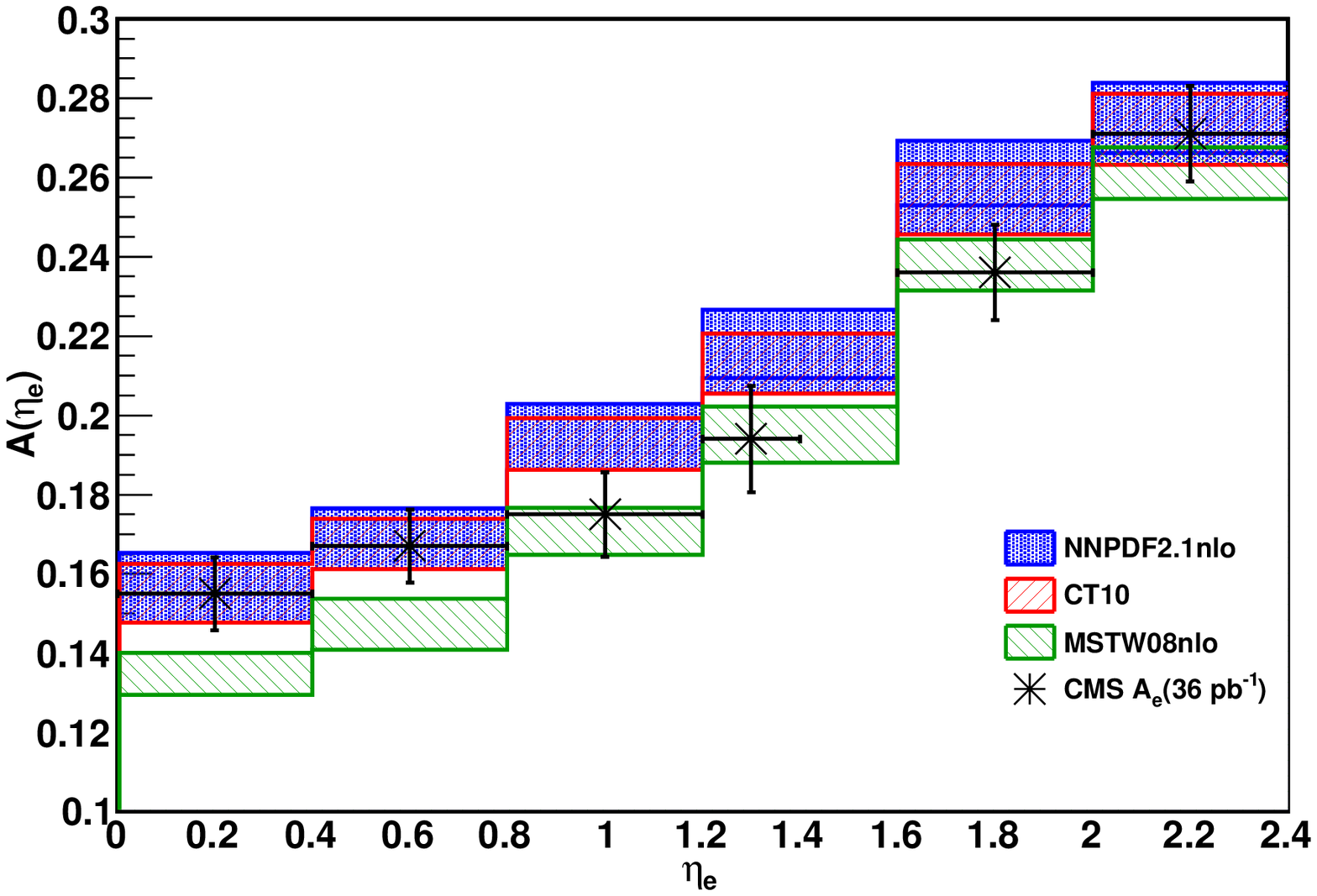}
    \epsfig{width=0.32\textwidth,figure=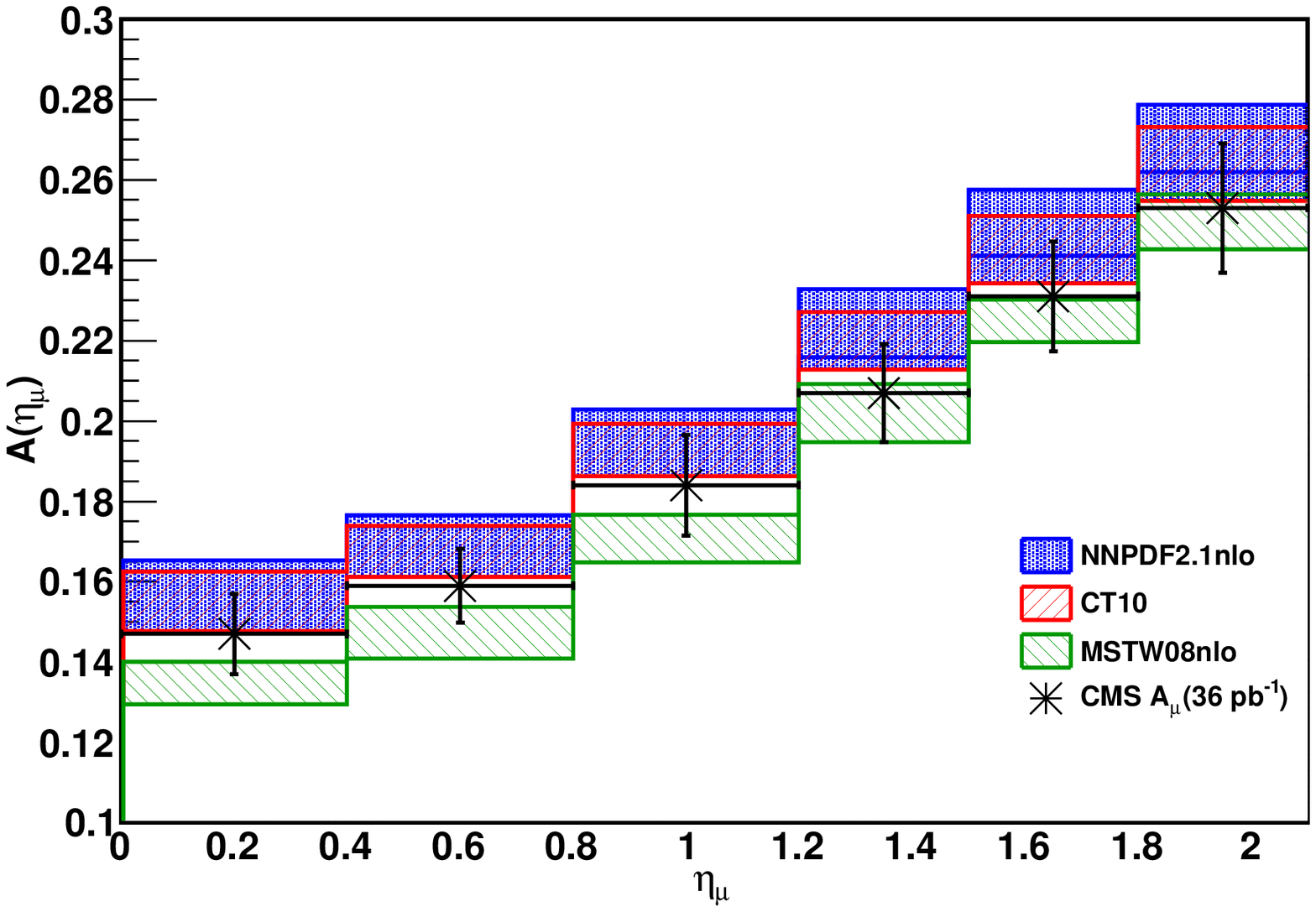}
    \label{fig:atlas-cms-chi2}
  \end{center}
  \caption{Predictions for the $W$ lepton asymmetry at NLO, obtained with 
    DYNNLO~\cite{dynnlo} using the CT10, MSTW08 and NNPDF2.1 parton sets, compared to
    measurements for the muon charge asymmetry from ATLAS~\cite{atlas}
    (left plot), and the electron (centre plot)
    and muon (right plot) charge asymmetries from CMS~\cite{cms}.}
\end{figure}

The ATLAS muon charge asymmetry data~\cite{atlas}
and CMS electron and muon data~\cite{cms} are compared to the predictions
obtained using three different NLO global fits, CT10~\cite{Lai:2010vv}, MSTW2008~\cite{Martin:2009iq} and
NNPDF2.1 in Fig.~\ref{fig:atlas-cms-chi2}.  The theoretical
predictions including NLO QCD corrections are obtained using the fully
differential Monte Carlo code DYNNLO~\cite{dynnlo} which allows for
the implementation of arbitrary experimental cuts.

\begin{table}[ht]
  \begin{center}
    \begin{tabular}{|c|c|c|c|c|}
      \hline
       & $N_{\mathrm{dat}}$& NNPDF2.1 & CT10 & MSTW08 \\
      \hline
      ATLAS(31pb$^{-1}$)                       & 11 & 0.76 & 0.77 & 3.32 \\
      \hline
      CMS(36pb$^{-1}$) electron $p_{T}>25$ GeV &  6 & 1.83 & 1.19  & 1.70 \\
      \hline
      CMS(36pb$^{-1}$) muon $p_{T}>25$ GeV     &  6 & 1.24 & 0.73  & 0.77 \\
      \hline
    \end{tabular}
  \end{center}
  \caption{Values of 
$\chi^{2}/N_{\rm dat}$ for the ATLAS and CMS lepton charge asymmetry data for different
    PDFs sets. Theory predictions are computed at NLO accuracy using the DYNNLO 
    code. Note that in Ref.~\cite{atlas} a somewhat lower value is
  quoted for MSTW08, due to the use of the MC@NLO code.}
  \label{tab:atlas-cms-chi2}
\end{table}

To give a more quantitative estimate of the level of agreement of the different predictions
with the experimental data, in Table ~\ref{tab:atlas-cms-chi2} we collect the $\chi^{2}$ 
per number of data points for each individual dataset. Since no covariance matrix is provided 
by the LHC experiments at this point, we add statistical and systematic uncertainties in 
quadrature in the computation of the $\chi^{2}$ values.

\begin{figure}[ht]
  \centering
    \epsfig{width=0.32\textwidth,figure=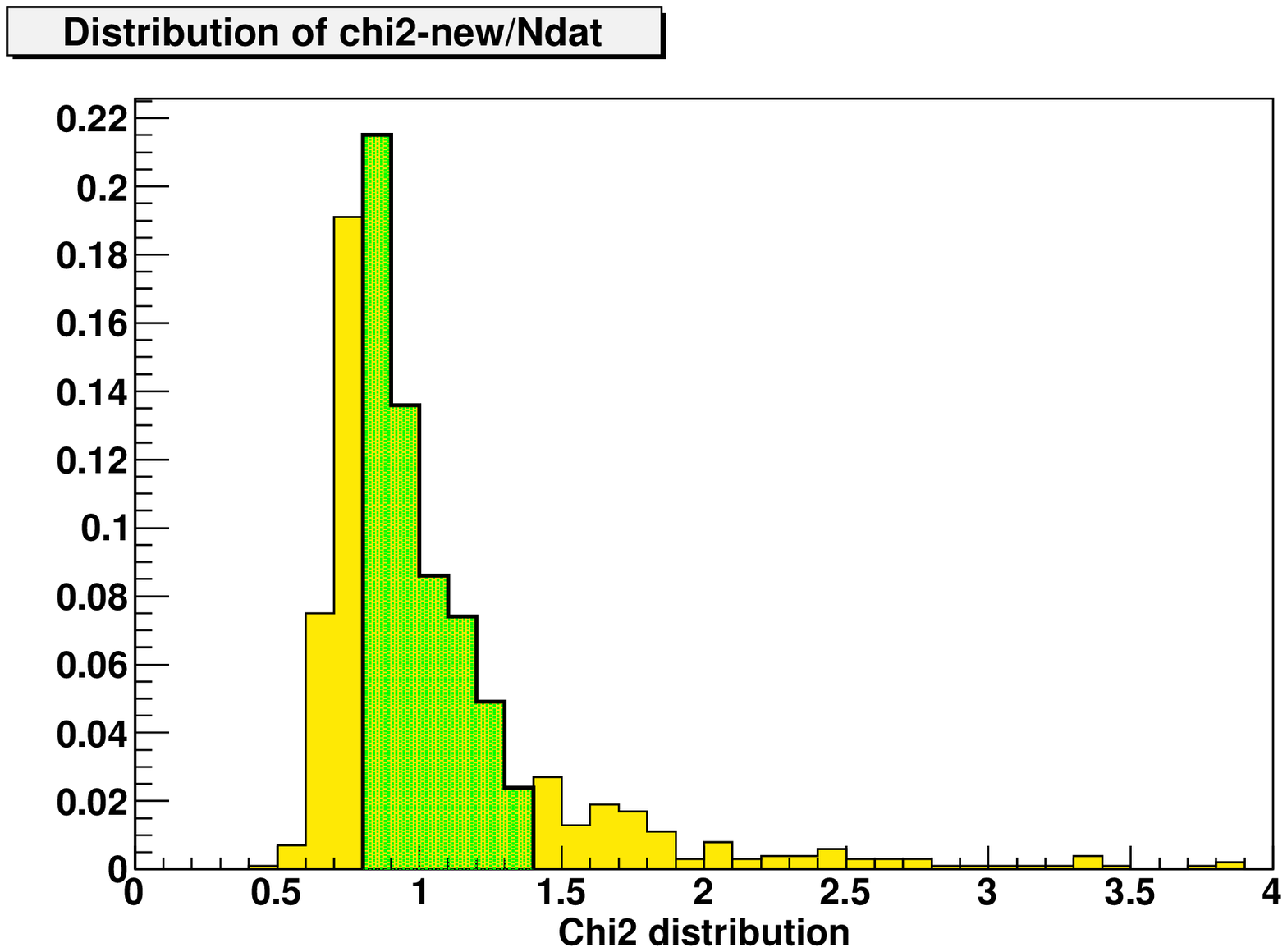}
    \epsfig{width=0.32\textwidth,figure=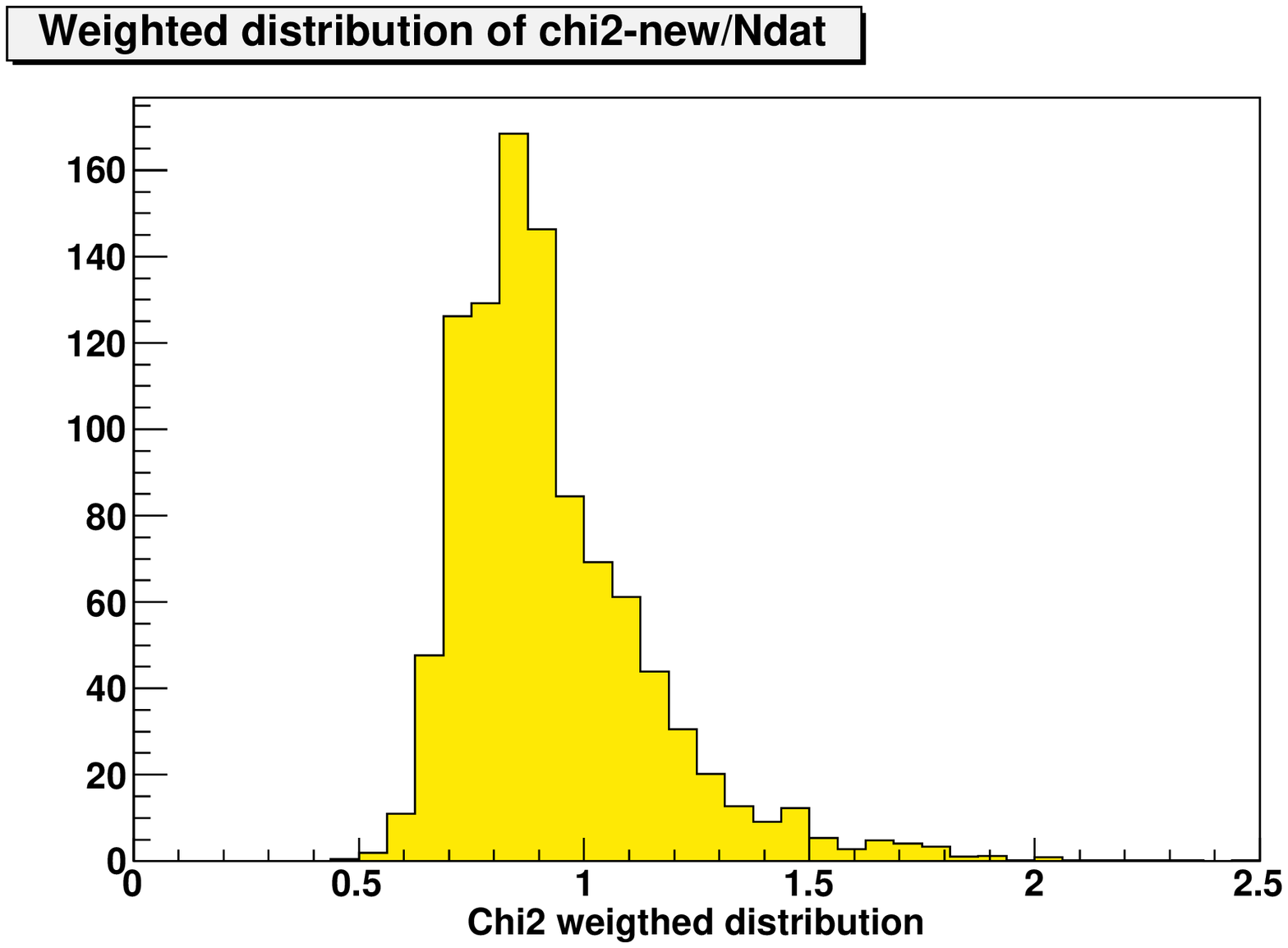}
    \epsfig{width=0.32\textwidth,figure=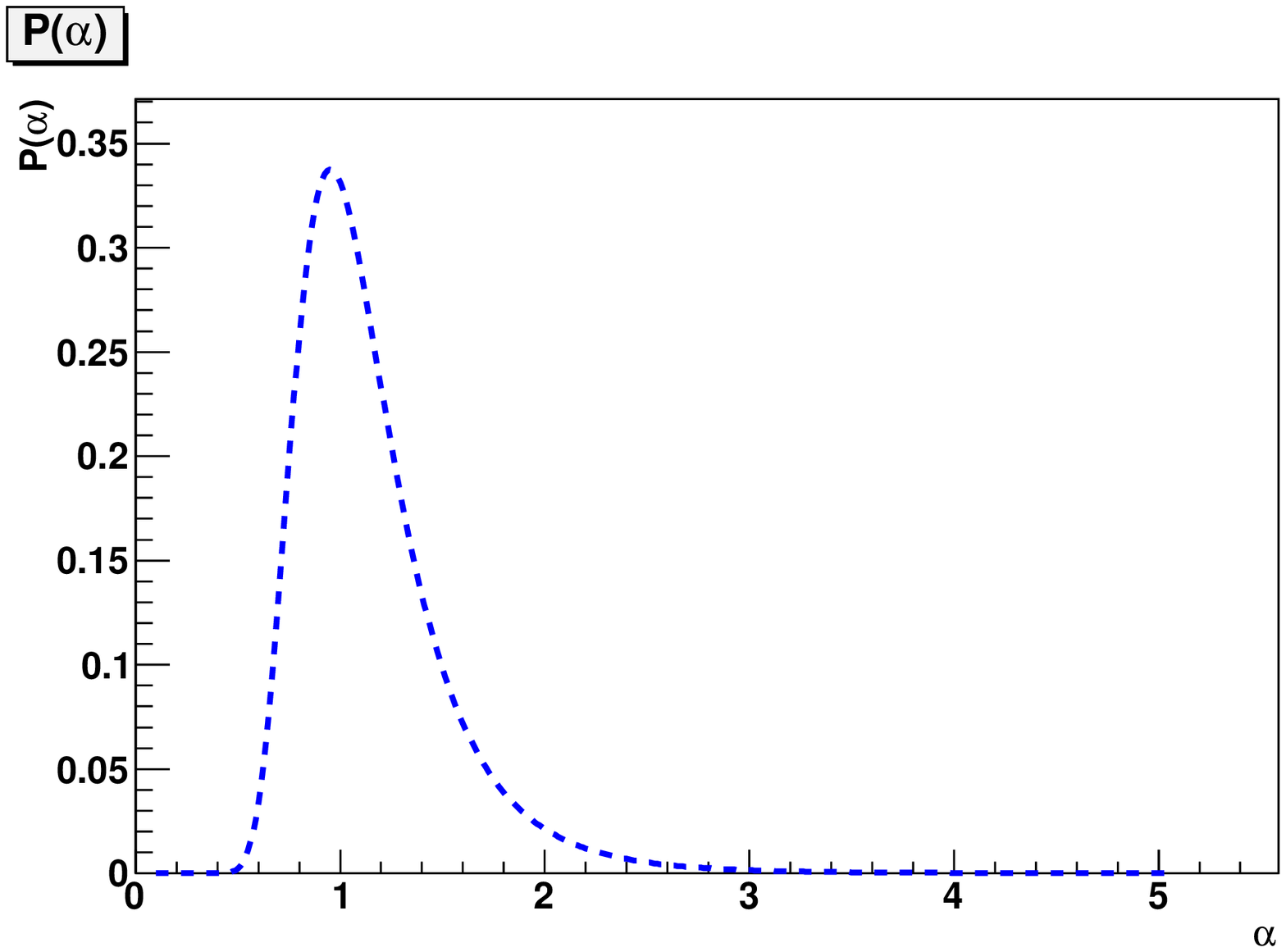}
    \caption{Distribution of $\chi^2/N_{\rm dat}$ for individual replicas prior (left)
      and after (middle) reweighting and ${\cal{P}}(\alpha)$ distribution (right) for the 
      ATLAS muon charge asymmetry data. In the left plot the shaded region corresponds to
    the central 68\% of the distribution.}
    \label{fig:chi2-atlas}
\end{figure}

The ATLAS muon charge asymmetry data are already very well described by the NNPDF2.1 
prediction before being included in the analysis. This is shown by the excellent 
$\chi^{2}/N_{\rm dat}=0.76$ reported in Table~\ref{tab:atlas-cms-chi2}
and
demonstrated by 
the distribution of $\chi^2$ for the individual replicas before reweighting shown  
in the left plot of Fig.~\ref{fig:chi2-atlas}, which has a sharp peak around one.
The compatibility of a new dataset with the data already included in a global 
analysis can be assessed by looking at the probability density for the parameter 
$\alpha$, ${\cal{P}}(\alpha)$ defined in Eq.~(12) of~\cite{Ball:2010gb}.
If this probability distribution peaks close to one, the new data are consistent
with the ones already included in the global fit. 
For the ATLAS data, the $P(\alpha)$ distribution, shown in the right plot of
Fig.~\ref{fig:chi2-atlas}, is peaked slightly below one, thereby showing
the good compatibility of these data with those included in the global 
analysis. Note that optimal values of $\chi^{2}/N_{\rm dat}$ are to
be expected because statistical and systematic errors have been added
in quadrature, thereby leading to an overestimation of uncertainties.

\begin{figure}[h!]
  \centering
    \epsfig{width=0.44\textwidth,figure=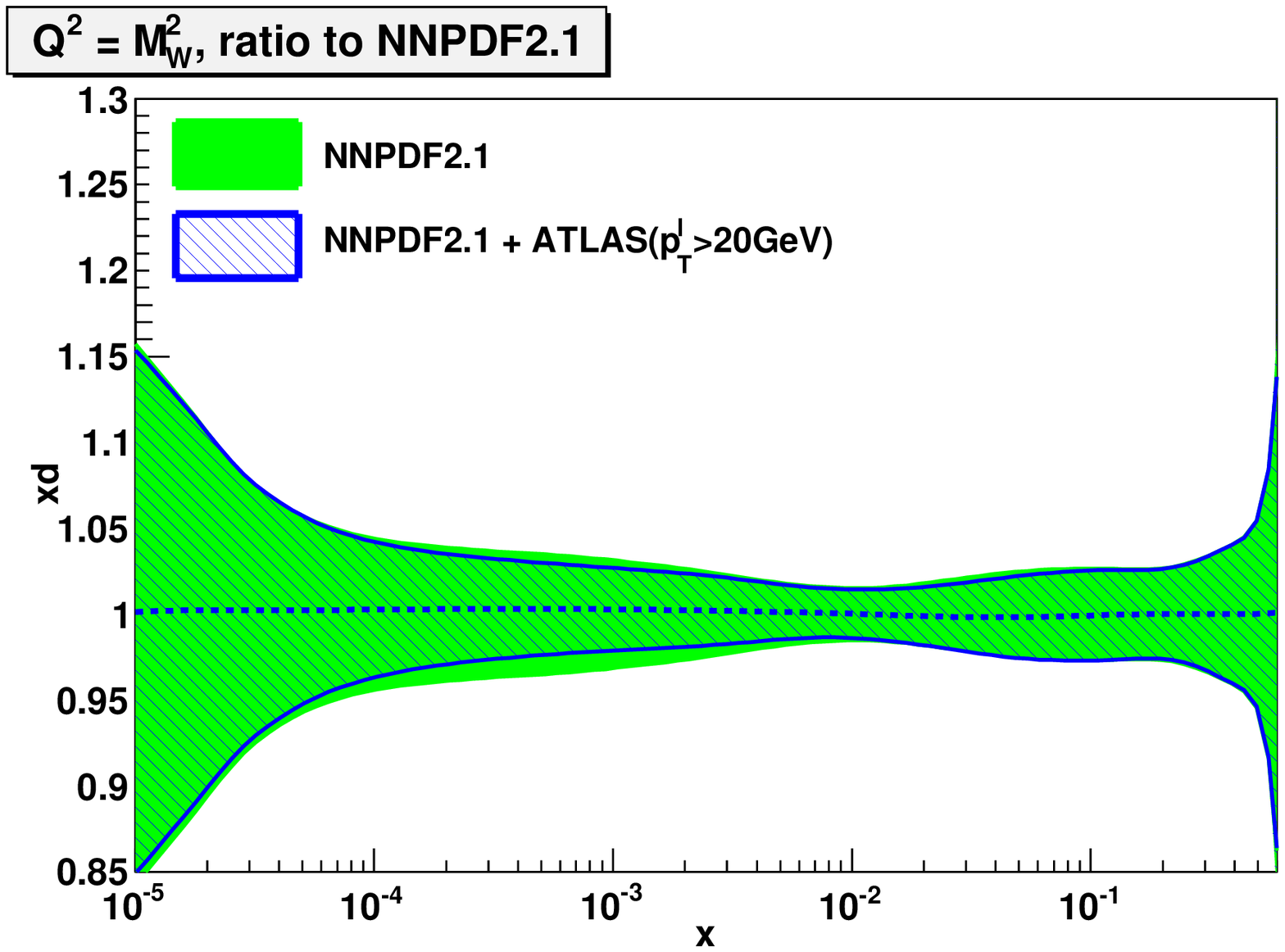}
    \epsfig{width=0.44\textwidth,figure=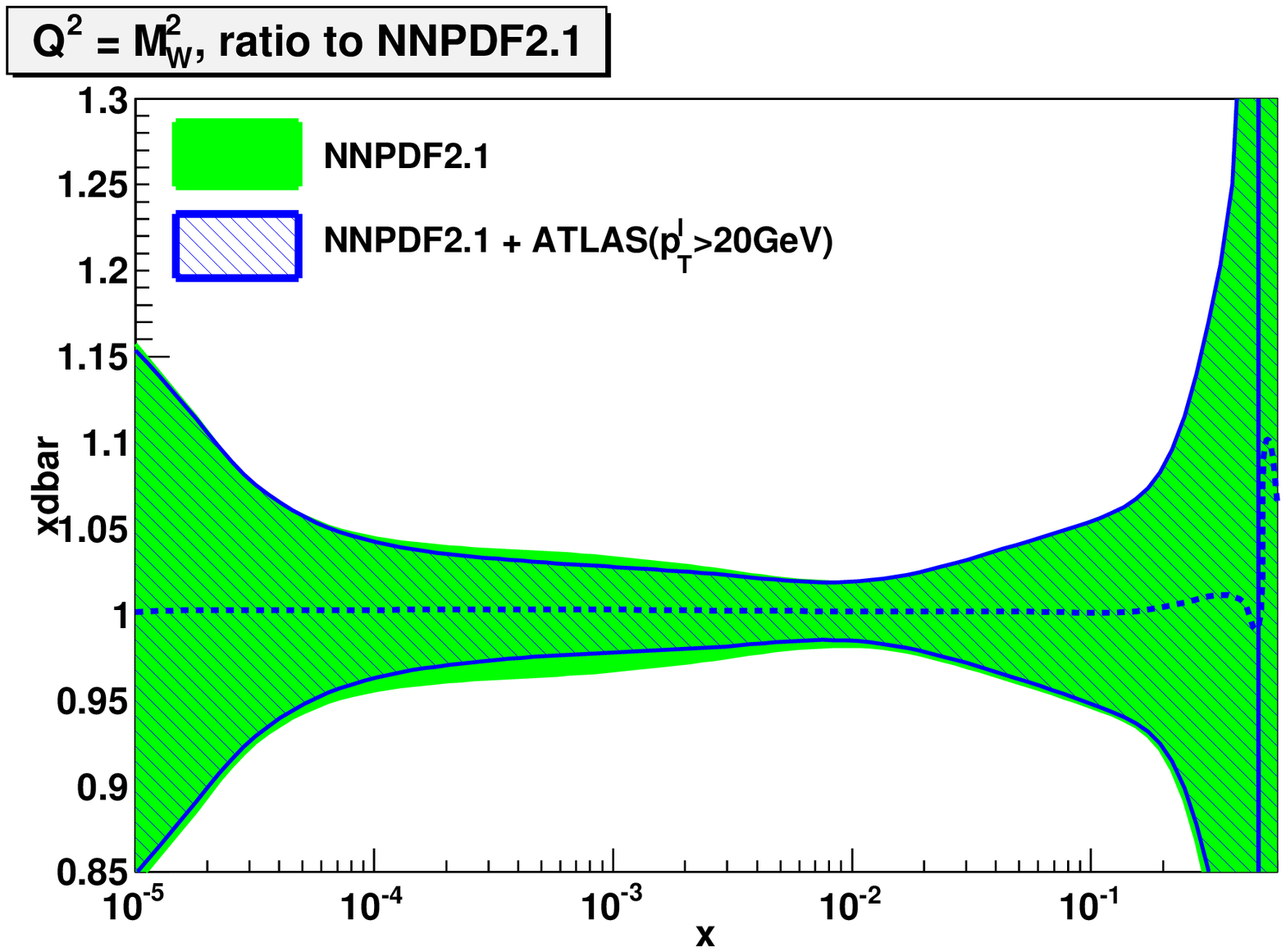}
    \epsfig{width=0.44\textwidth,figure=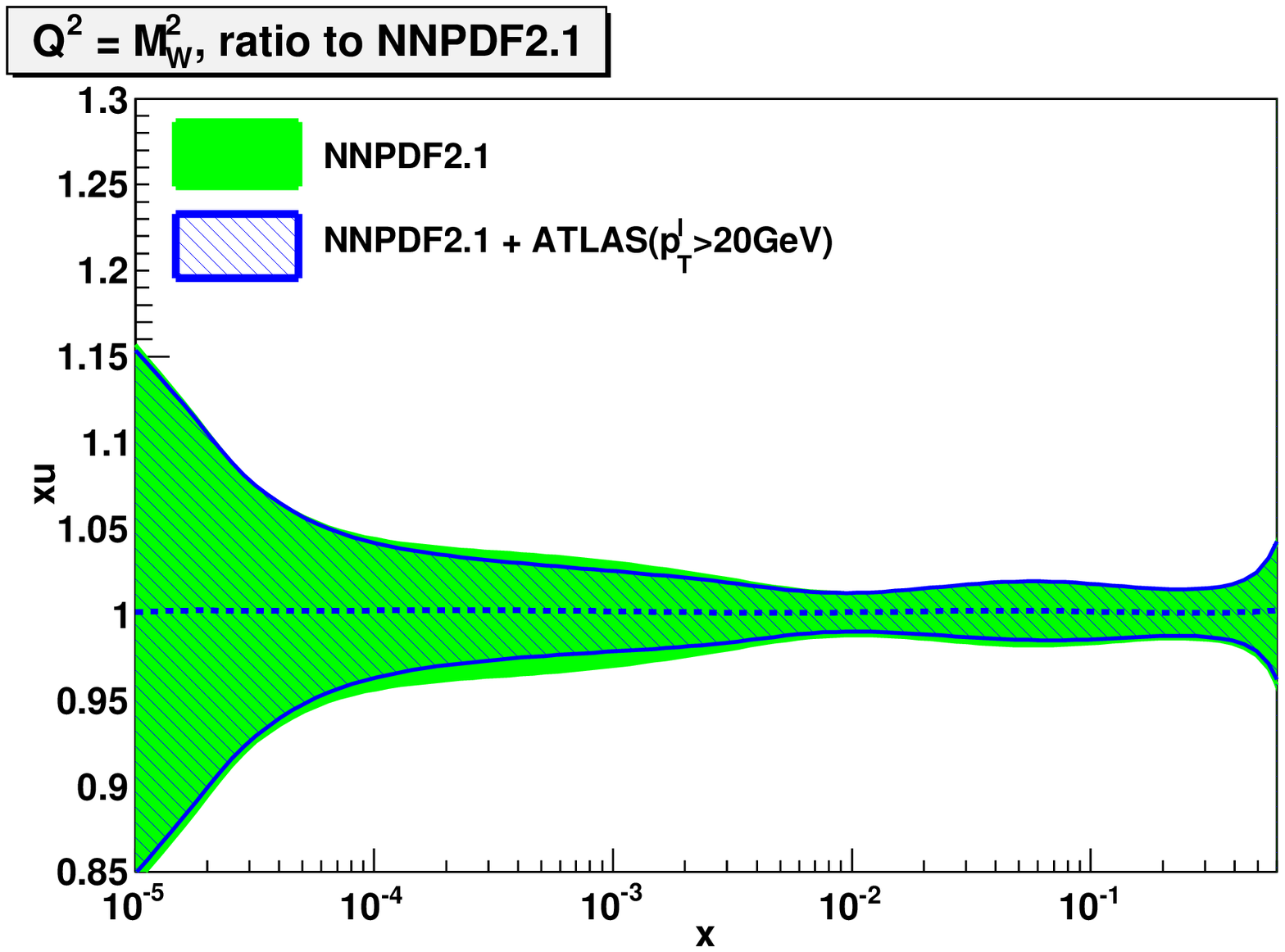}
    \epsfig{width=0.44\textwidth,figure=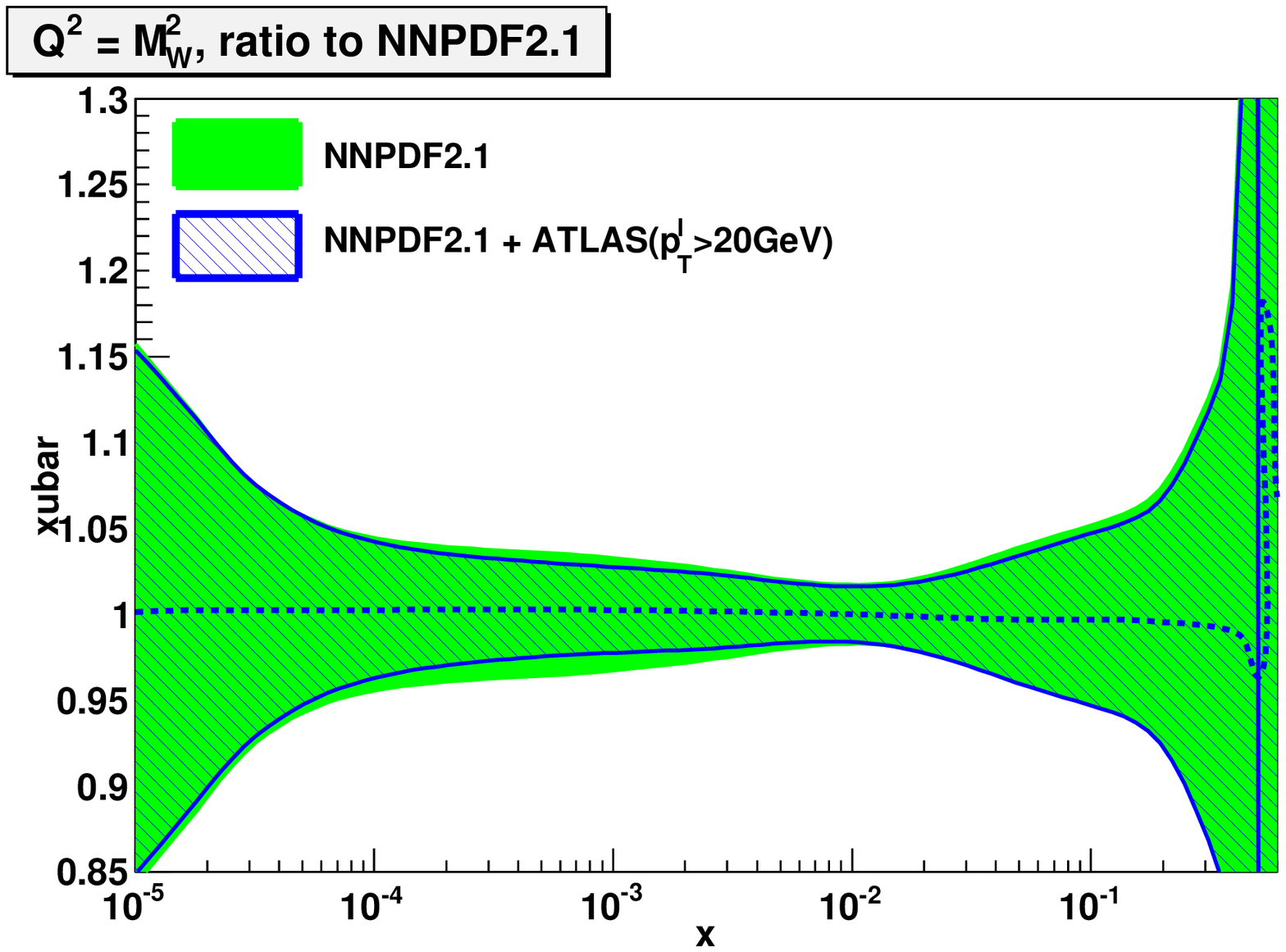}
    \caption{Comparison of light quark and antiquark distributions at the scale
      $Q^2=M_W^2$ from the global NNPDF2.1 NLO global fit and the same distributions
      obtained after adding ATLAS muon charge asymmetry data via reweighting.
      Parton densities are plotted normalized to the NNPDF2.1 central value.}
    \label{fig:pdfs-atlas}
\end{figure}

After reweighting NNPDF2.1 with the ATLAS data the quality of
their description remains substantially unchanged, with the value
$\chi^2_{\rm rw}/N_{\rm dat}=0.72$.  The number of effective
replicas of the reweighted sets computed according to Eq.~(42) in
Appendix of~\cite{Ball:2010gb} is $N_{\rm eff}=928$, out of the
initial number of $N_{\rm rep}=1000$ replicas in the prior. The
distribution of the $\chi^{2}/N_{\rm dat}$ for the weighted
replicas, shown in the center plot of Fig.~\ref{fig:chi2-atlas}, peaks
just below one, again confirming the very good description of these data
also after reweighting.

Given the outcome of the previous statistical analysis -- a very good
description of the data by the prior set to start with, 
resulting in a large number of
surviving replicas ($N_{\rm eff}=928$) -- it is easy to
predict that the ATLAS data alone will impose only mild constraints
on the underlying PDFs.  This is in fact what is seen in
Fig.~\ref{fig:pdfs-atlas} where we compare the NNPDF2.1 light
(anti)flavour densities at the scale $Q^2=M_W^2$ to the ones obtained
after reweighting with the ATLAS data. The most noticeable effect is
a reduction of the uncertainties on these PDFs in the
medium-small $x$ region, around $x\sim 10^{-3}$, by up to $20\%$.

\begin{figure}[ht]
  \centering
    \epsfig{width=0.30\textwidth,figure=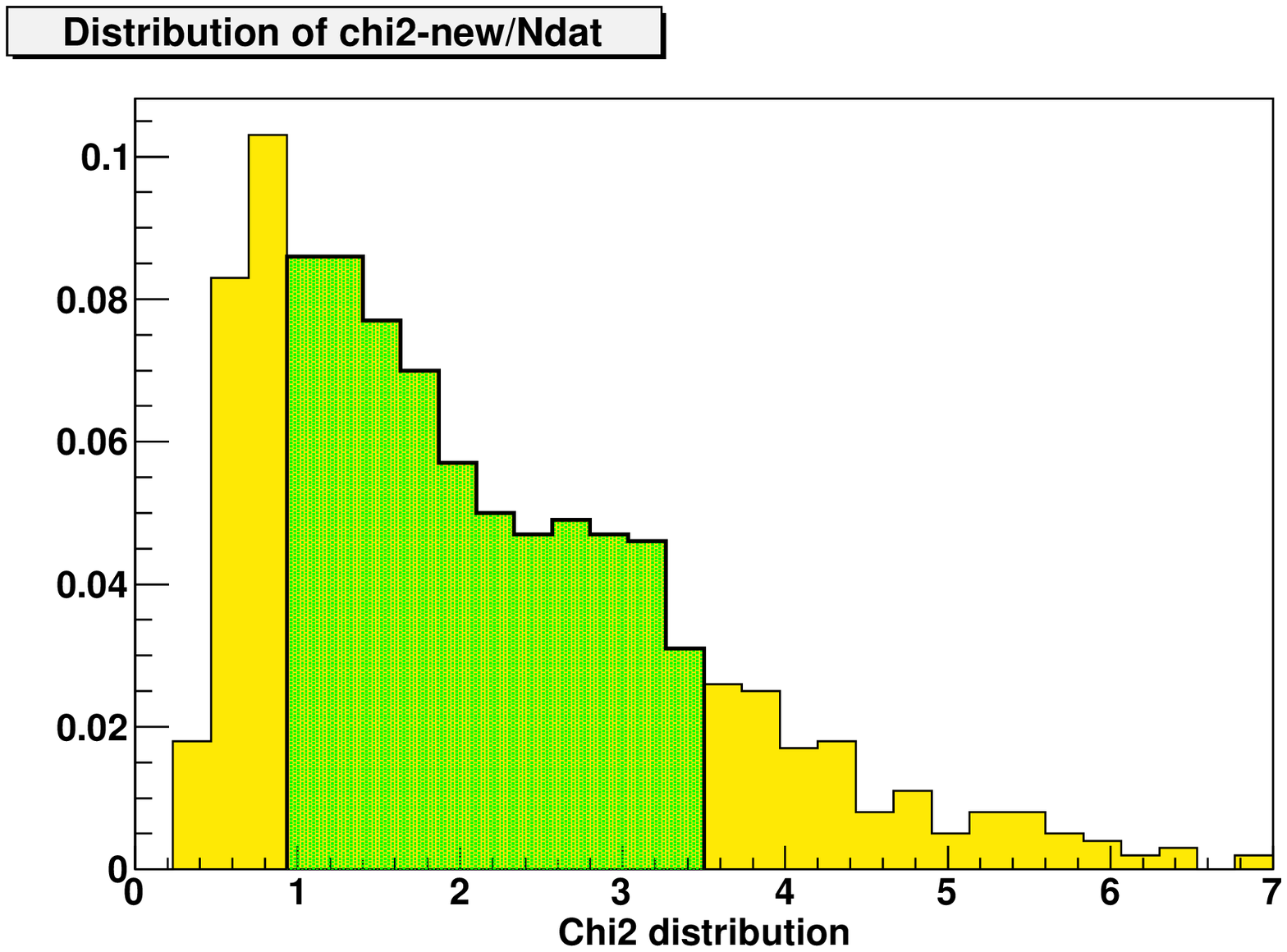}
    \epsfig{width=0.30\textwidth,figure=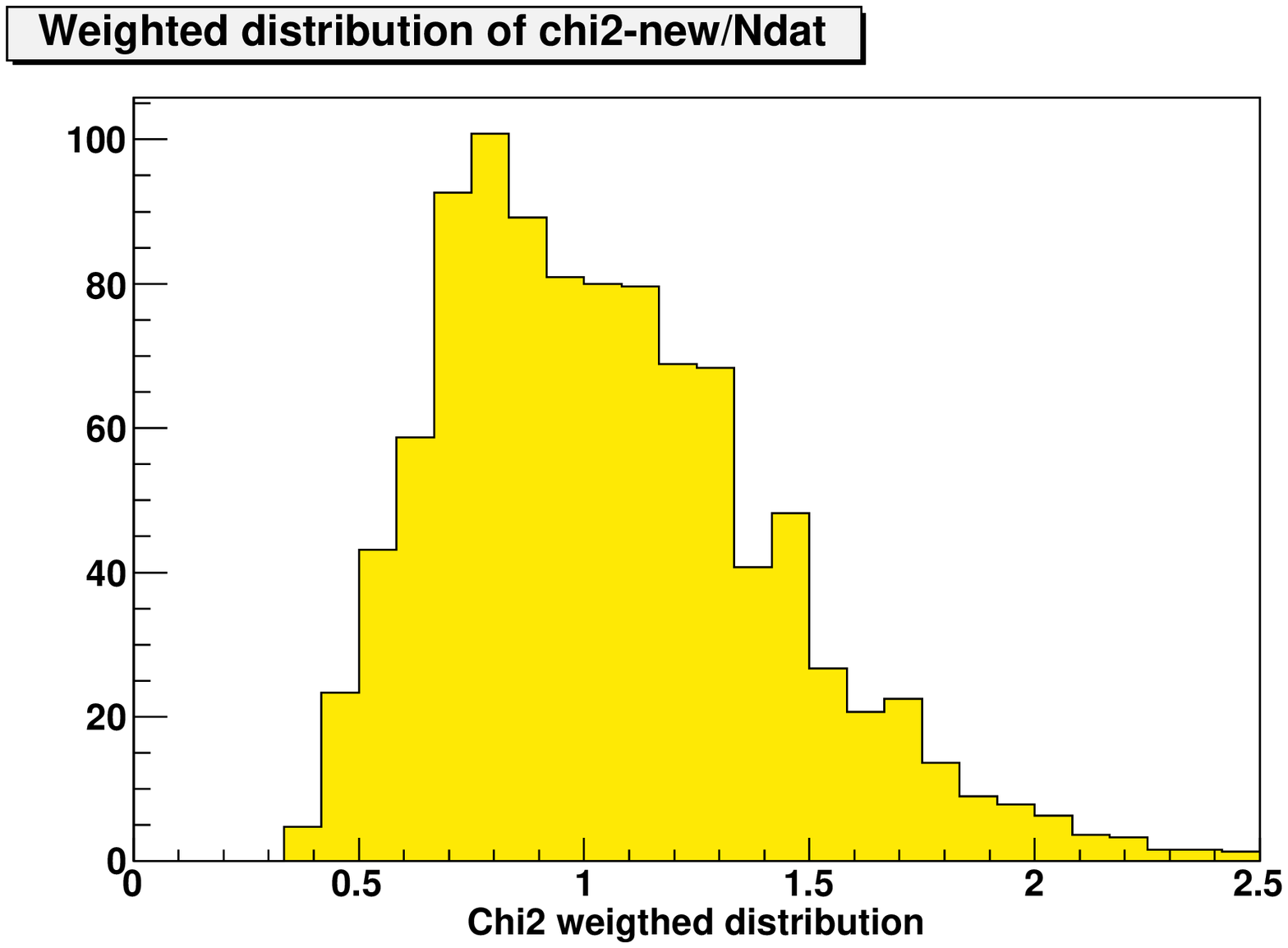}
    \epsfig{width=0.30\textwidth,figure=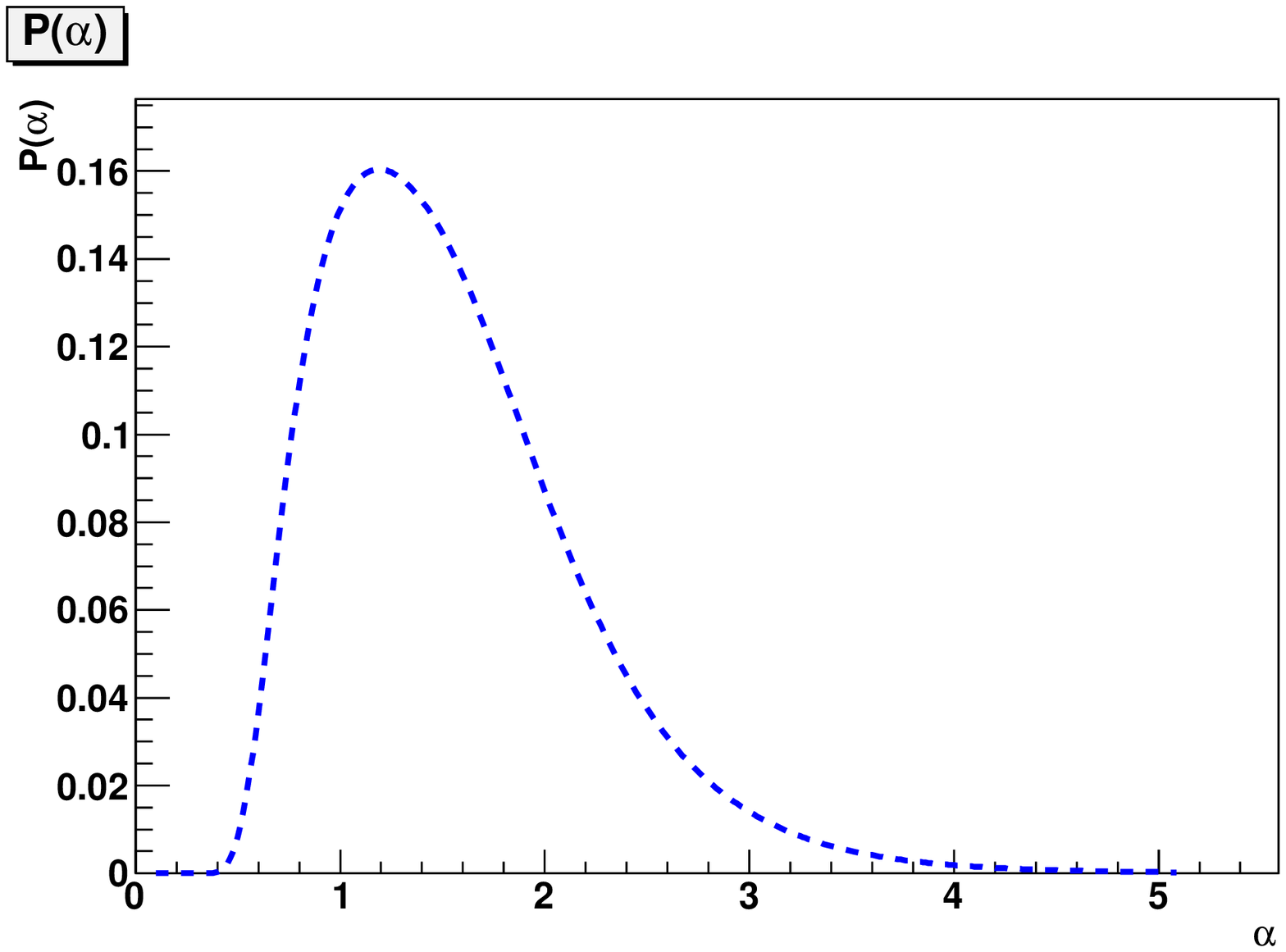}\\
    \epsfig{width=0.30\textwidth,figure=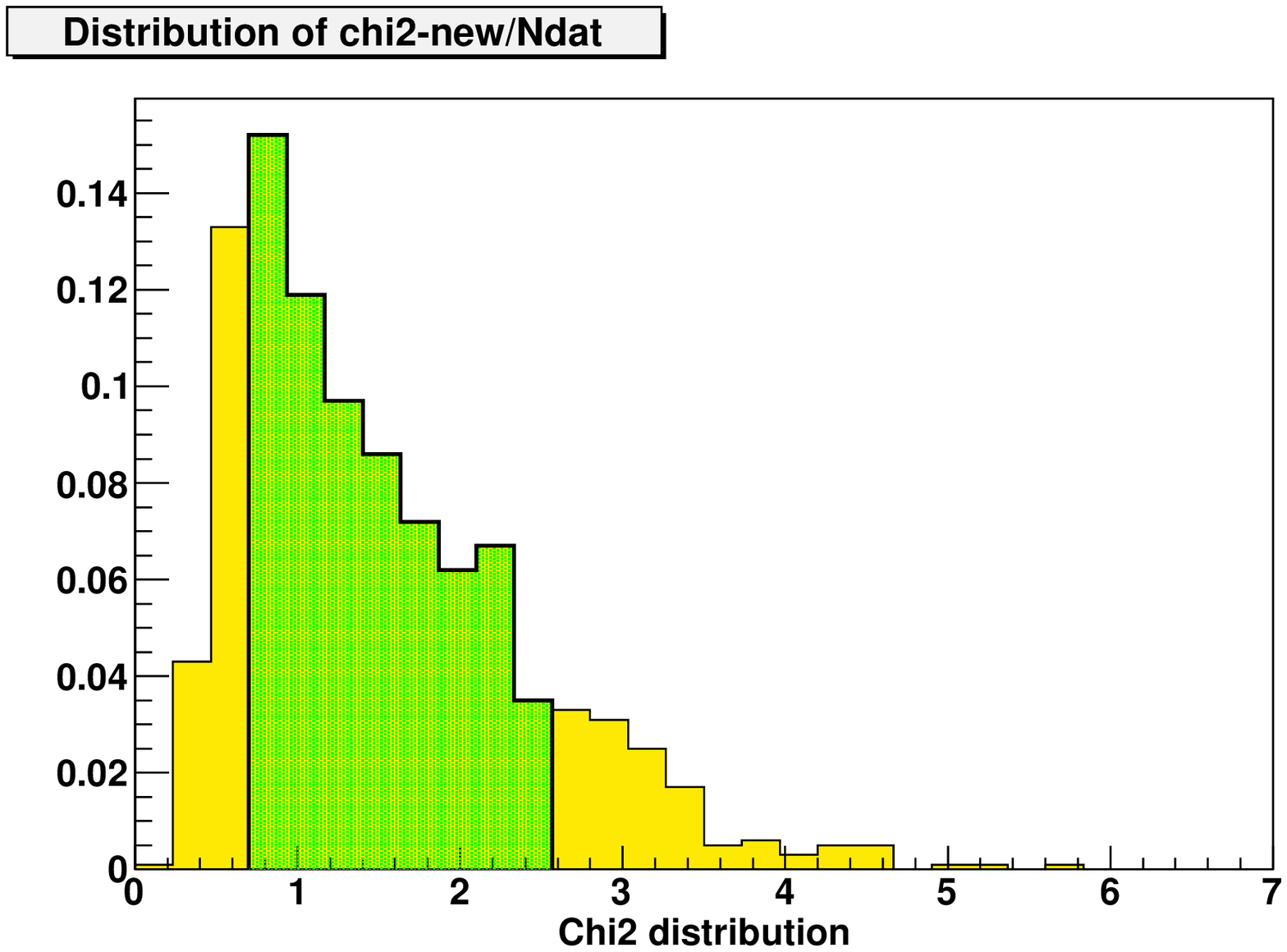}
    \epsfig{width=0.30\textwidth,figure=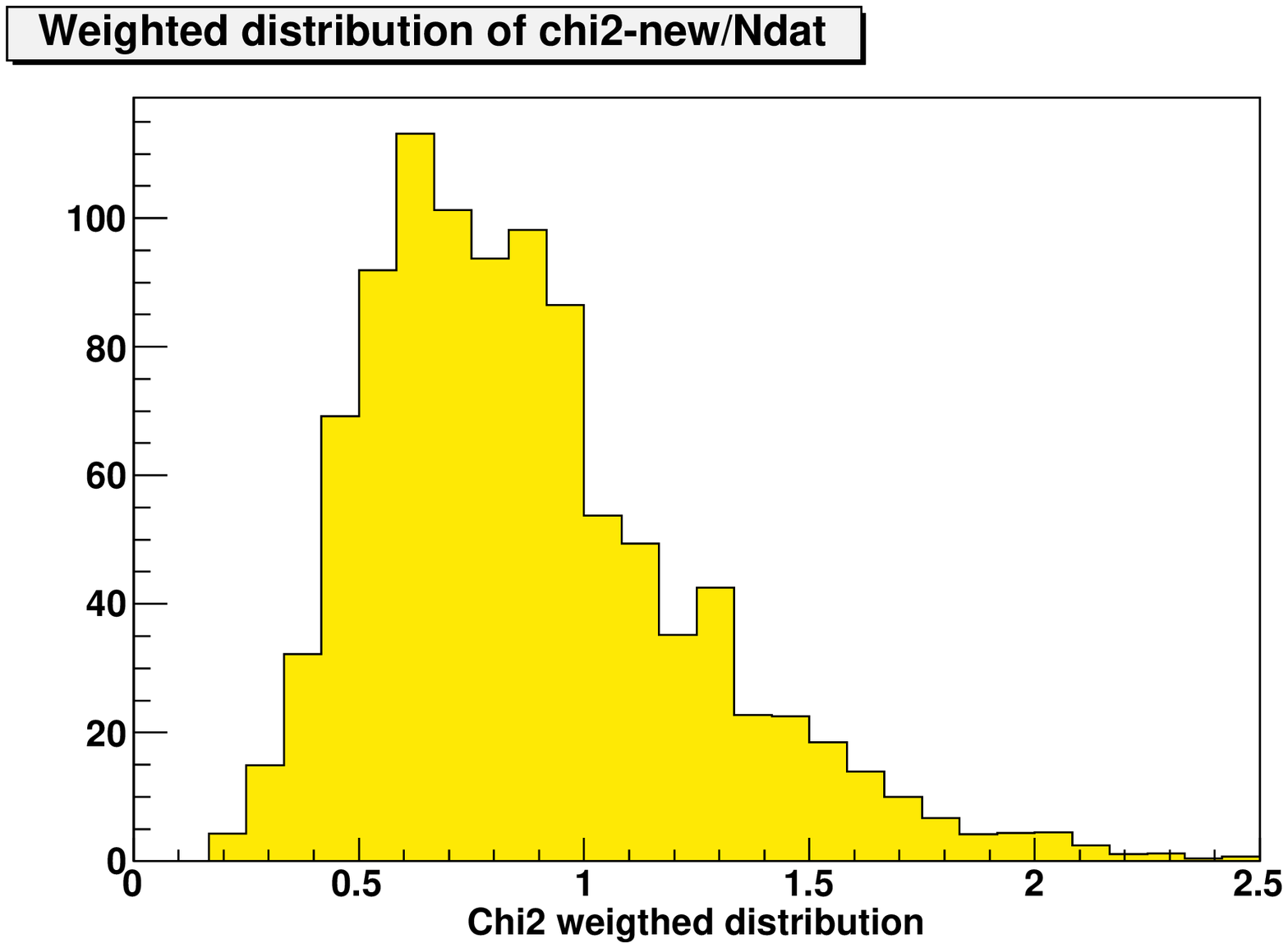}
    \epsfig{width=0.30\textwidth,figure=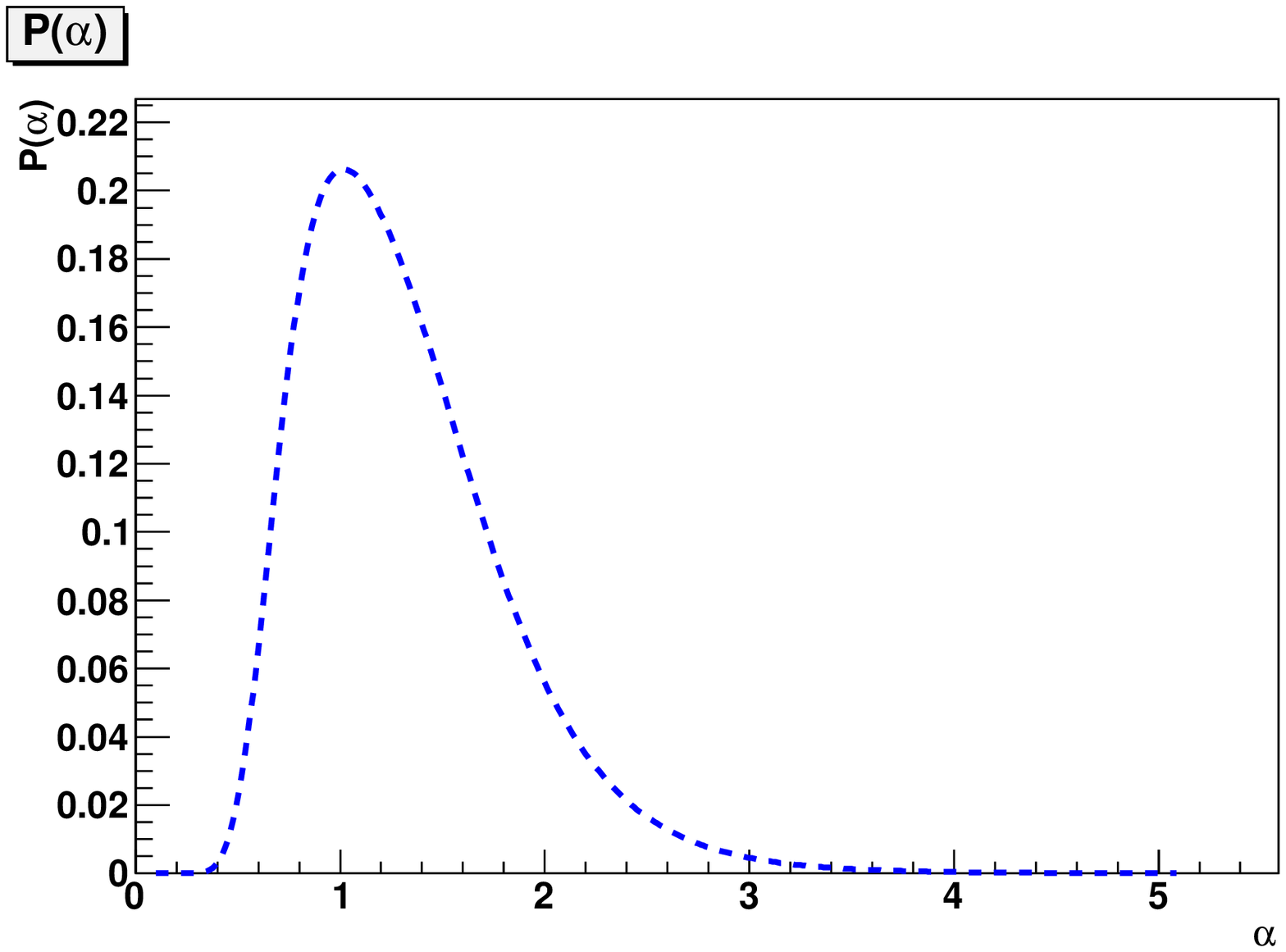}
 \caption{Same as Fig.~\ref{fig:chi2-atlas} for  the 
    CMS($p_{T}>25$ GeV) (top) and CMS($p_{T}>30$ GeV) (bottom) lepton charge 
    asymmetry data. }
  \label{fig:chi2-cms}
\end{figure}
\begin{figure}[h!]
  \centering
  \epsfig{width=0.44\textwidth,figure=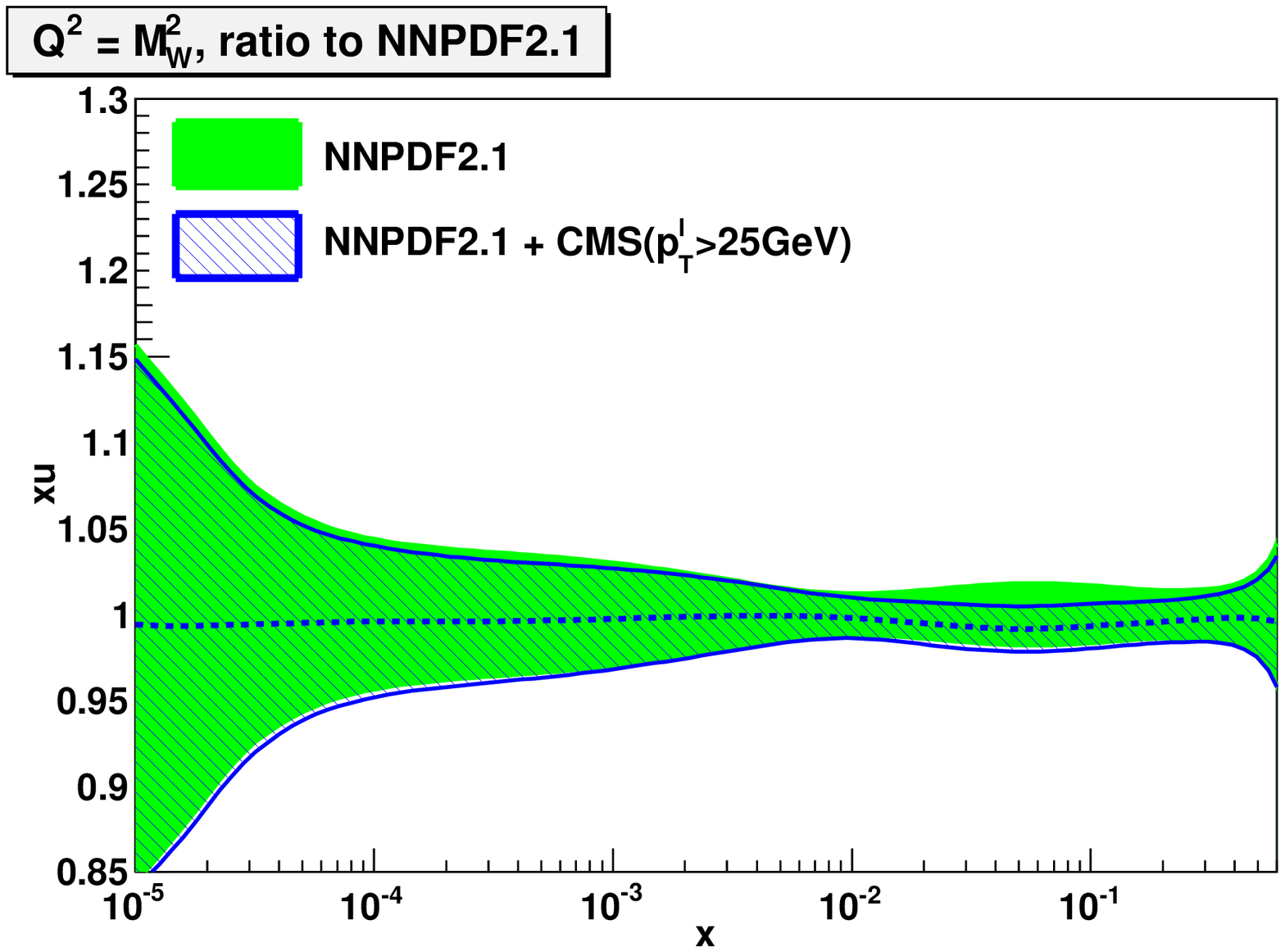}
  \epsfig{width=0.44\textwidth,figure=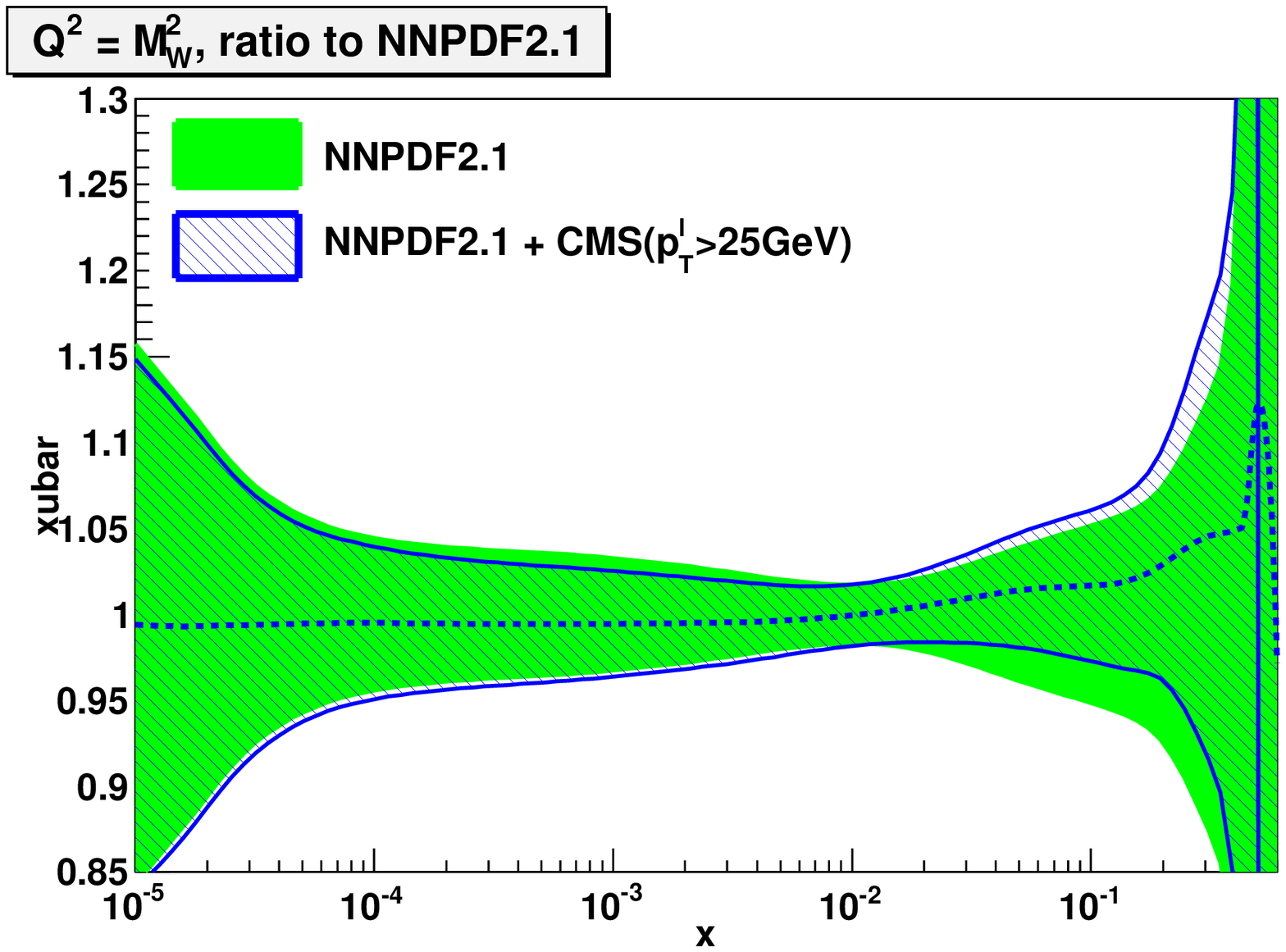}
  \epsfig{width=0.44\textwidth,figure=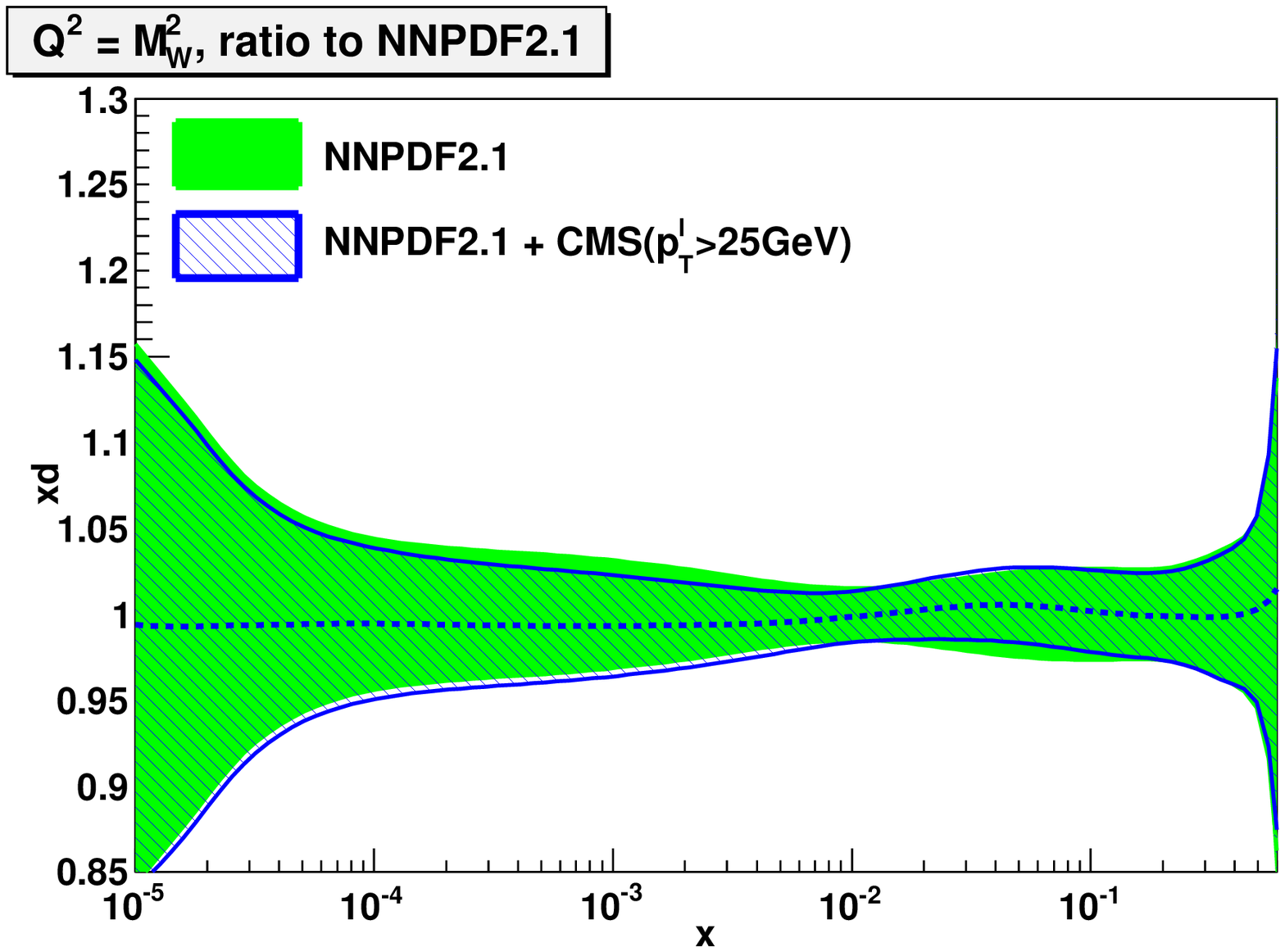}
  \epsfig{width=0.44\textwidth,figure=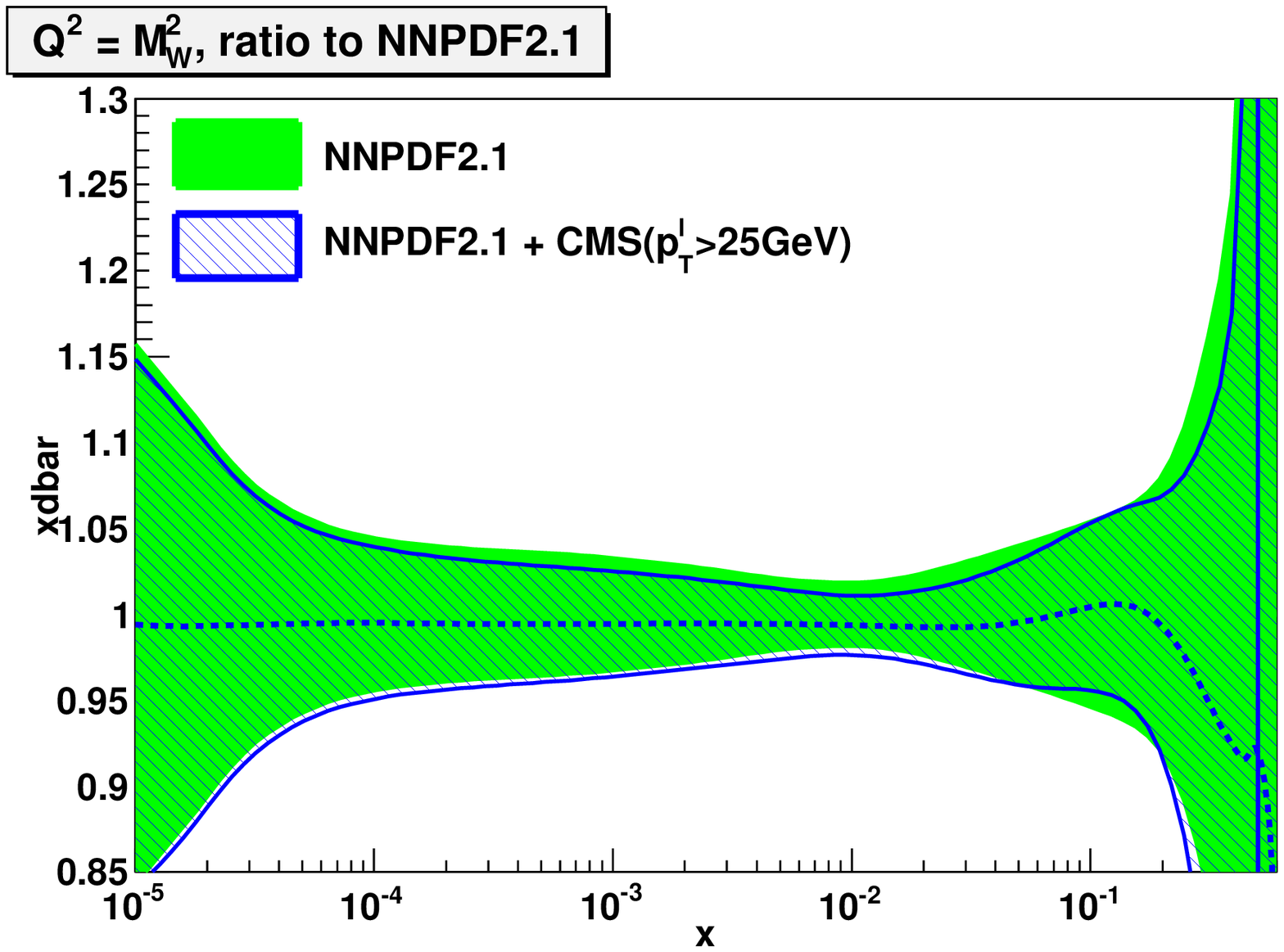}
  \caption{Same as Fig.~\ref{fig:pdfs-atlas} but after adding CMS lepton charge asymmetry data.}
  \label{fig:pdf-cms-25}
\end{figure}

We now turn to the CMS measurements described in~\cite{cms}. CMS presented data
for both the electron and muon charge asymmetries from $W$ decays with two 
different cuts on the transverse momentum of the detected lepton: $p_\perp>25$ GeV and
$p_\perp>30$ GeV. From the values for $\chi^2/N_{\rm dat}$ obtained using the NNPDF2.1
global set reported in Table~\ref{tab:atlas-cms-chi2}, and the plots of the distribution
of $\chi^2/N_{\rm dat}$ for individual replicas and of the ${\cal{P}(\alpha)}$ 
distribution shown in Fig.~\ref{fig:chi2-cms}, we see that both sets are equally well
described by the NNPDF2.1 set and thus compatible with the data included in the global 
analysis.
Since the two datasets are not independent we have to choose which one to use in our
reweighting analysis and thus we only
consider the dataset with the 
looser cut $p_{T}>25$ GeV, which proves to be more constraining of the PDFs. 
We perform our reweighting analysis including the muon and electron data as a single 
dataset.

The NNPDF2.1 prediction provides a good, though not optimal, description of the CMS 
data, as shown by the $\chi^{2}/N_{\rm dat}=1.51$ obtained combining the values for the
electron and muon data collected in Table~\ref{tab:atlas-cms-chi2}.
After reweighting, the description of these data improves significantly with 
$\chi^2_{\rm rw}/N_{\rm dat}=0.77$.
The number of effective replicas computed as above is roughly half the initial 
number of replicas, $N_{\rm eff}=531$ out of $N_{\rm rep}=1000$, suggesting that 
these data will have have a significant impact on the PDFs. 
The distribution of the $\chi^{2}/N_{\rm dat}$ of 
individual replicas after reweighting is centered around one, as shown
in the middle-upper plot of Fig.~\ref{fig:chi2-cms}.

The impact of the CMS data on light (anti)flavour PDFs, is 
shown in Fig.~\ref{fig:pdf-cms-25} where we observe a reduction of 
uncertainties in the medium $x$ region smaller than that due to the ATLAS data, 
but also a change in the shape of the $\bar{u}$ and $\bar{d}$ distributions at 
relatively large $x\sim 0.1$, pushing up the central value a little and reducing the 
uncertainties by around $10\%$ for the down distributions and as much as $25\%$ for the up.

\begin{figure}[ht]
  \centering
  \epsfig{width=0.44\textwidth,figure=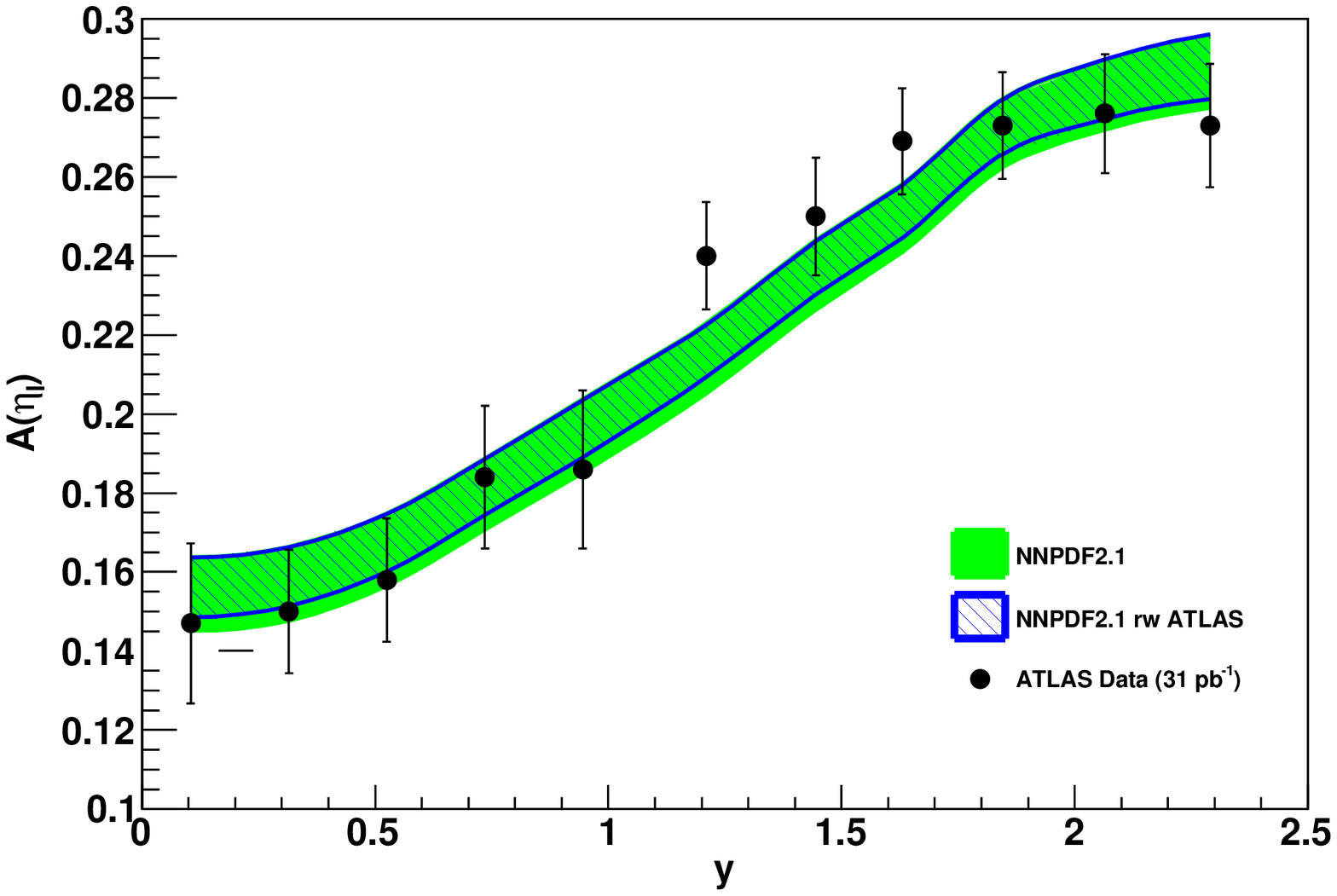}
  \epsfig{width=0.44\textwidth,figure=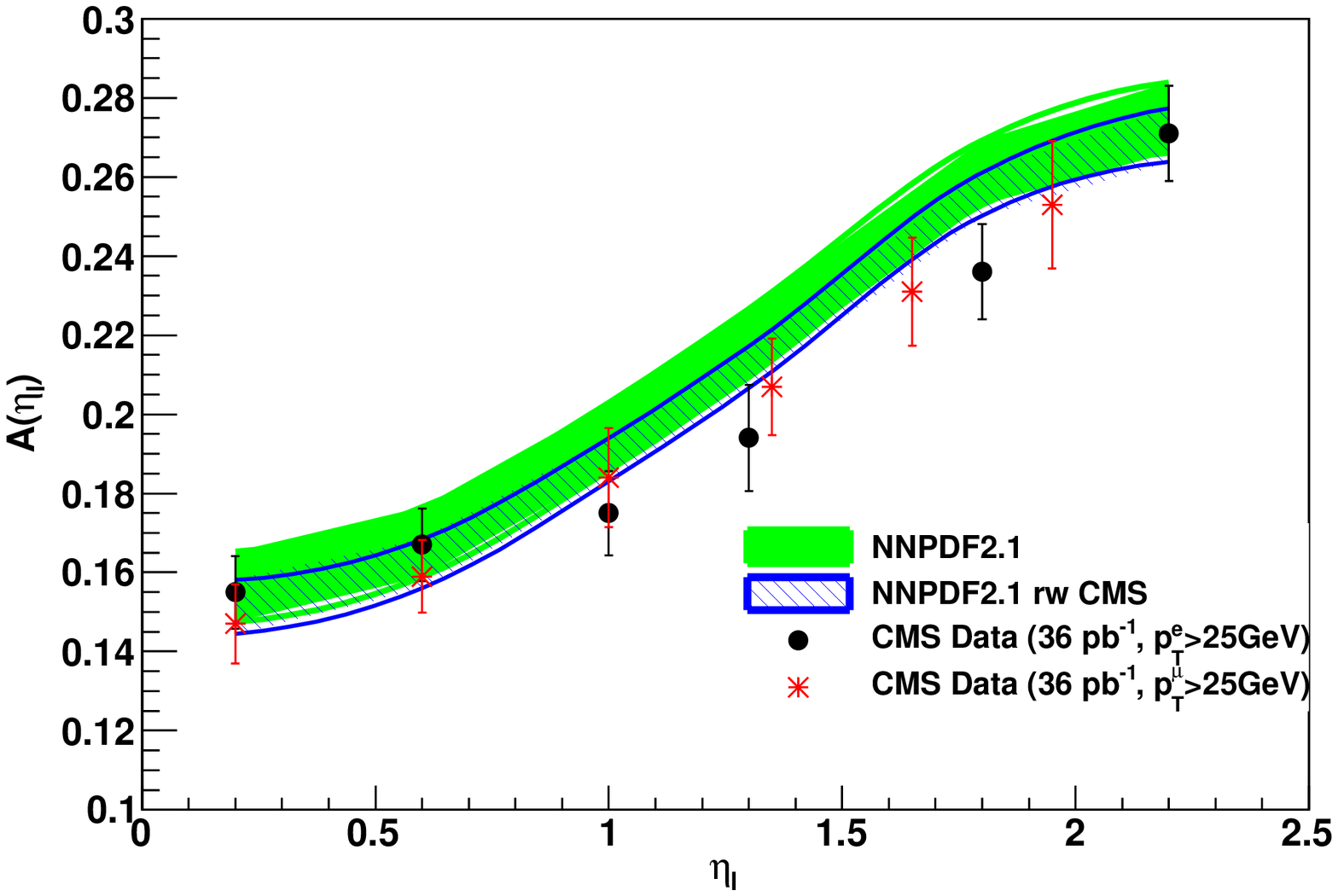}
  \caption{Comparison of the lepton charge asymmetry from $W$ boson production computed
    with the NNPDF2.1 NLO PDF set and sets where ATLAS (left) and CMS (right) lepton charge
    asymmetry data have been included using reweighting.}
  \label{fig:wasy_rw-atlas-cms}
\end{figure}

We conclude this Section by comparing the predictions for the charge asymmetry computed
with NNPDF2.1 and NNPDF2.1 after reweighting with the ATLAS and CMS data respectively in 
Fig.~\ref{fig:wasy_rw-atlas-cms}. The effect on the prediction for the CMS data is more substantial, 
because the data undershoot the NNPDF2.1 NLO prediction in most of the higher rapidity bins. 

\subsection{Combination of ATLAS and CMS data}

We now consider adding the ATLAS and CMS lepton charge asymmetry data as a single 
dataset to the NNPDF2.1 NLO global fit using reweighting.

\begin{figure}[ht]
  \centering
  \epsfig{width=0.30\textwidth,figure=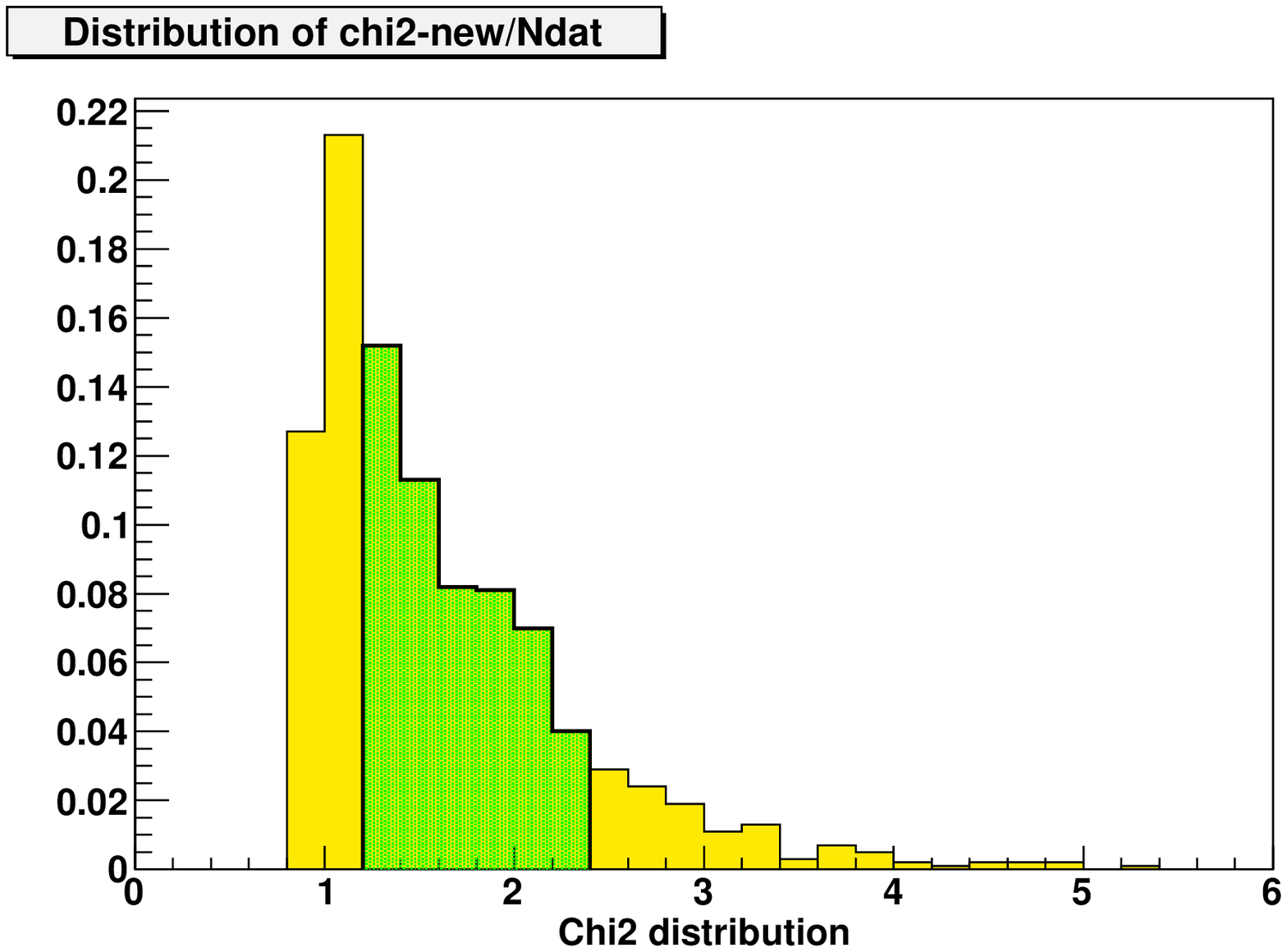}
  \epsfig{width=0.30\textwidth,figure=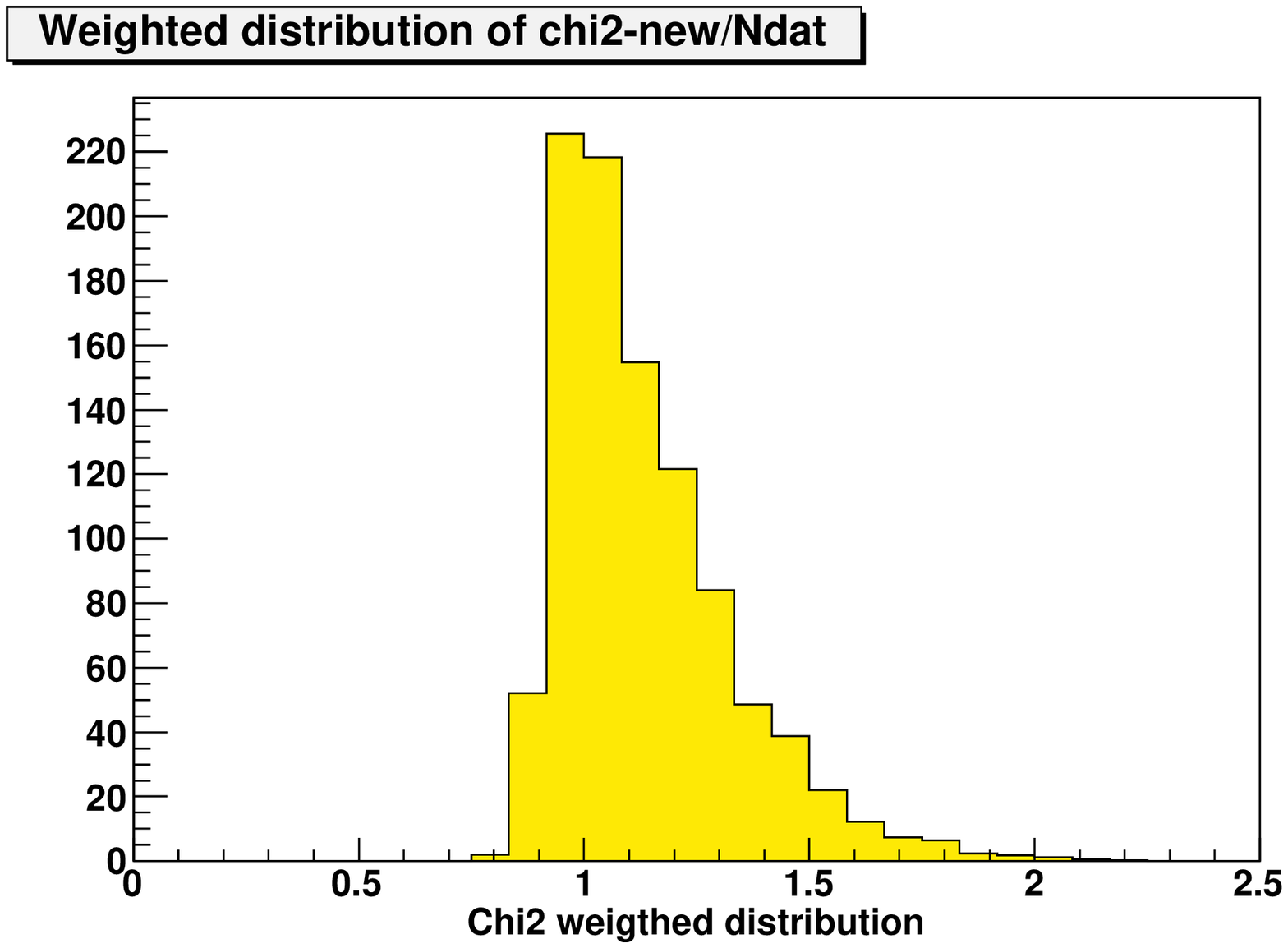}
  \epsfig{width=0.30\textwidth,figure=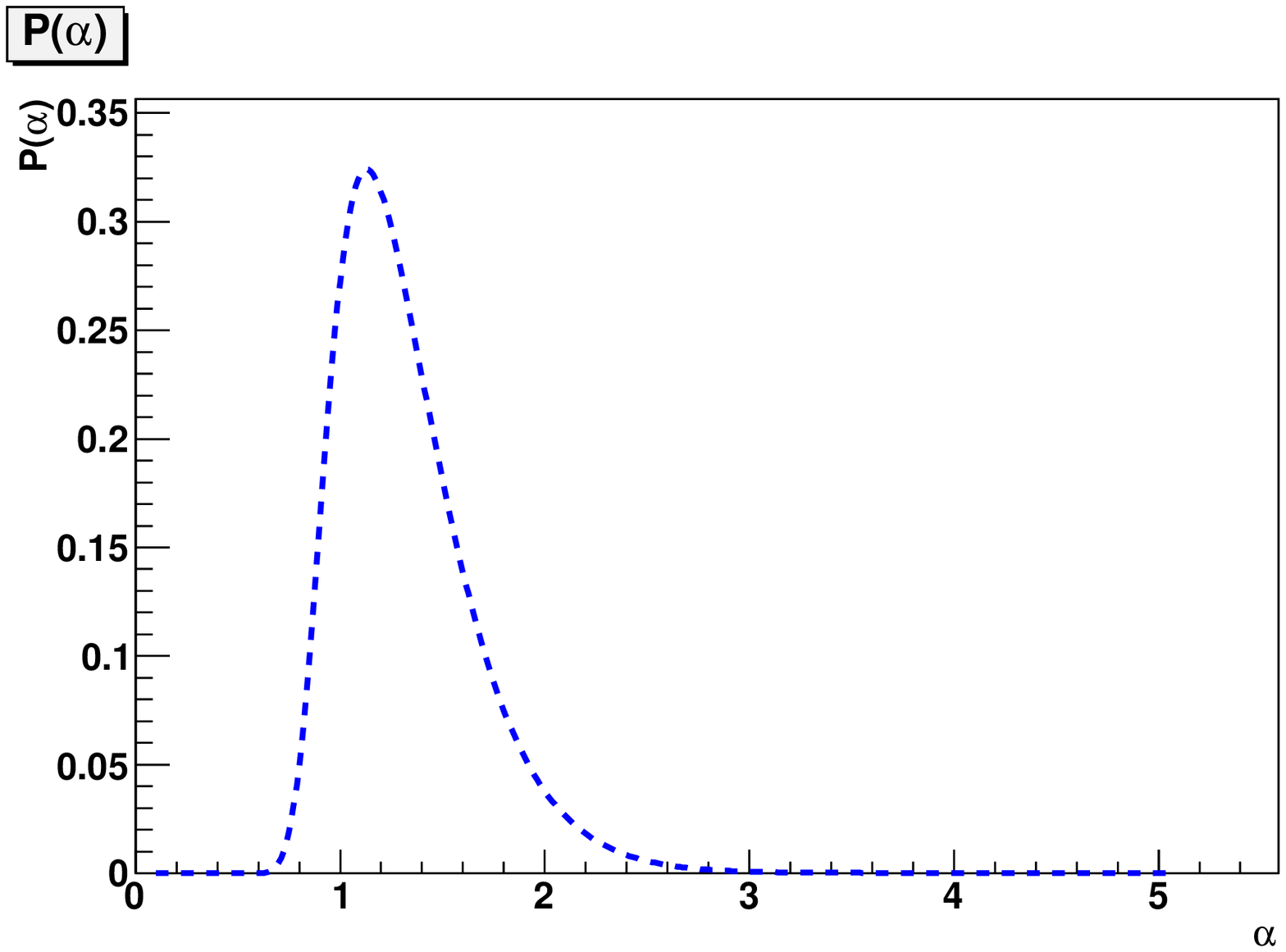}
  \caption{Same as Fig.~\ref{fig:chi2-atlas} for the combined
    ATLAS+CMS~($p_{T}>25$ GeV) lepton charge asymmetry data.}
  \label{fig:chi2-lhc25}
\end{figure}

The whole dataset is already well described by the NNPDF2.1 NLO dataset with 
$\chi^{2}/N_{\rm dat}=1.17$ and the distributions of $\chi^{2}/N_{\rm dat}$ for individual replicas
having a sharp peak around one, as shown by the left plot in Fig.~\ref{fig:chi2-lhc25}. 
The compatibility of the ATLAS+CMS data  with the data included in the global analysis and
among the two experiments is also good, as can be deduced by looking at the 
${\cal{P}}(\alpha)$ distribution shown in the right plot in Fig.~\ref{fig:chi2-lhc25}, 
which is nicely peaked around one. 

\begin{figure}[h!]
  \centering
  \epsfig{width=0.44\textwidth,figure=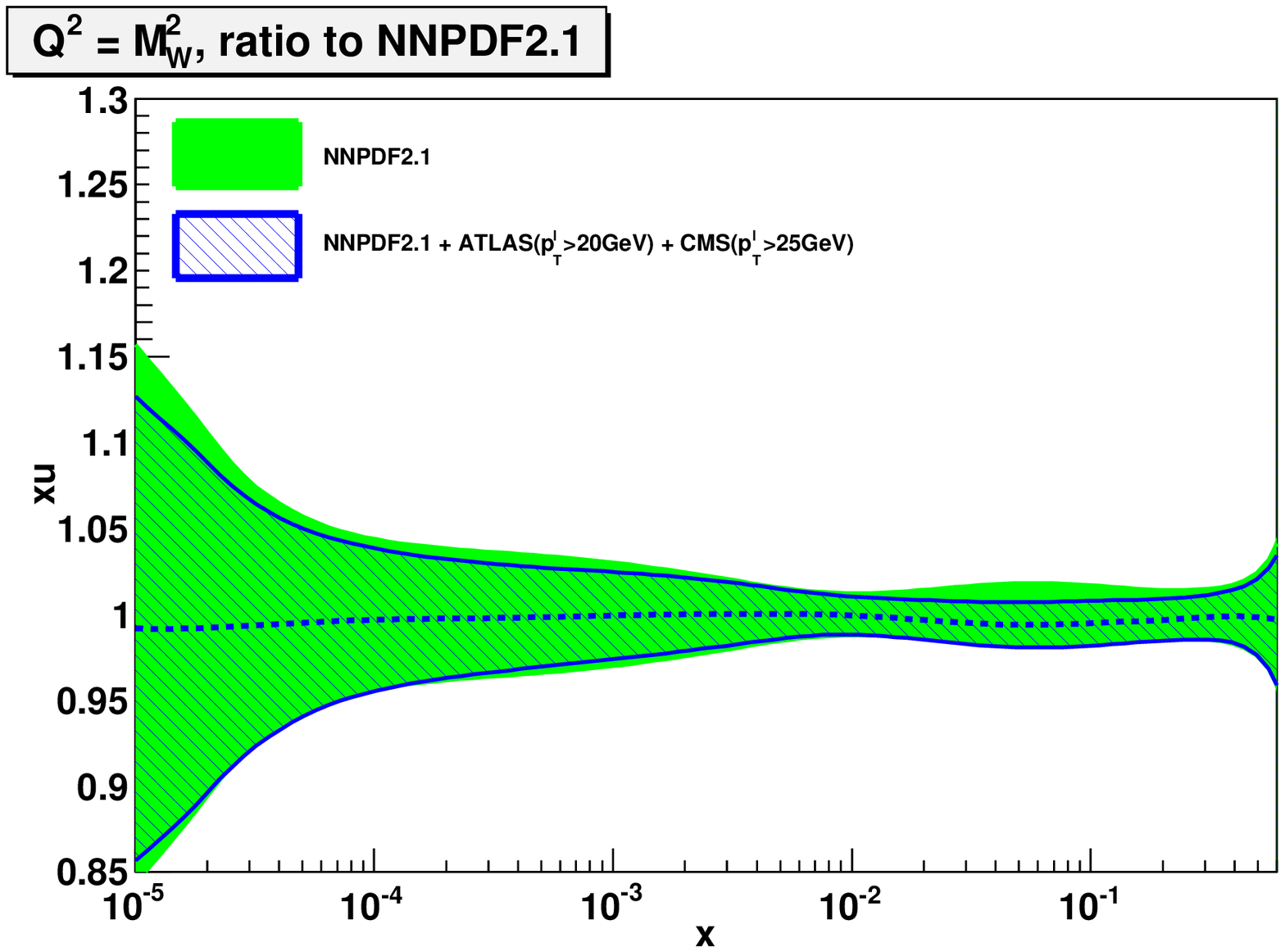}
  \epsfig{width=0.44\textwidth,figure=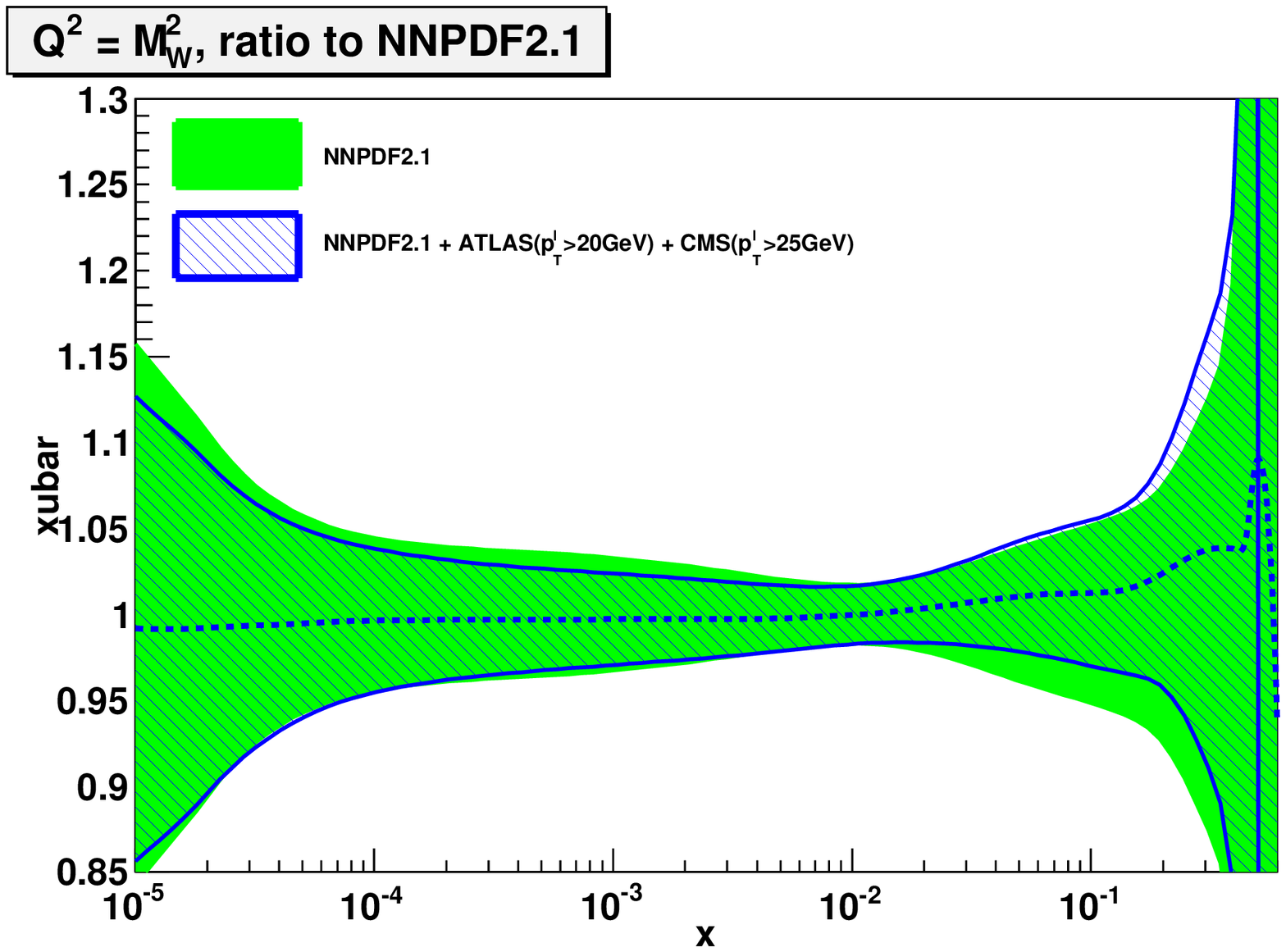}
  \epsfig{width=0.44\textwidth,figure=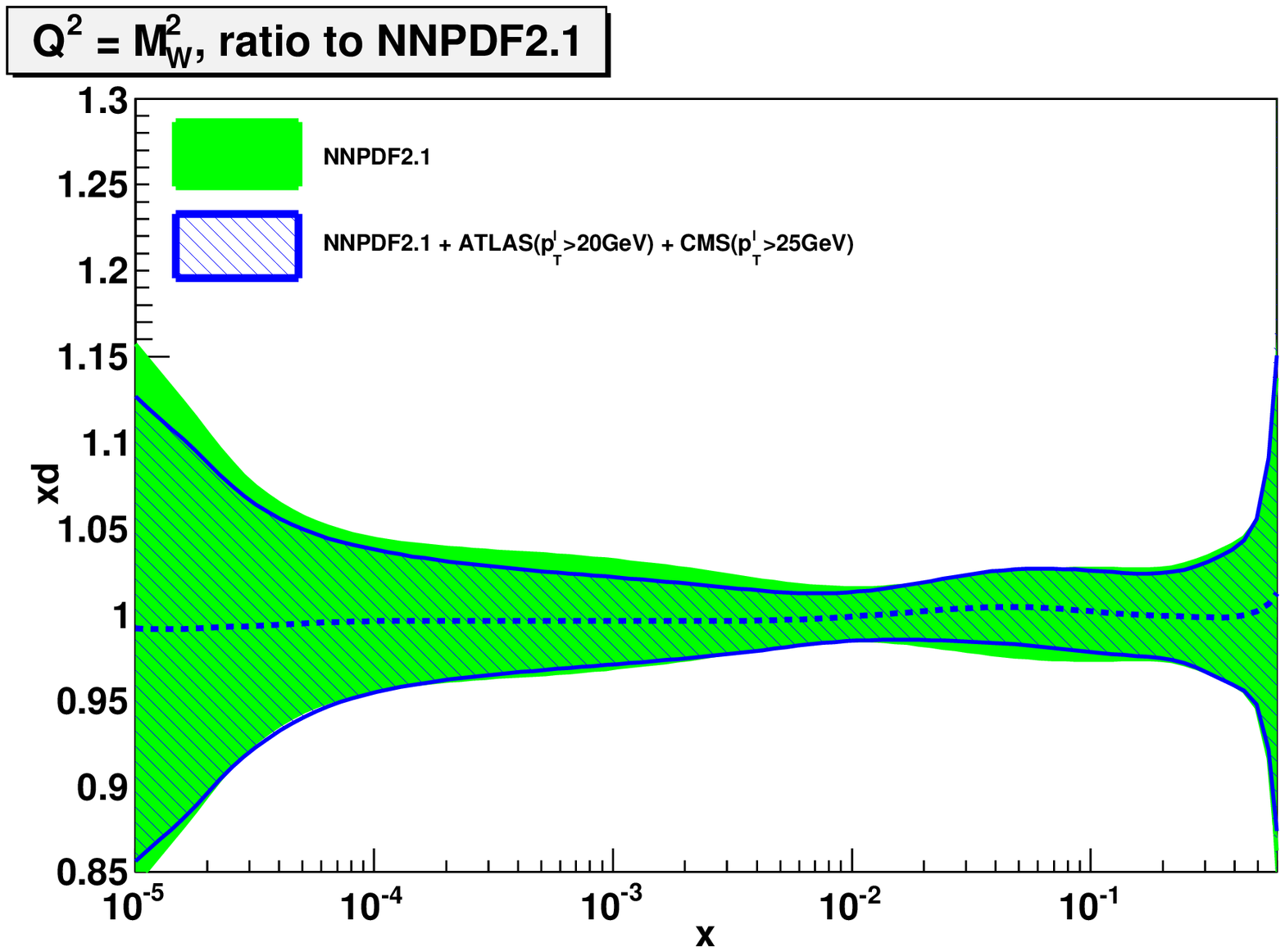}
  \epsfig{width=0.44\textwidth,figure=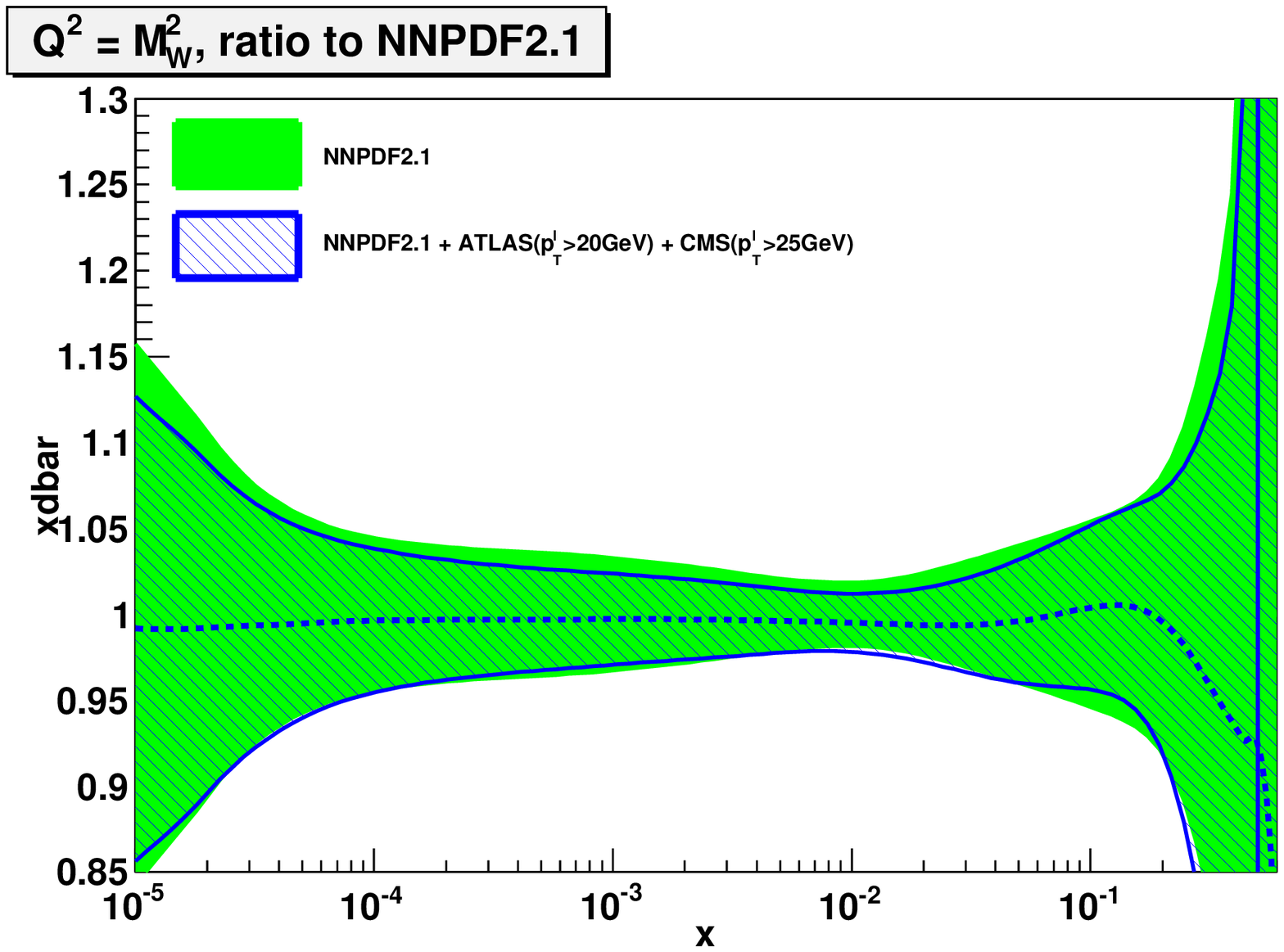}
  \caption{Same as Fig.~\ref{fig:pdfs-atlas} but after adding both the ATLAS and 
CMS lepton charge asymmetry data.}
  \label{fig:pdf-lhc-25}
\end{figure}

After reweighting the description of the data improves, with 
$\chi^2 _{\rm rw}/N_{\rm dat}=0.95$ with the distribution of 
$\chi^2 _{\rm rw}/N_{\rm dat}$ for individual replicas shown in the middle plot of
Fig.~\ref{fig:chi2-lhc25} showing a sharp peak around one.
These results, combined with the number of effective replicas surviving after 
reweighting, namely $N_{\rm eff}=619$ out of the initial $N_{\rm rep}=1000$, show that the 
use of the ATLAS and CMS data together in the fit is not only possible but imposes
a moderate constraint on PDFs. However the constraint is not quite so great as with the CMS data alone, 
suggesting a mild incompatibility particularly in the high rapidity bins.

The impact of the data on the light flavour and anti-flavour distributions is shown in 
Fig.~\ref{fig:pdf-lhc-25}, where we compare the $u$ and $d$ quark and antiquark
distributions at the scale $Q^2=M_W^2$ from the NNPDF2.1 global fit and the ones obtained 
after adding the ATLAS and CMS lepton charge asymmetry data using reweighting. There is around 
$20\%$ reduction in uncertainties around $x\sim 10^{-3}$, mainly due to the ATLAS data,
complemented by a reduction of between $10\%$ and $25\%$ at larger $x$, mainly due to the CMS data.

\section{Global PDFs including LHC data}
\label{sec-tev+lhc}

In this section we will check the consistency of the D0 and ATLAS+CMS
datasets among themselves, and use both datasets to reweight the
NNPDF2.1 NLO PDFs. The unweighting method presented in
Sect.~\ref{sec-unw} is then used to produce a set of 100 unweighted
replicas.
The final product of this analysis is a new set of
NNPDF parton distribution functions, NNPDF2.2 NLO, which includes,
together with all the datasets already included in the NNPDF2.1 NLO
global set, the D0, ATLAS and CMS lepton charge asymmetry data
described above.

\subsection{Tevatron $W$ asymmetry data}
\label{sec-d0}

In Ref.~\cite{Ball:2010gb} we used the reweighting technique to study
the compatibility of the D0 $W$ lepton charge asymmetry data with the
data included in the NNPDF2.0 NLO global fit and to assess their
impact on the fitted parton densities.  The conclusion of this study
was that the data that are inclusive in the $p_\perp$ of the
identified lepton, namely the muon charge asymmetry data presented
in~\cite{d0muon} and electron charge asymmetry data with $p_\perp >
25$ GeV released in~\cite{d0electron}, are consistent with each other
and with all the other datasets included in NNPDF2.0, in particular
with the CDF $W$ asymmetry data~\cite{cdfboson} and the fixed-target
DIS deuteron data.  When included in the fit they have a moderate
impact on PDFs, providing a reduction of the uncertainty of the
valence quark distributions in the medium-high $x$ region ($x\sim 10^{-2}$).

Less inclusive electron charge asymmetry data were also presented
in~\cite{d0electron}. They are binned in $p_\perp$, divided into two
sets with $25{\rm GeV} <p_\perp < 35{\rm GeV}$ and $p_\perp > 35{\rm
  GeV}$ respectively. We observed~\cite{Ball:2010gb} that these data, which could have
potentially more impact on the PDFs, are inconsistent with some of the
DIS data included in the global analysis and have problems of internal
consistency.  Similar conclusions have been reported by the
MSTW~\cite{Thorne:2010kj} and CTEQ~\cite{Lai:2010vv} collaborations,
as they tried to include these datasets in the context of a PDF global
analysis. We will thus not use these datasets here.

These results, though obtained using the NNPDF2.0 global fit, remain
substantially unchanged if we use instead the NNPDF2.1 NLO global set as a
prior fit to start the reweighting analysis. The muon charge asymmetry~\cite{d0muon}
and inclusive electron charge asymmetry data (with $p_\perp >
25$ GeV)~\cite{d0electron} can thus provide additional information 
to that from the ATLAS and CMS data considered in
the previous section. We thus proceed directly to a combined fit of these data 
together with the LHC data.

\begin{table}
  \begin{center}
    \begin{tabular}{|c|c|c|c|c|}
      \hline 
      Experiment & $N_{\mathrm{dat}} $ & NNPDF2.1 & NNPDF2.1 LHC & NNPDF2.2 \\
      \hline
      \hline
      NMC-pd     &  132 &  0.97 &  0.95 &  0.97 \\
      \hline
      NMC        &  221 &  1.73 &  1.72 &  1.72 \\
      \hline
      SLAC       &   74 &  1.33 &  1.26 &  1.28 \\
      \hline
      BCDMS      &  581 &  1.24 &  1.23 &  1.23 \\
      \hline
      HERAI-AV   &  592 &  1.07 &  1.07 &  1.07 \\
      \hline
      CHORUS     &  862 &  1.15 &  1.15 &  1.15 \\
      \hline
      FLH108     &    8 &  1.37 &  1.37 &  1.37 \\
      \hline
      NTVDMN     &   79 &  0.79 &  0.74 &  0.70 \\
      \hline
      ZEUS-H2    &  127 &  1.29 &  1.28 &  1.28 \\
      \hline
      ZEUSF2C    &   50 &  0.78 &  0.79 &  0.78 \\
      \hline
      H1F2C      &   38 &  1.51 &  1.52 &  1.51 \\
      \hline
      DYE605     &  119 &  0.84 &  0.84 &  0.86 \\
      \hline
      DYE886     &  199 &  1.25 &  1.23 &  1.27 \\
      \hline
      CDFWASY    &   13 &  1.85 &  1.81 &  1.81 \\
      \hline
      CDFZRAP    &   29 &  1.66 &  1.61 &  1.70 \\
      \hline
      D0ZRAP     &   28 &  0.60 &  0.60 &  0.58 \\
      \hline
      CDFR2KT    &   76 &  0.98 &  0.98 &  0.96 \\
      \hline
      D0R2CON    &  110 &  0.84 &  0.84 &  0.83 \\
      \hline
      \hline
      ATLASmuASY &   11 & [0.77]  &  0.97    &  1.07  \\
      \hline
      CMSeASY    &   6 &  [1.83]  &  1.23    &  1.08  \\
      \hline
      CMSmuASY   &   6 &  [1.24]  &  0.63    &  0.56  \\
      \hline
      D0eASY     &   12 & [4.39]  &  [3.46]  &  1.38  \\
      \hline
      D0muASY    &   10 & [1.48]  &  [1.17]  &  0.35  \\
      \hline
      \hline
      Total      &      &  1.165 &  1.158 &  1.157 \\
      \hline
    \end{tabular}
  \end{center}
  \caption{Table of $\chi^2/N_{\mathrm{dat}}$ values for the experiments 
    included in the
    NNPDF2.1 NLO fit, the NNPDF2.1 LHC fit discussed in 
    Section~\ref{sec-lhc} and the NNPDF2.2 NLO fit. The numbers in square 
    brackets correspond to the experiments which are not included in the fit.
    The three fits thus have respectively $N_{\mathrm{dat}}=$ 3338, 3361 and 
    3383.}
  \label{tab:chisq22}
\end{table}

\subsection{Combining LHC and Tevatron $W$ asymmetry data}
\label{sec-nnpdf22}

\begin{figure}[ht]
  \centering
  \epsfig{width=0.30\textwidth,figure=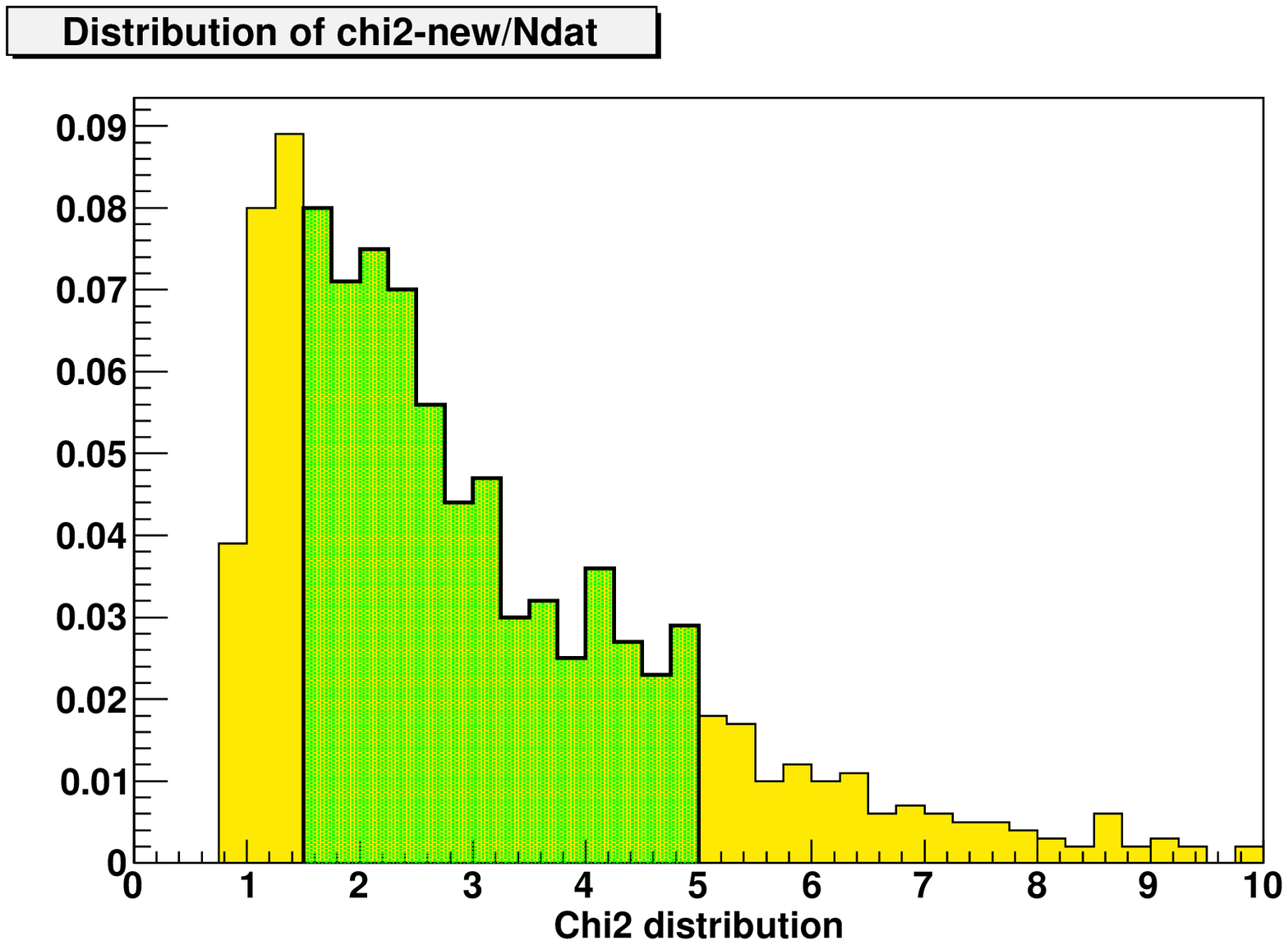}
  \epsfig{width=0.30\textwidth,figure=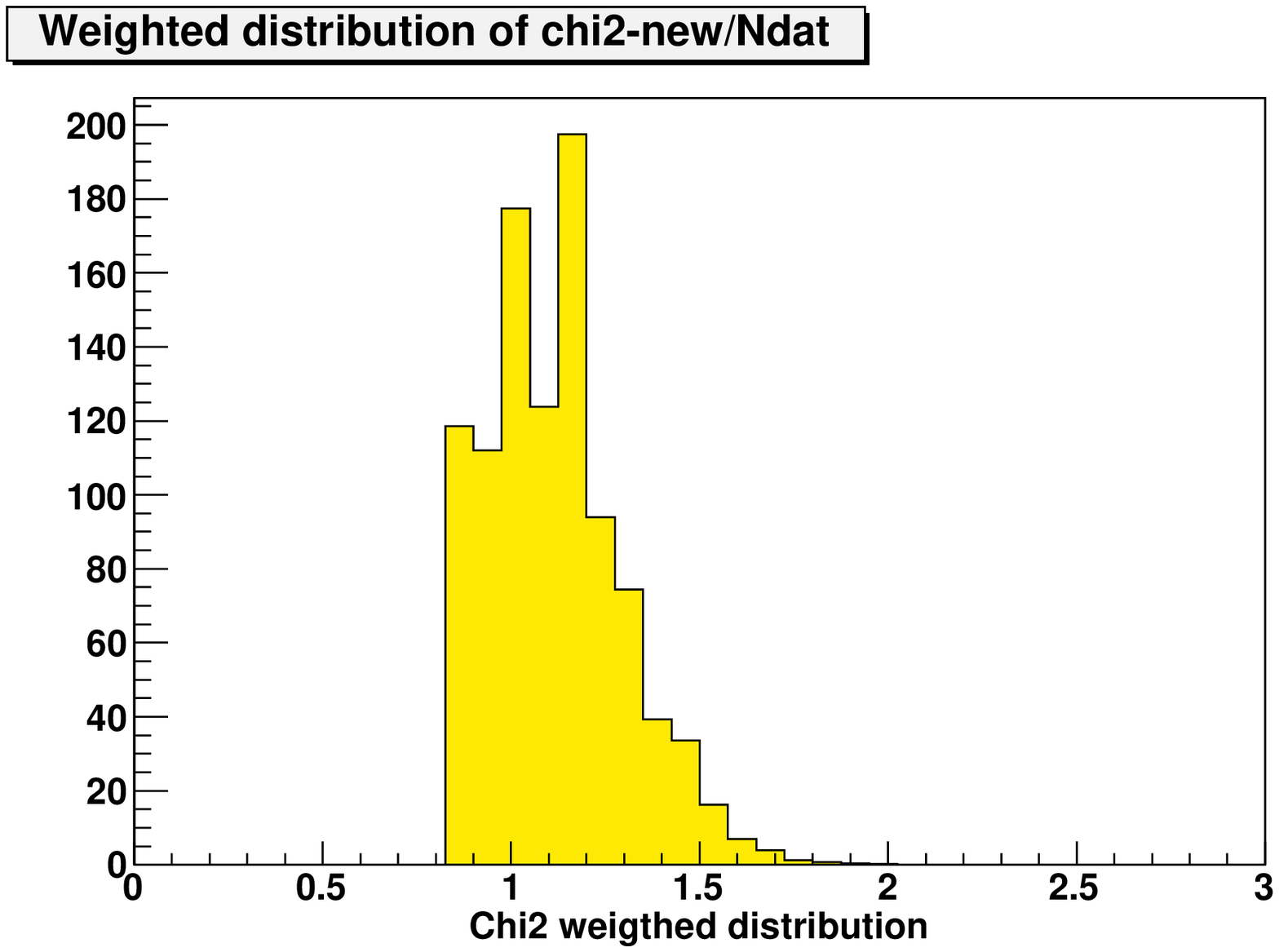}
  \epsfig{width=0.30\textwidth,figure=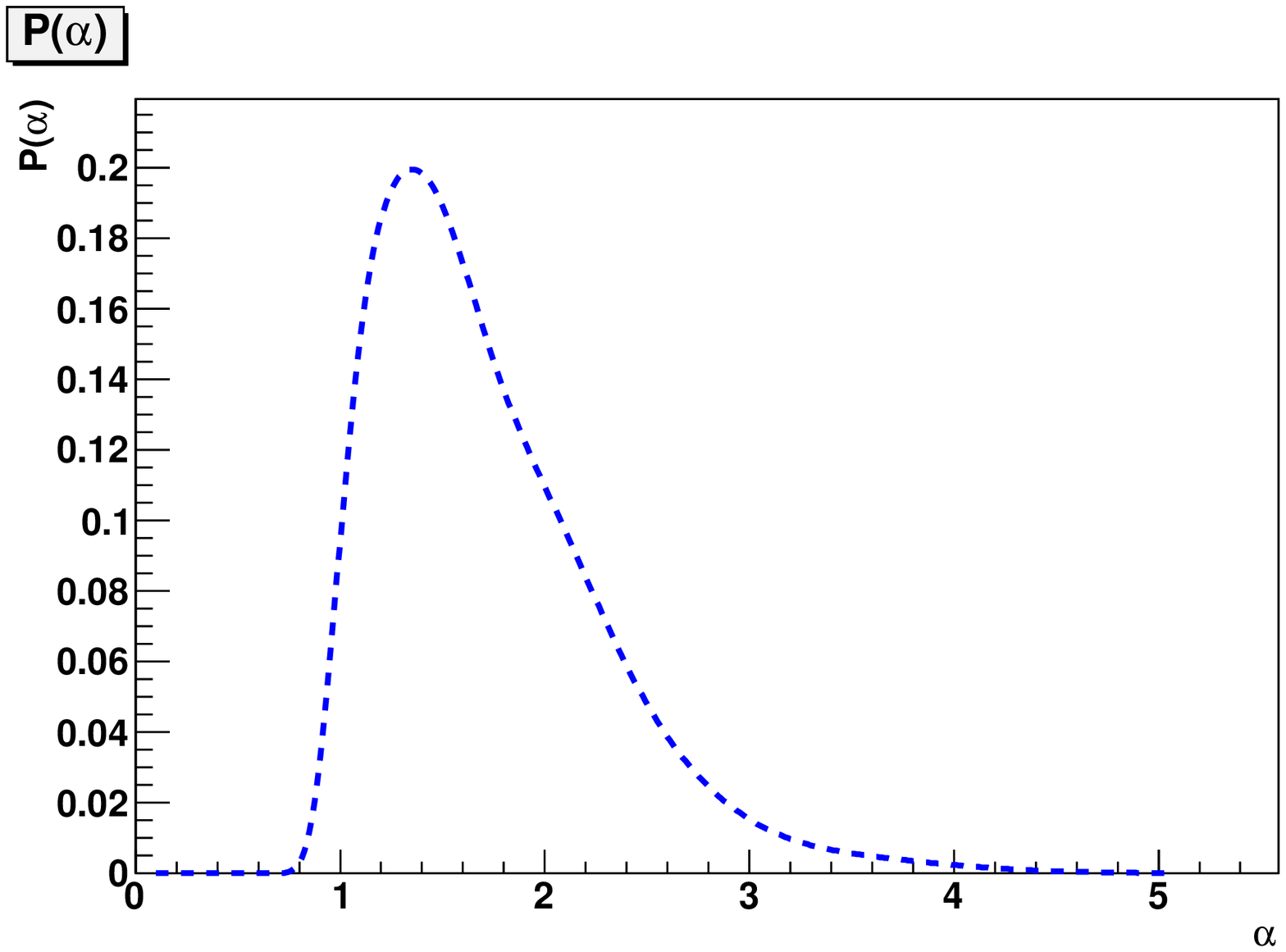}
  \caption{ Same as Fig.~\ref{fig:chi2-atlas} for the combined
 D0+ATLAS+CMS $p_{T}>25$~GeV dataset. }
  \label{fig:chi2-tev+lhc25}
\end{figure}

The description of the combined ATLAS, CMS and D0 
charge asymmetry datasets obtained using the 
NNPDF2.1 NLO global fit, in which they were not included, is reasonably good
but not optimal, with $\chi^{2}/N_{\rm dat}=2.22$: a detailed comparison 
is shown in 
Table~\ref{tab:chisq22}. The distribution of the combined 
$\chi^{2}/N_{\rm dat}$ for 
individual replicas before and after reweighting, 
and the $P(\alpha)$ distribution, shown in 
Fig.~\ref{fig:chi2-tev+lhc25}, indicate however that these data are reasonably 
compatible with the data already included in the NNPDF2.1 analysis 
and would provide a significant constraint on the PDFs.

\begin{figure}[h!]
  \centering
  \epsfig{width=0.44\textwidth,figure=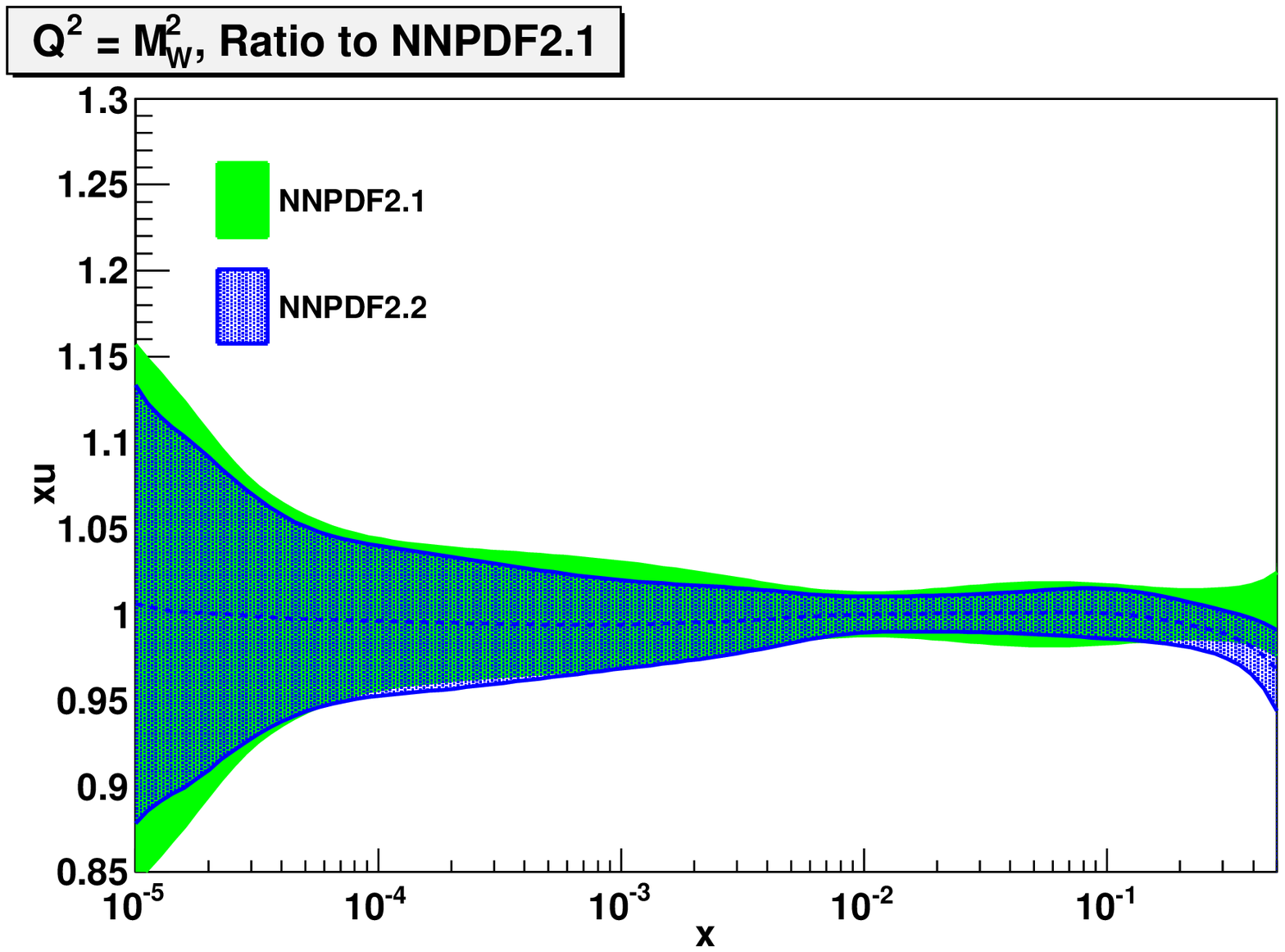}
  \epsfig{width=0.44\textwidth,figure=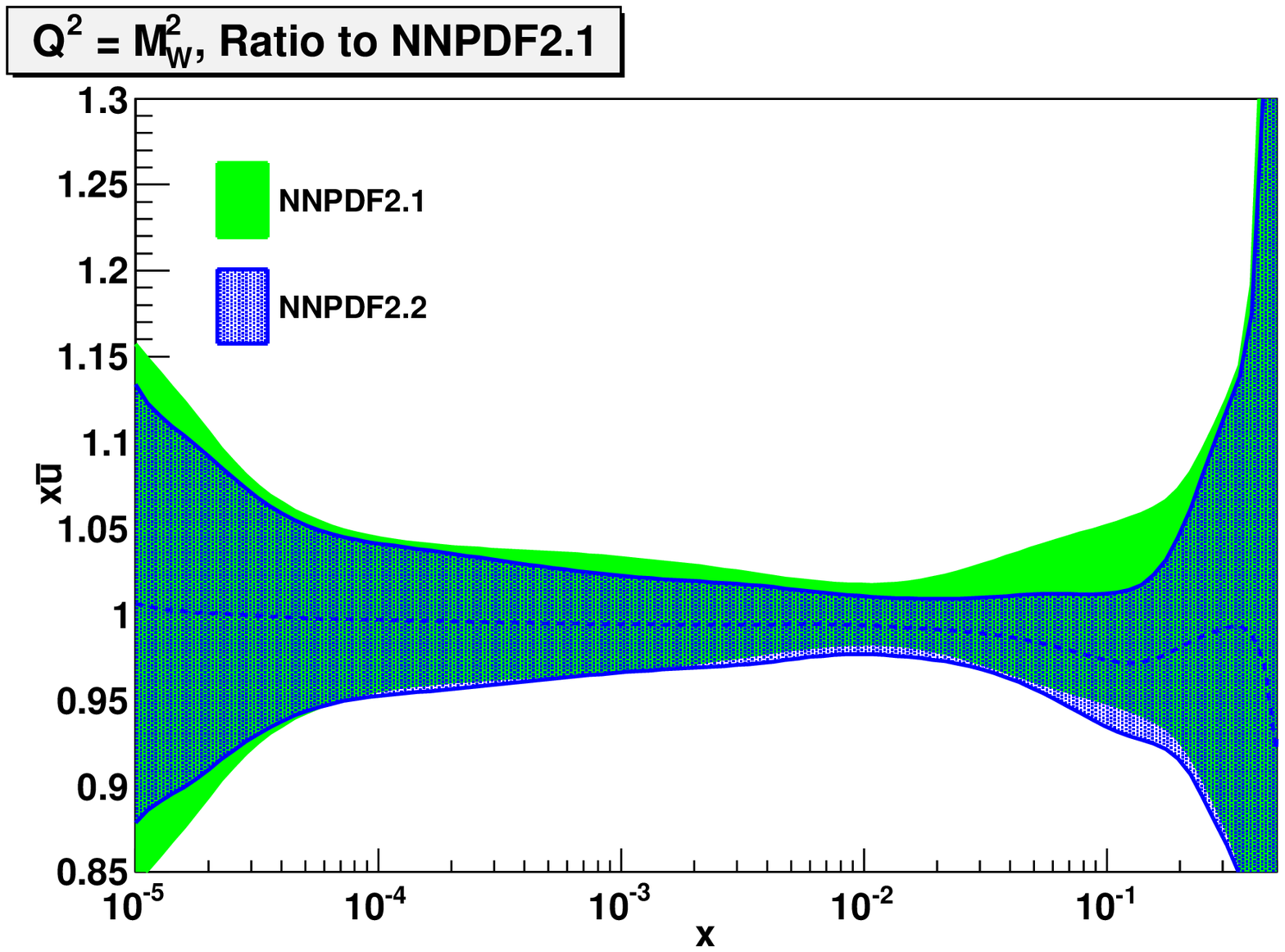}
  \epsfig{width=0.44\textwidth,figure=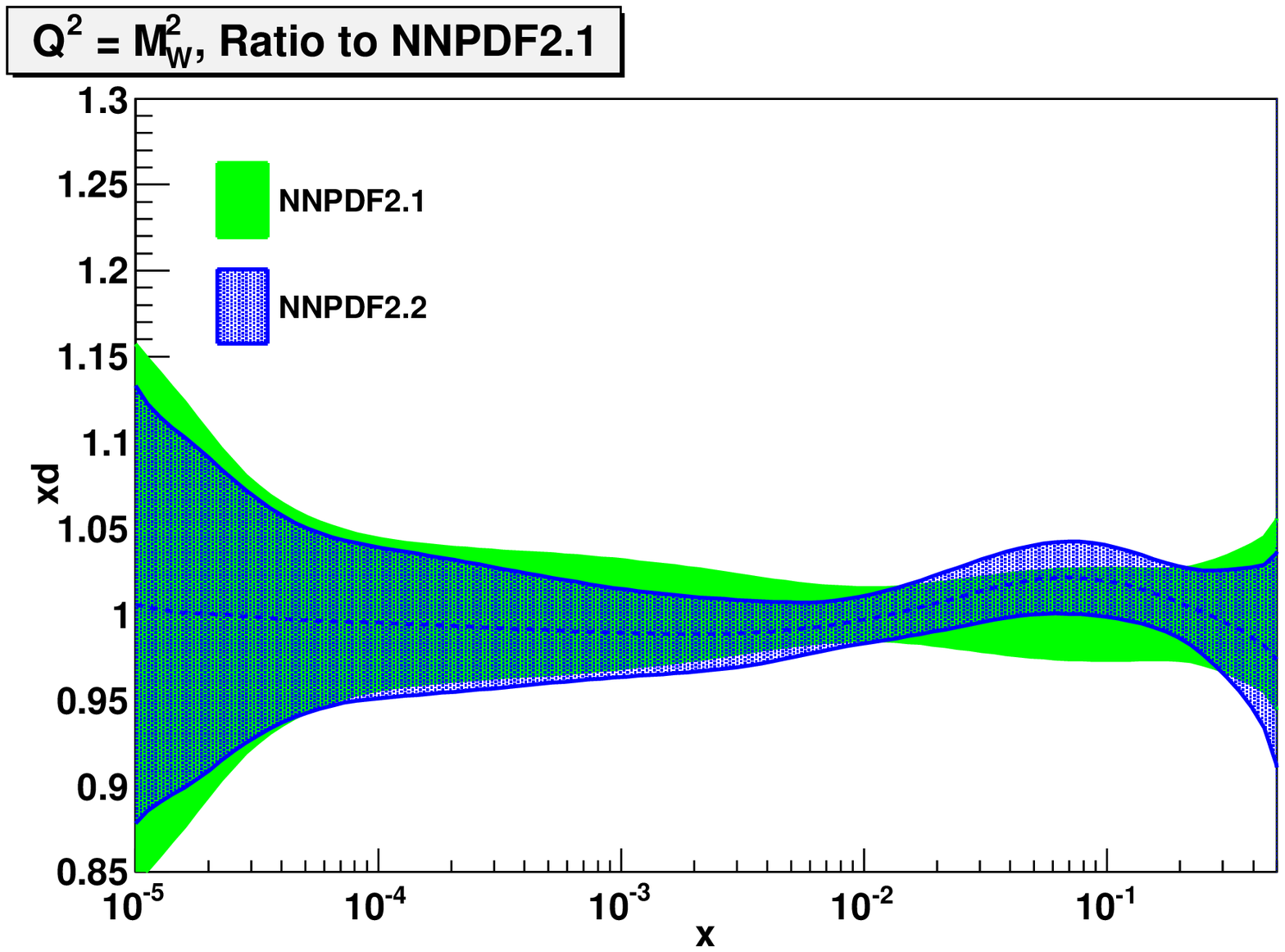}
  \epsfig{width=0.44\textwidth,figure=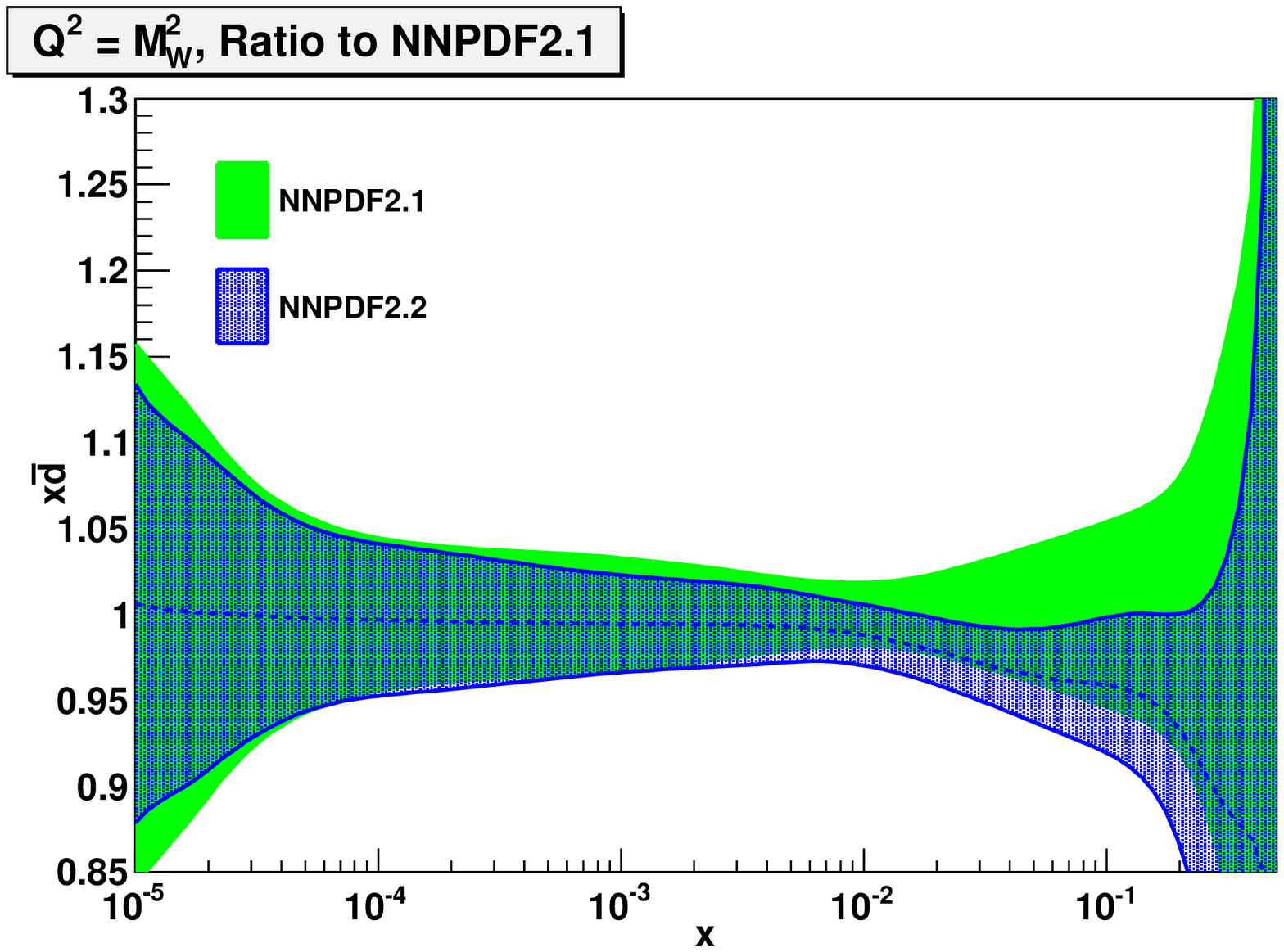}
  \caption{Comparison of light quark and antiquark distributions at the scale
    $Q^2=M_W^2$ from the global NNPDF2.1 and NNPDF2.2 global fits. 
    Parton densities are plotted normalized to the NNPDF2.1 central value.}
 \label{fig:pdf1-tev+lhc25}
\end{figure}

These conclusions are indeed confirmed when the effect of the ATLAS, CMS and 
D0 data is included using the reweighting technique.
After reweighting their overall description improves significantly, with a combined  
$\chi^2_{\rm rw}/N_{\rm dat}=0.81$. This is due to a significant improvement in the 
fit to the CMS and the D0 data: the fit to the ATLAS data deteriorates a little, again showing 
that there is some tension. 
The number of effective replicas is now $N_{\rm eff}=181$ 
out of the initial $N_{\rm rep}=1000$, showing that the $W$ lepton asymmetry data indeed introduce 
very significant constraints on the PDFs.
The distribution of the $\chi^{2}/N_{\rm dat}$ for the individual replicas after
reweighting, shown in the middle plot of Fig.~\ref{fig:chi2-tev+lhc25}, 
is peaked around one, 
confirming the compatibility of these data with the other datasets included in the global
analysis. 

\begin{figure}[ht]
  \centering
  \epsfig{width=0.44\textwidth,figure=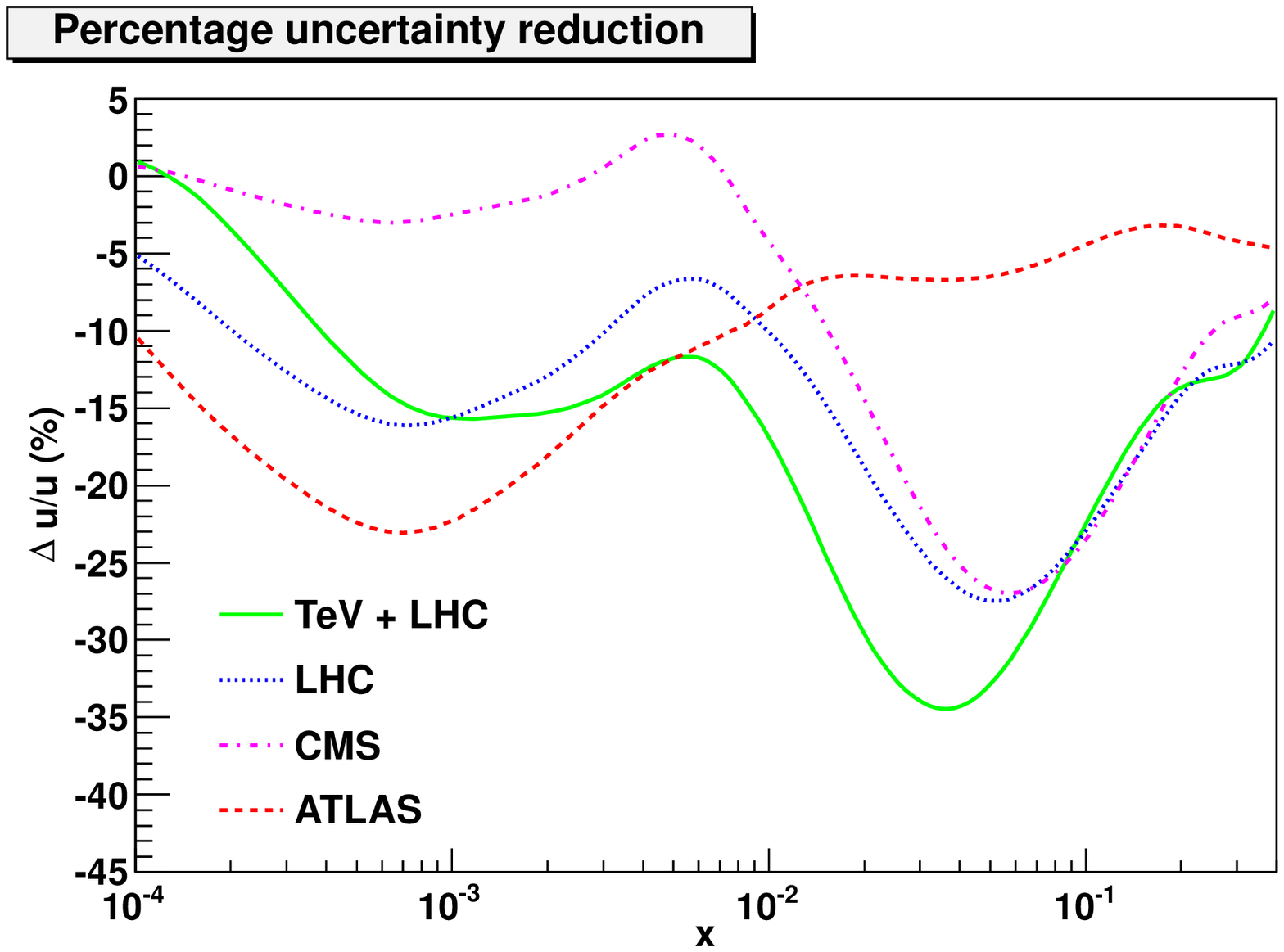}
  \epsfig{width=0.44\textwidth,figure=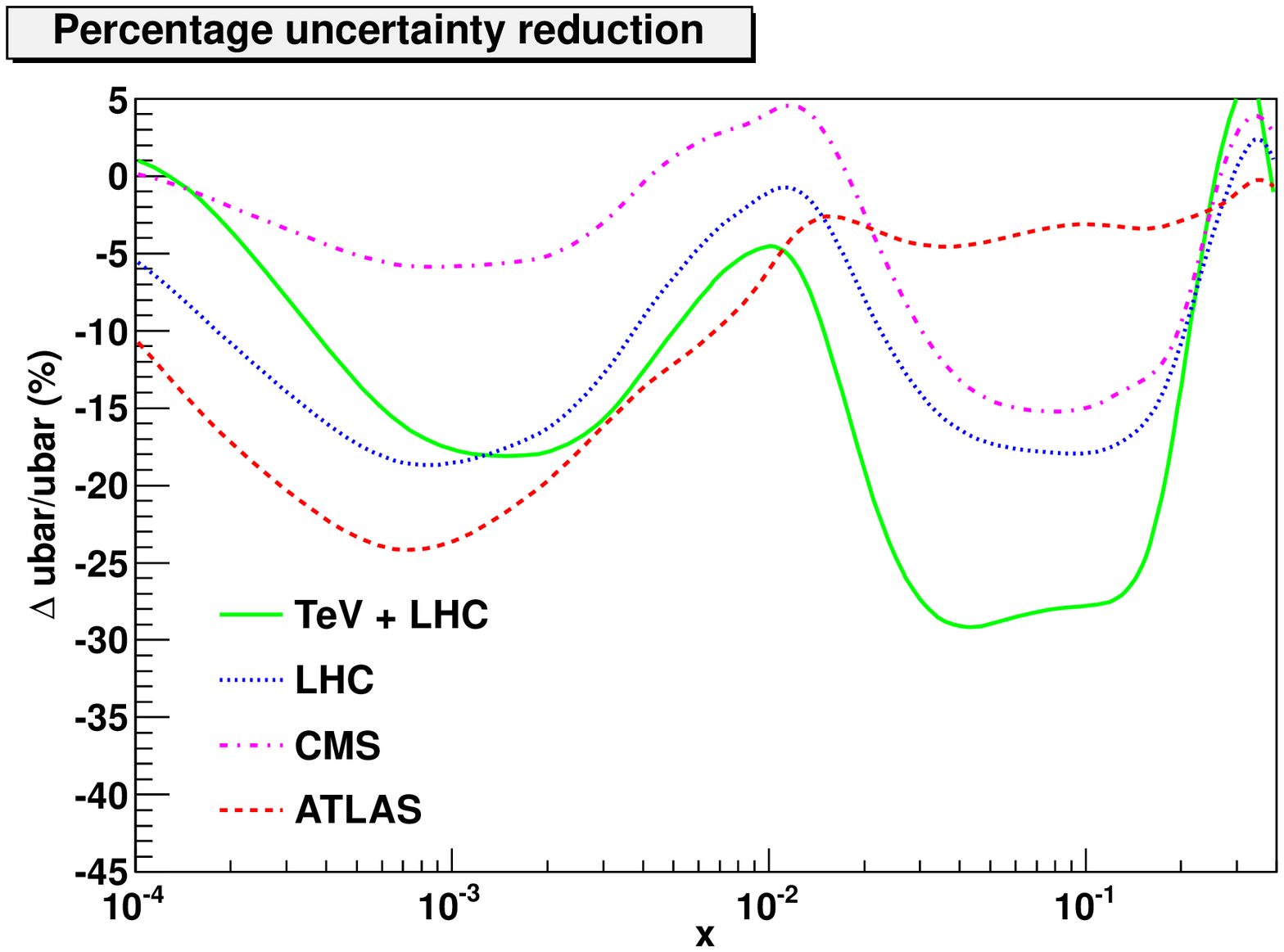}
  \epsfig{width=0.44\textwidth,figure=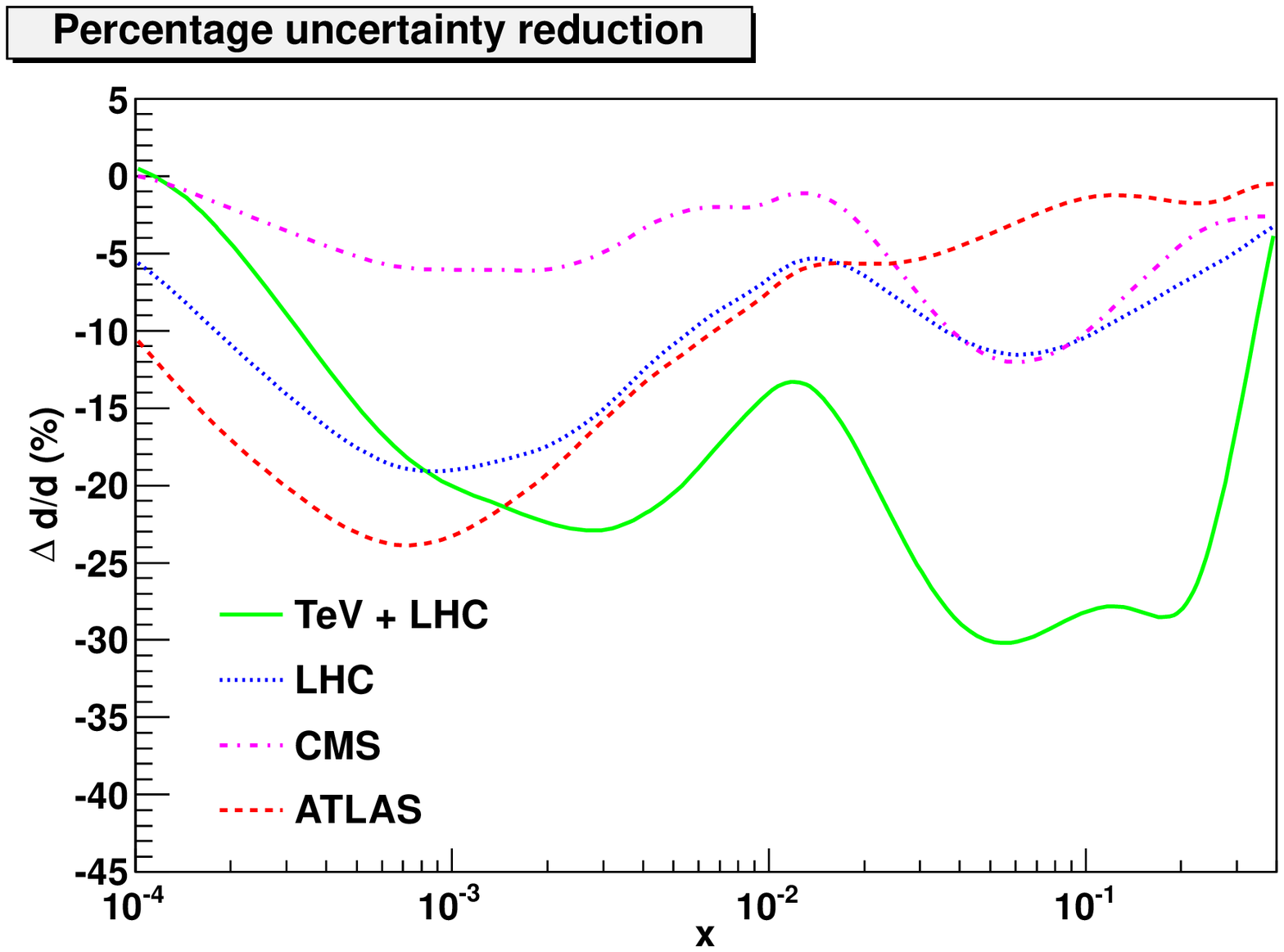}
  \epsfig{width=0.44\textwidth,figure=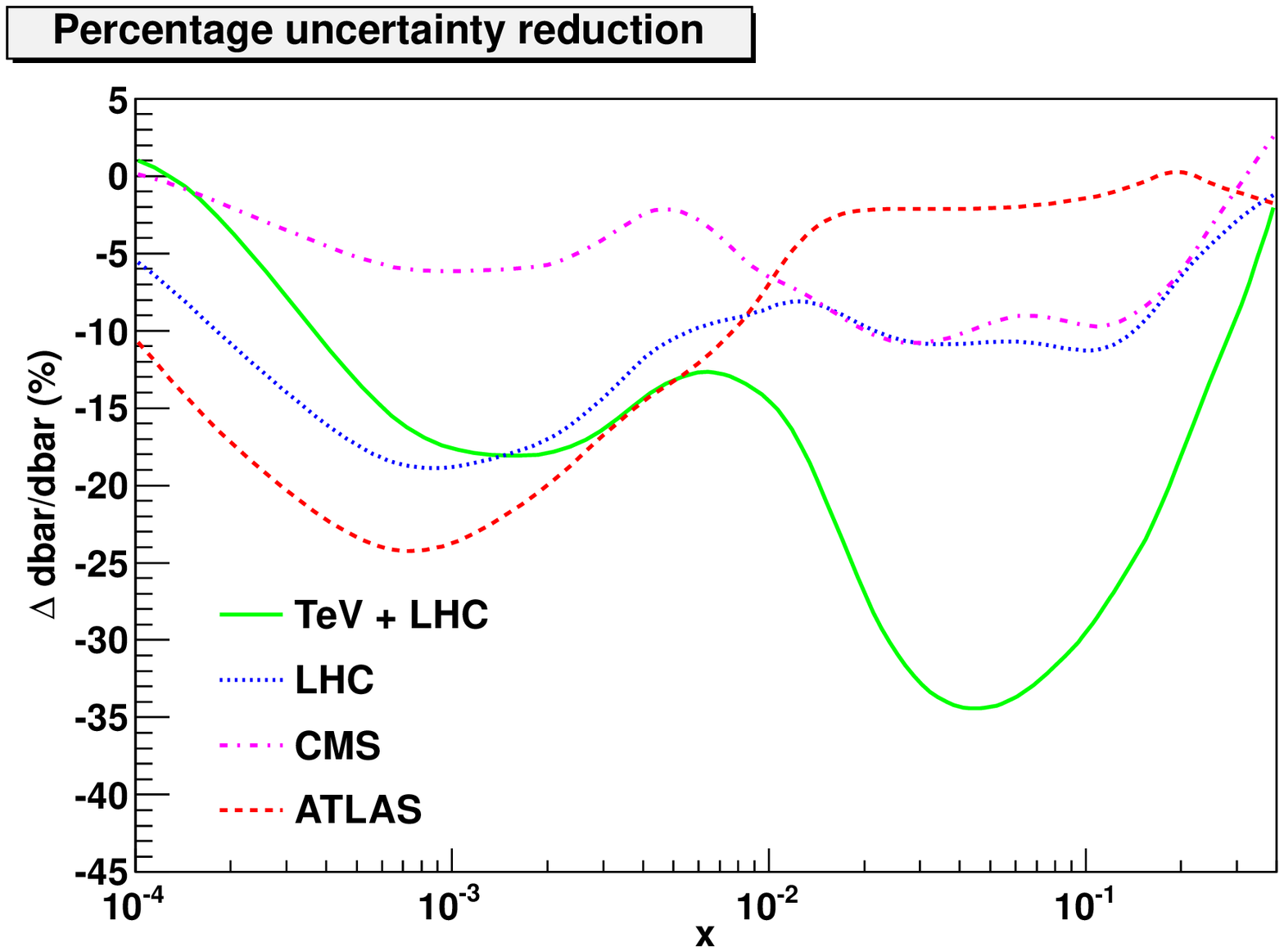}
  \caption{The percentage change in the uncertainty in the light quark and antiquark distributions 
    at the scale $Q^2=M_W^2$ in the global NNPDF2.1 NLO global fit, after adding
    ATLAS, CMS and D0 lepton charge asymmetry data via reweighting. The four curves show in each case 
    the effect of ATLAS (red) and CMS (pink) only, together (blue), and then together with the D0 data (green), i.e. NNPDF2.2.}
  \label{fig:perc}
\end{figure}

After reweighting, the unweighting procedure of Sec.~\ref{sec-unw} may be used to 
give a $100$ replica set of PDFs equivalent to a global fit which includes all the data 
already included in NNPDF2.1, plus the ATLAS, CMS and D0 $W$ asymmetry data. We call this 
new NLO PDF set NNPDF2.2. The quality of the data to all the sets used in this new fit 
is shown in Tab.~\ref{tab:chisq22}. There is no 
significant deterioration in the $\chi^2$ in any of other datasets included in the 
global fit, and the fit to the NuTeV dimuon data improves significantly. 
The overall $\chi^2_{\rm tot}/N_{\rm dat}$ thus also improves a little.

The impact on light flavour and anti-flavour PDFs is shown in 
Fig.~\ref{fig:pdf1-tev+lhc25},
where we compare the $u$ and $d$ quark and antiquark distributions at the scale 
$Q^2=M_W^2$ from the NNPDF2.1 NLO set to the ones obtained for the NNPDF2.2 NLO set.
The most noticeable effects of the inclusion of the new data are
concentrated in two separate regions of $x$, namely, the  $x\sim
10^{-3}$ region, which is mostly affected by the ATLAS data, and the $x\sim
10^{-2}-10^{-1}$ region, which is mostly affected by the CMS and D0 data.
In each of these regions, the $W$ asymmetry data leads to a reduction
of uncertainties on the light flavour and
antiflavour distribution, or around $20\%$ in the low $x$ region, and up to 
$30\%$ at higher $x$ when CMS and D0 are combined (see Fig.~\ref{fig:perc}). 
At higher $x$ changes in the central values for these PDFs by up to one sigma are also 
observed: these are mainly due to the D0 data (compare Fig.~\ref{fig:pdf1-tev+lhc25} with
Fig.~\ref{fig:pdf-lhc-25}).

As recently shown in the extensive studies carried out in the
context of the PDF4LHC Working Group~\cite{Alekhin:2011sk}, 
there is rather good agreement among NLO parton
distributions determined from the widest global datasets, specifically
by the NNPDF, MSTW and CTEQ groups. However, 
there still are some significant differences, notably
in the flavour separation at medium-large $x$.  Since this is the
region which is directly probed by the Tevatron and LHC lepton charge
asymmetry data studied here, these data might
help in resolving some of these outstanding incompatibilities. 

\begin{figure}[ht]
  \centering
    \epsfig{width=0.44\textwidth,figure=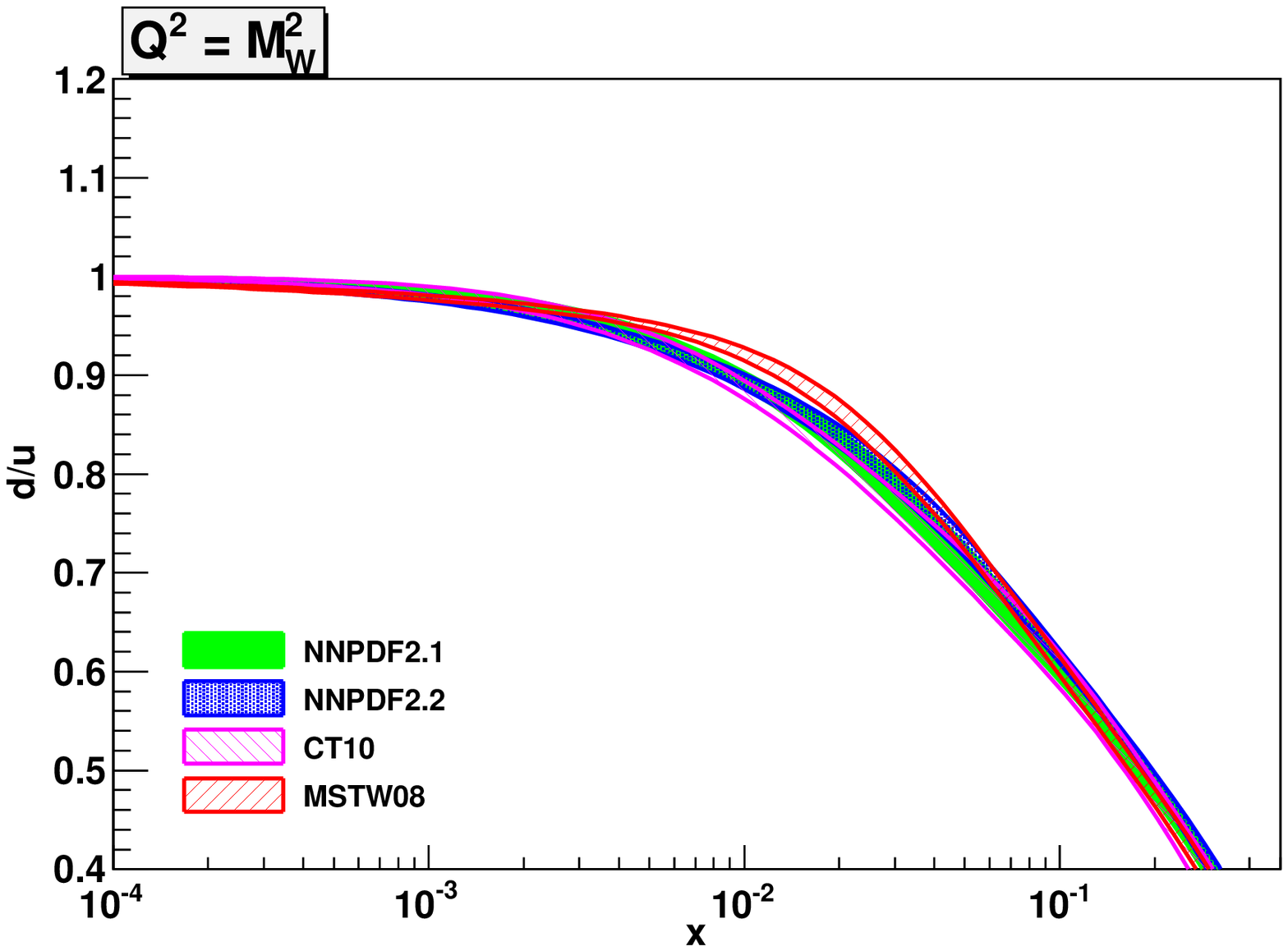}
    \epsfig{width=0.44\textwidth,figure=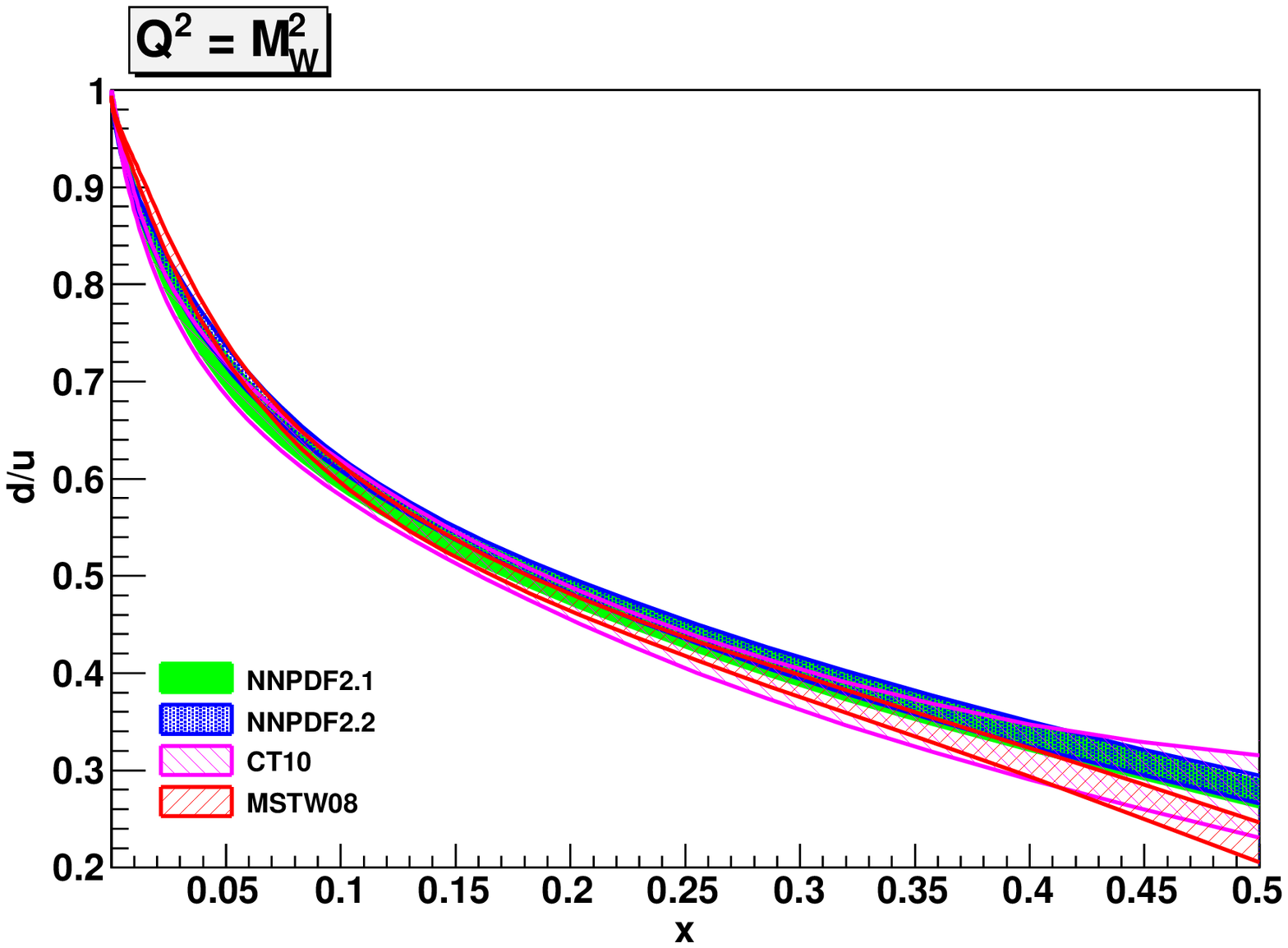}
    \epsfig{width=0.44\textwidth,figure=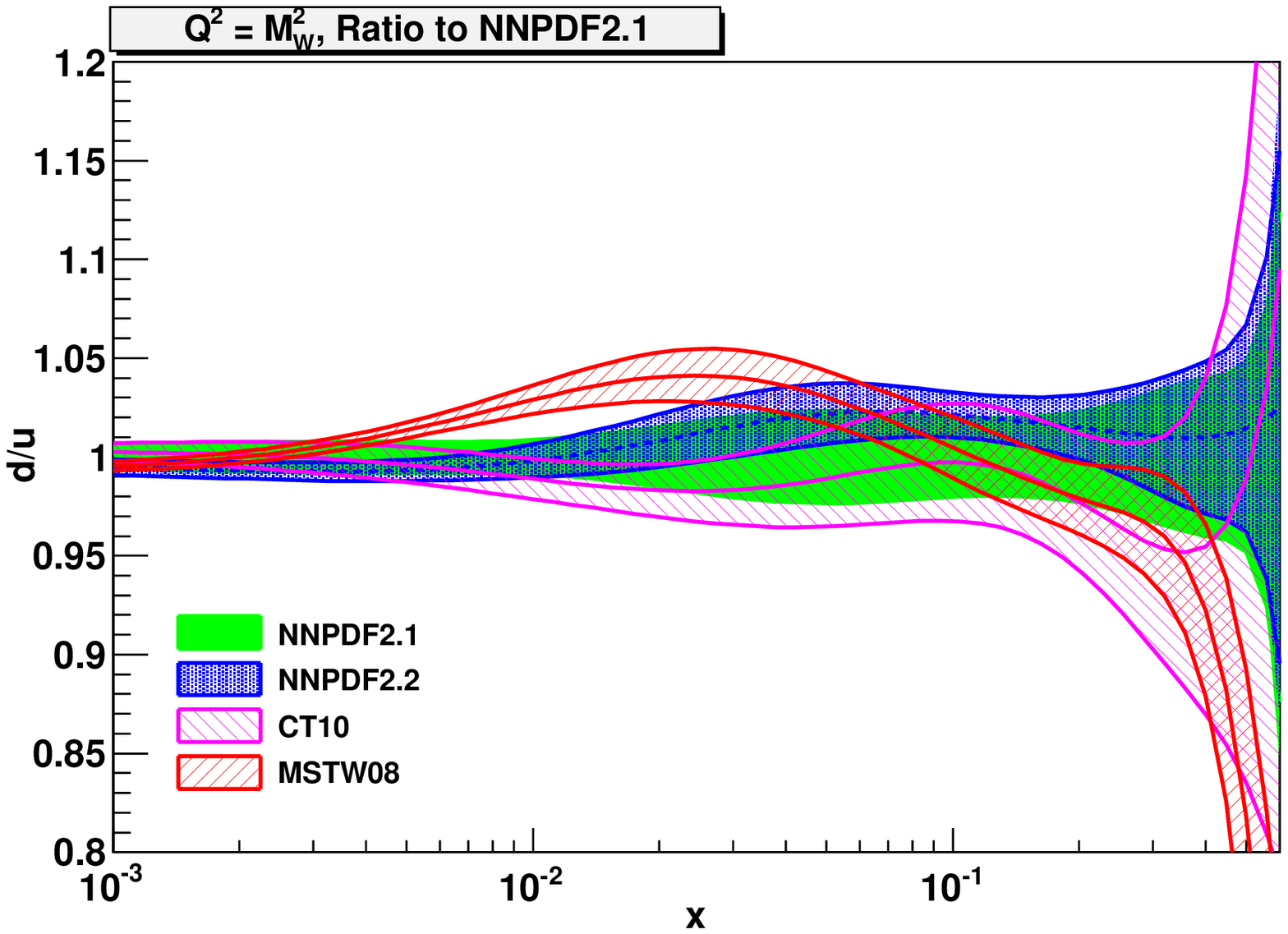}
    \epsfig{width=0.44\textwidth,figure=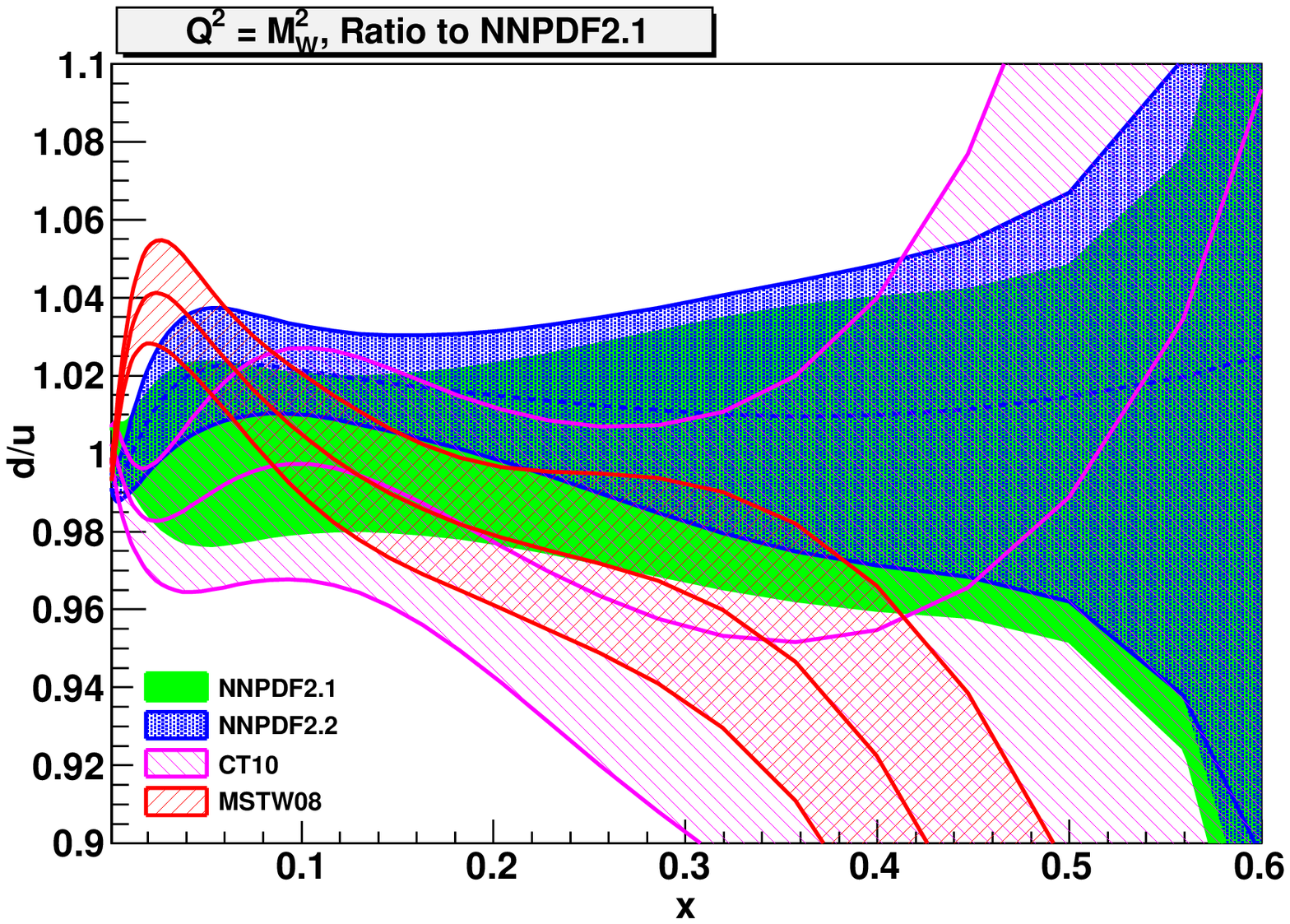}
 \caption{Comparison of the $d/u$ ratio at $Q^2=M_W^2$ in  NNPDF2.1,
  CT10, MSTW08 and NNPDF2.2. Upper plots show absolute values, while 
the lower plots show the ratio to NNPDF2.1}
 \label{fig:pdf3-tev+lhc25}
\end{figure}

\begin{figure}[h!]
  \centering
  \epsfig{width=0.44\textwidth,figure=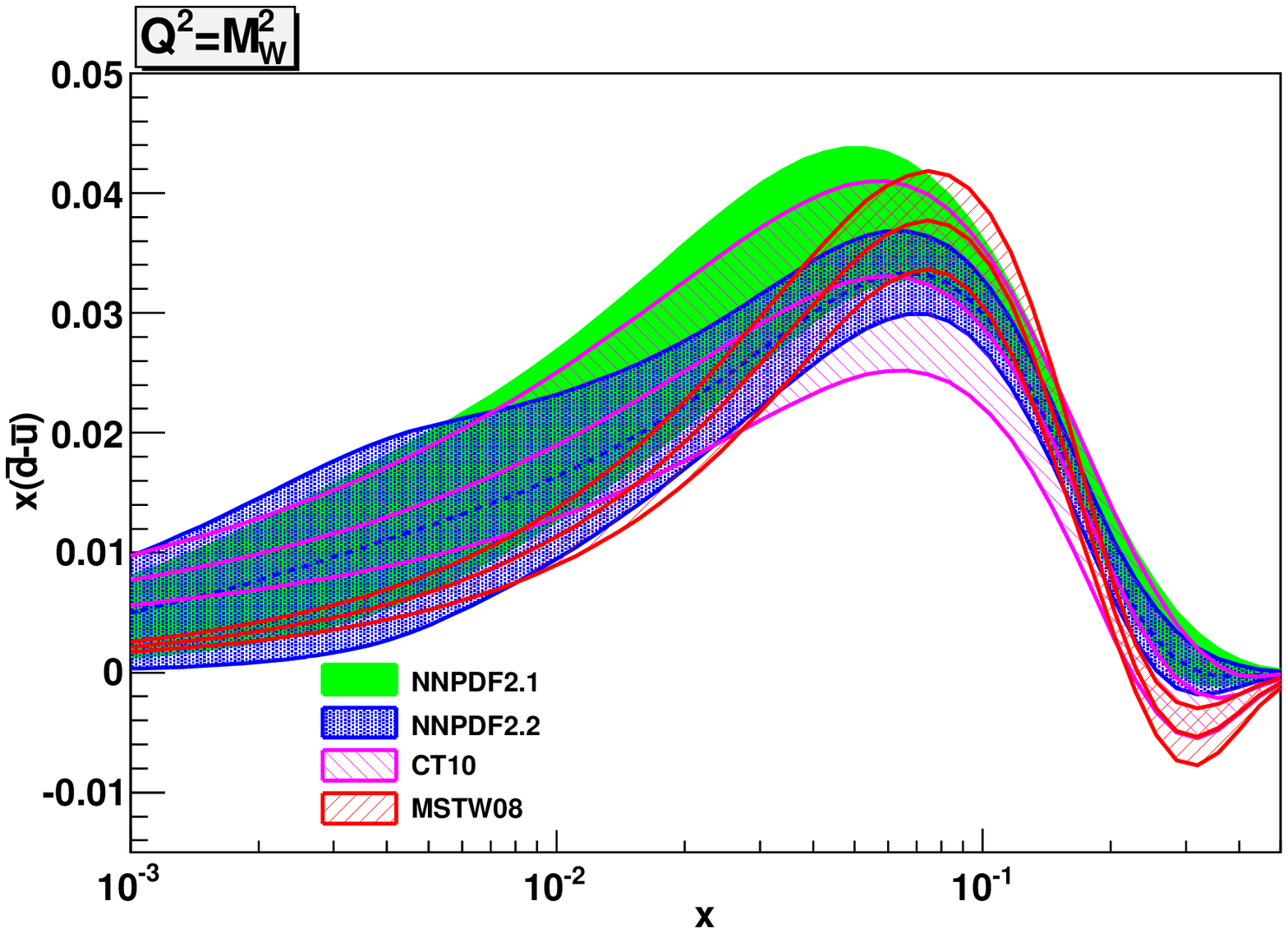}
  \epsfig{width=0.44\textwidth,figure=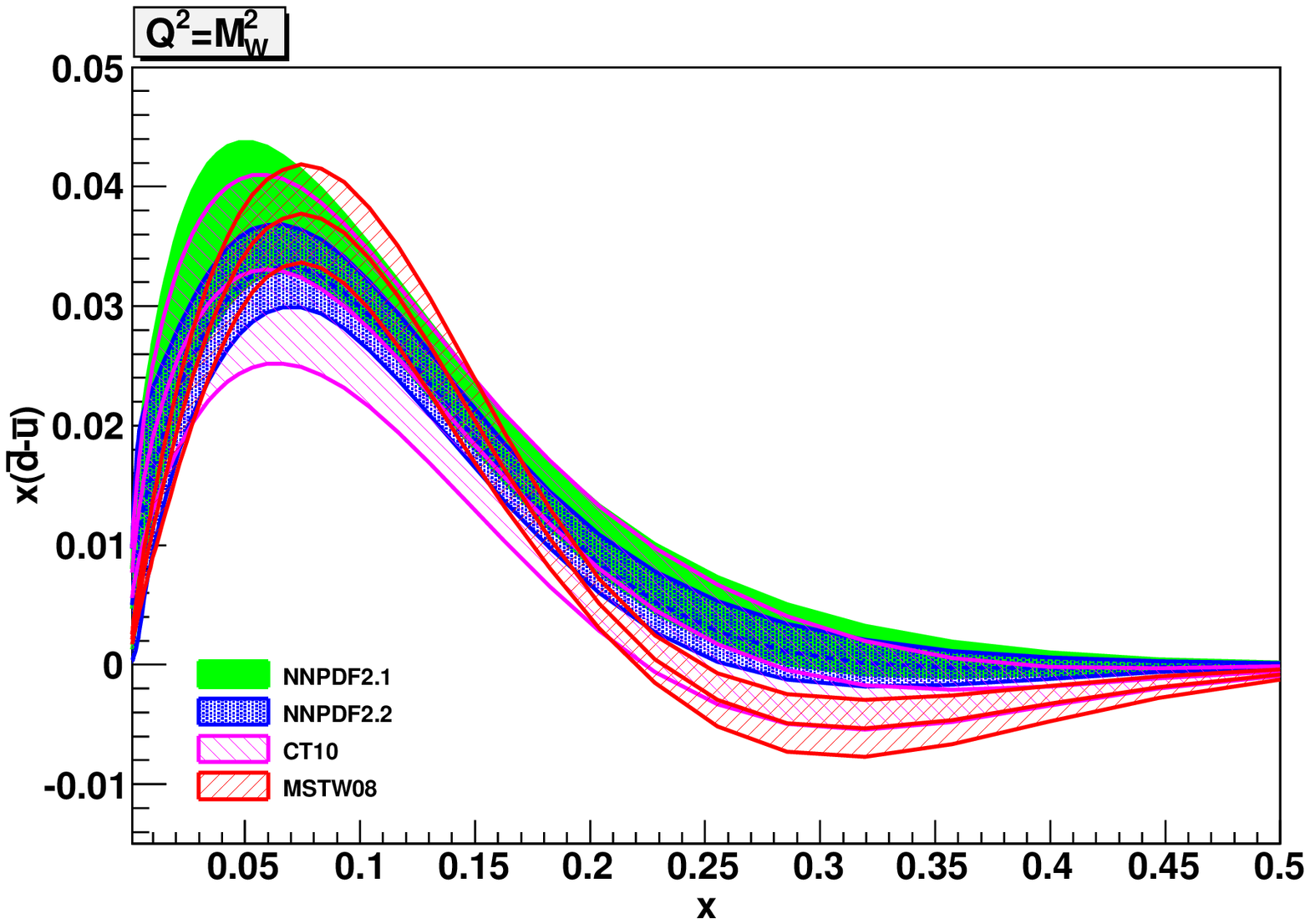}
  \epsfig{width=0.44\textwidth,figure=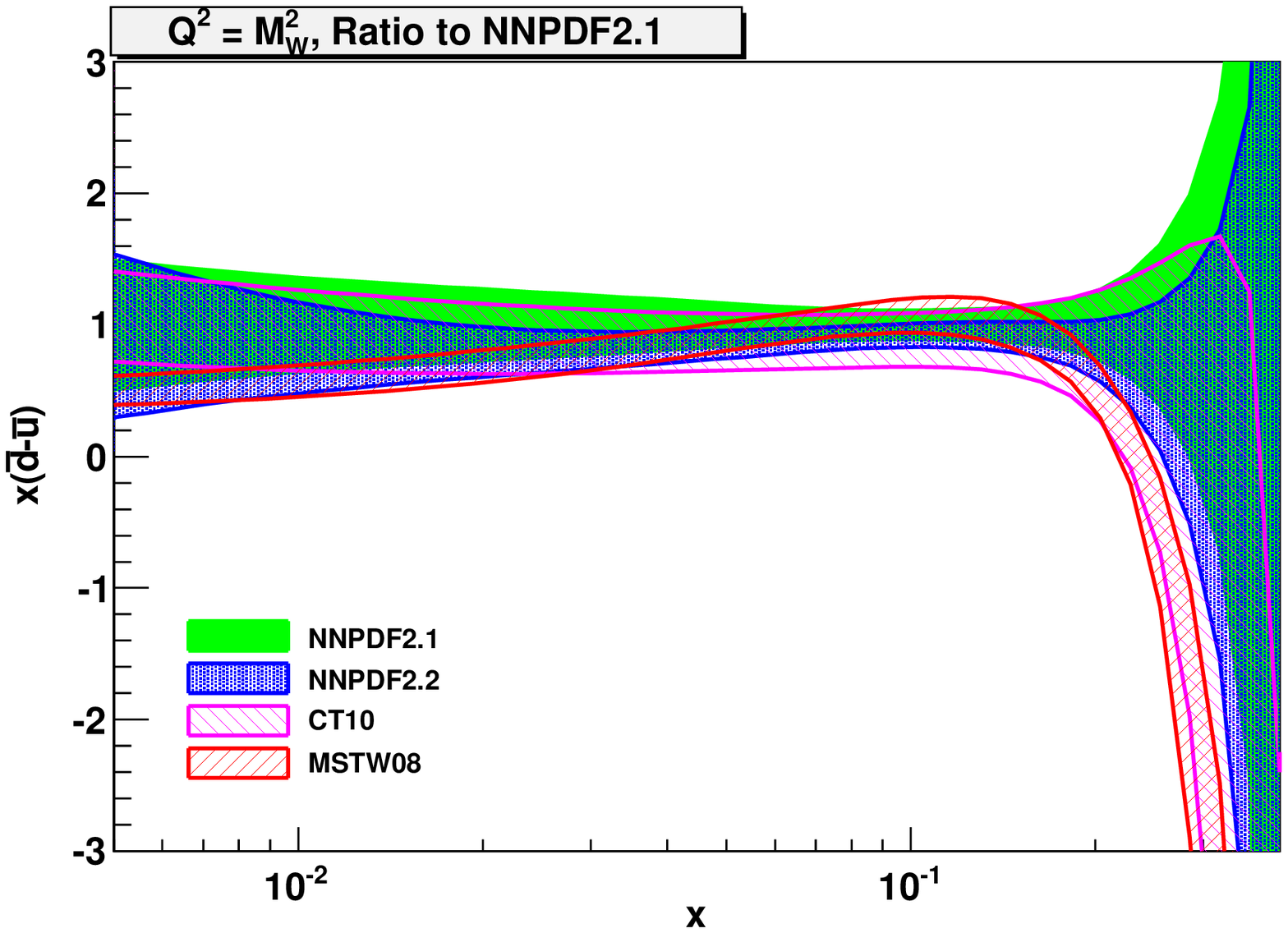}
  \caption{As Fig.~\ref{fig:pdf3-tev+lhc25}, but showing $(\overline{d}-\overline{u})$ at $Q^2=M_W^2$.}
 \label{fig:pdf4-tev+lhc25}
\end{figure}

To this end,
in Figs.~\ref{fig:pdf3-tev+lhc25}
and~\ref{fig:pdf4-tev+lhc25} we compare the $d/u$ and
$(\overline{d}-\overline{u})$ combinations at the scale $Q^2=M_W^2$
obtained in the NNPDF2.1 and MSTW08 NLO global analyses, which
do not include any of the $W$ asymmetry data, the CT10 analysis, which includes only
the D0 data, and the new NNPDF2.2 fit, which also includes the
ATLAS and CMS data. The new data lie in a region of $x$ where the compatibility between the
results obtained by different collaborations is at best
marginal: in particular the $d/u$ ratio given by MSTW08 is too low at large $x$ and too high at medium $x$.
The reduction of uncertainty when going from NNPDF2.1 to
NNPDF2.2 is quite visible: the NNPDF2.2 prediction
should thus be taken as the most reliable at present.
Future LHC data will constrain the light quark PDFs in this region even more.

\section{Conclusions and outlook}
\label{sec:conclusions}

The reweighting method which we have reviewed, re-derived and refined
in this paper is a powerful techinque which enables one both to
preform interesting studies of the statistical properties of parton
distributions viewed as probability distributions in a space of
functions, and to rapidly and effectively include new experimental
information in parton sets.  Coupled to the unweighting method that we
have presented and tested here it allows one to quickly upgrade 
existing Monte Carlo replica PDF sets to new sets which, while
retaining the same format, include new experimental information. 

The method has been used here to construct the NNPDF2.2 NLO PDF set --- the
first PDF set to include LHC data. This will doubtless be the first of many such sets: 
the quantity, quality and diversity of LHC measurements potentially relevant for PDF 
determination is now growing at an impressive rate.

\bigskip
\bigskip
\begin{center}
\rule{5cm}{.1pt}
\end{center}
\bigskip
\bigskip

The NNPDF2.2 NLO  LO PDF set that has been presented in 
Section~\ref{sec-tev+lhc} is available from the NNPDF web site,
\begin{center}
{\bf \url{http://sophia.ecm.ub.es/nnpdf}~}
\end{center}
and will be also available
through the LHAPDF interface~\cite{Bourilkov:2006cj}:

\begin{itemize}

\item NNPDF2.2 NLO, set of $N_{\rm rep}=100$ replicas:\\
{\tt NNPDF22\_nlo\_100.LHgrid}

\end{itemize}

\bigskip
\bigskip
\begin{center}
\rule{5cm}{.1pt}
\end{center}
\bigskip
\bigskip

{\bf\noindent  Acknowledgments \\}

We are especially grateful to John Collins and Jon Pumplin for 
detailed questions and a critique of the reweighting method which
largely stimulated this investigation.
We thank Georgios Daskalakis, Gautier Hamel de
Monchenault, Michele Pioppi, Michael Schmitt and Ping Tan for help
with the LHC $W$ asymmetry data, and Giancarlo Ferrera for help with
the DYNNLO code. 
LDD acknowledges the warm hospitality of the theory group at KMI,
Nagoya, during the final stages of this work. RDB would likewise like to thank
the Discovery Center at the NBI, Copenhagen. 
MU is supported by the Bundesministerium f\"ur Bildung and Forschung (BmBF) of the Federal 
Republic of Germany (project code 05H09PAE).
We would like to acknowledge the use of the computing resources provided 
by the Black Forest Grid Initiative in Freiburg and by the Edinburgh Compute 
and Data Facility (ECDF) (http://www.ecdf.ed.ac.uk/). The ECDF is partially 
supported by the eDIKT initiative (http://www.edikt.org.uk).



\end{document}